\documentclass[10pt,journal,twoside,final]{IEEEtran}
\usepackage{graphicx}
\usepackage{hyperref}
\usepackage{graphicx}
\graphicspath{{./Pictures/}}
\usepackage{tablefootnote}
\usepackage{amsmath}
\usepackage{url}
\usepackage{mathrsfs}
\usepackage{amssymb}
\usepackage{threeparttable}
\usepackage{tikz}

\usepackage{multirow}
\usepackage{array}
\newcolumntype{L}[1]{>{\raggedright\let\newline\\\arraybackslash\hspace{0pt}}m{#1}}
\newcolumntype{C}[1]{>{\centering\let\newline\\\arraybackslash\hspace{0pt}}m{#1}}
\newcolumntype{R}[1]{>{\raggedleft\let\newline\\\arraybackslash\hspace{0pt}}m{#1}}
\usepackage{bigstrut}
\usepackage{cite}

\ifCLASSOPTIONcompsoc
\usepackage[caption=false,font=normalsize,labelfont=sf,textfont=sf,subrefformat=parens,labelformat=parens]{subfig}
\else
\usepackage[caption=false,font=footnotesize,subrefformat=parens,labelformat=parens]{subfig}
\fi
\usepackage{color,xcolor,colortbl}

\newcommand{\mmj}[1]{{\color{black}#1}}

\usepackage{algorithm}
\usepackage{algpseudocode}
\algnewcommand\algorithmicinput{\textbf{INPUT:}}
\algnewcommand\INPUT{\item[\algorithmicinput]}
\algnewcommand{\algorithmicoutput}{\textbf{OUTPUT:}}
\algnewcommand\OUTPUT{\item[\algorithmicoutput]}
\algrenewcommand{\algorithmiccomment}[1]{\hskip3em$\%$ #1}
\usepackage{booktabs}      % For better table formatting
\usepackage{multirow}      % For cells spanning multiple rows
\usepackage{makecell}      % For better header cells
\usepackage{xcolor}        % For cell coloring
\usepackage{colortbl}      % For cell coloring
\usepackage{pdflscape}     % For landscape orientation

\definecolor{lopycolor}{RGB}{180, 200, 255}        % Light blue
\definecolor{mbed1261color}{RGB}{255, 200, 200}    % Light pink
\definecolor{mbed1272color}{RGB}{200, 255, 200}    % Light green
\definecolor{seeeduinocolor}{RGB}{253, 233, 180}    % Sand/yellow
\definecolor{draginocolor}{RGB}{220, 200, 255}     % Light purple

\usepackage{multirow}
\usepackage{colortbl}
\usepackage{xcolor}
\usepackage{graphicx}
\usepackage[margin=0.5in]{geometry}
\begin{document}

\title{Open-Set Radio Frequency Fingerprint Identification with Feature Fusion and Vision Transformer}

\author{Jie~Ma,~\IEEEmembership{Member,~IEEE,}
	Junqing~Zhang,~\IEEEmembership{Senior~Member,~IEEE,}
    Guanxiong~Shen,~\IEEEmembership{Member,~IEEE,}
    Linning~Peng,~\IEEEmembership{Member,~IEEE,}
	Alan~Marshall,~\IEEEmembership{Senior Member,~IEEE}
	
\thanks{Manuscript received xxx; revised xxx; accepted xxx. Date of publication xxx; date of current version xxx.  The work of J. Zhang was supported in part by the UK Engineering and Physical Sciences Research Council (EPSRC) under grant ID EP/Y037197/1, UKRI2109, and UKRI3210, as well as in part by Royal Society Research Grants under grant ID RGS/R1/231435.
For the purpose of open access, the authors have applied a Creative Commons Attribution (CC BY) licence to any Accepted Manuscript version arising.
The review of this paper was coordinated by xxx. 
\textit{(Corresponding author: Junqing Zhang.)}}
\thanks{J. Ma, J.~Zhang, and A.~Marshall are with the School of Computer Science and Informatics, University of Liverpool, Liverpool, L69 3DR, United Kingdom. (email: Jie.Ma2@liverpool.ac.uk; Junqing.Zhang@liverpool.ac.uk; Alan.Marshall@liverpool.ac.uk)}
\thanks{G.~Shen and L.~Peng are with the School of Cyber Science and Engineering, Southeast University, Nanjing, China. (email: gxshen@seu.edu.cn; pengln@seu.edu.cn)}
\thanks{Digital Object Identifier xxx}	
}

\maketitle	

%\IEEEtitleabstractindextext{%
\begin{abstract}
Radio frequency fingerprint identification (RFFI) is a promising device authentication technique for the Internet of Things (IoT) that leverages hardware impairments to uniquely identify devices. However, most existing RFFI systems operate under a closed-set assumption and lack the ability to detect rogue devices, which poses significant security risks.
In this paper, we propose an open-set RFFI approach that jointly performs rogue device detection and legitimate device classification. Specifically, we designed a multi-task RFFI system that integrates an autoencoder (AE) for rogue device detection and a fully connected (FC) layer for classification. To enrich the feature representation, we generated two channel-independent spectrograms (CISs) with distinct window lengths from the I/Q signals, capturing complementary time–frequency characteristics. These CISs were processed in parallel by two Vision Transformer (ViT) networks, and their outputs were fused via feature concatenation to construct a device fingerprint for both tasks.
We built a testbed involving ten legitimate LoRa devices and five rogue devices. We conducted a comprehensive experimental evaluation and compared our approach with several benchmark algorithms. Experimental results show that our fused feature multi-task RFFI consistently outperformed the baselines in both legitimate device classification and rogue device detection.
Our proposed approach achieved an accuracy of 91.1\% and an area under the curve (AUC) of over 0.72 in a realistic wireless scenario where the signal-to-noise ratio (SNR) was 20 to 30~dB.
\end{abstract}


\section{Introduction}

\IEEEPARstart{D}{evice}  authentication is fundamental to safeguarding Internet of Things (IoT) networks against unauthorized access and malicious attacks~\cite{zhang2025tifs}.
Existing authentication schemes rely on 
cryptographic challenge-response methods and software-based addresses.
Although cryptographic methods provide strong security guarantees, sharing secret keys among legitimate devices is challenging in IoT networks~\cite{zhang2020new}.
In addition, software addresses, such as media access control (MAC) addresses, are susceptible to tampering and spoofing attacks. 

Radio frequency fingerprint identification (RFFI) has emerged as a promising physical layer authentication technique for IoT devices~\cite{rfdna}.
RFFI is non-cryptographic and does not rely on a key. Instead, RFFI employs intrinsic hardware impairments, such as in-phase and quadrature (I/Q) imbalance, carrier frequency offset (CFO), and power amplifier nonlinearity. These impairments are unintentionally introduced during the manufacturing process~\cite{zhang2021tifs} and are unique and stable. Hence, they can act as unique physical fingerprints and be exploited for device authentication~\cite{he2020cooperative, al2020infocom}.
RFFI can be implemented on the receiver side and does not impose additional overhead on resource-constrained devices, making it particularly well suited for IoT applications~\cite{shen2022towards}. 

Deep learning (DL)-based approaches, such as convolutional neural networks (CNNs), long short-term memory (LSTM) networks, and transformers are prevalent in the RFFI community thanks to their superior capability for automatic feature extraction~\cite{zhao2024explanation, shen2023towards, yang2023infocom, gan2024tim,zhang2023real,li2025jsac}. 
The vision transformer (ViT), proposed in~\cite{dosovitskiy2020vit}, adapts the transformer architecture to visual recognition tasks by representing images as patch tokens. 
% It exhibits strong modeling capacity and scalability, achieving state-of-the-art performance across multiple benchmark datasets~\cite{liu2021swin, chen2021crossvit, song2023image}. 
Due to its superior feature extraction capability, ViT achieves competitive performance across multiple benchmark datasets and has attracted widespread attention~\cite{chen2021crossvit}. 
% Its application has extended to wireless communication domains recently, including modulation classification~\cite{kong2023transformer} and RFFI~\cite{deng2023lightweight, parpart2024transformer, zhang2024self}.
ViT has recently been applied to RFFI, demonstrating promising performance~\cite{deng2023lightweight, parpart2024transformer, yang2025tmc}. 
However, these works directly utilized raw I/Q signals as input and did not account for the impact of wireless channel effects, which limits system performance.

Multi-channel spectrogram representations have been investigated in wireless signal recognition to enhance feature diversity~\cite{tang2021sensor, wang2025detection, zeng2024, aerosp2026, yang2023slnet, ding2025enhanced}.
In~\cite{tang2021sensor}, the short-time Fourier transform (STFT), smoothed pseudo Wigner-Ville distribution (SPWVD), and reduced interference distribution with a Hanning kernel (RIDHK) were used to generate spectrograms from radar signals for human activity recognition. 
Wang \textit{et al.}~\cite{wang2025detection} constructed a three-channel input by combining the amplitude and phase of the STFT-based spectrogram with the wavelet transform (WT) representation.
Zeng \textit{et al.}~\cite{zeng2024} utilized coefficients of the fast Fourier transform (FFT) and STFT for feature extraction.
However, the choice of the spectrograms should be further clarified, particularly regarding how they complement each other and contribute to the overall performance. 
Multi-channel spectrograms generated using different window lengths were investigated in \cite{yang2023slnet, ding2025enhanced}, demonstrating their effectiveness in wireless sensing.
Nevertheless, their performance in the RFFI context has not been evaluated.

Fusion methods offer enhanced performance by leveraging more informative inputs from multiple sources or methods~\cite{zhang2024multisource, fusion2019, fusion2022iotj, he2025adaptive, shen2023towards,he2020cooperative, xie2025infocom}.
These fusion strategies can be categorized into decision fusion and feature-level fusion~\cite{he2025adaptive}.
Decision fusion aggregates the output probabilities from multiple sources~\cite{he2020cooperative, xie2025infocom}. For instance, Shen \textit{et al.}~\cite{shen2023towards} propose an adaptive soft fusion framework that integrates the prediction outputs from multiple receivers to enable collaborative identification, leading to improved performance. 
However, decision fusion operates at the output level, relying on prediction probabilities, which suffer from information loss.
Feature fusion combines different extracted features, and the final decision is made based on the fused feature~\cite{fusion2022iotj,zhang2024multisource}.
For example, Zhang \textit{et al.}~\cite{zhang2024multisource} present a multisource heterogeneous specific emitter identification (MH-SEI) method for RFF extraction and an attention-based fusion module for feature integration, demonstrating robustness in scenarios involving heterogeneous receivers.

Most existing RFFI methods are designed for closed-set classification, assuming that all devices encountered during inference are devices previously seen in training~\cite{zhang2021tifs, shen2024federated,shen2023towards, shen2021radio}. 
As a result, such approaches fail to detect rogue devices, thereby introducing potential security vulnerabilities.
Specifically, a rogue device will be misclassified as a legitimate device with a similar RFF, enabling unauthorized access to the system.
It undermines the security of the system and opens the door for attackers to carry out malicious activities. 
In wireless communication scenarios, rogue devices are likely to appear during inference due to the open nature of the wireless medium~\cite{shen2022towards}.
To this end, it is essential for RFFI systems to handle open-set scenarios, enabling not only device classification but also the identification of rogue devices.

Existing approaches to open-set recognition can generally be categorized into three types: distance-based, score-based, and reconstruction-based methods~\cite{xi2025}.
% Distance-based approaches, such as KNN~\cite{shen2022towards}, predicts an input with its $K$ nearest neighbors in the training dataset. 
Distance-based approaches are commonly adopted to evaluate the similarity between test samples and known classes in a learned feature space. 
For instance, K-nearest neighbors (KNN)~\cite{shen2022towards} identifies an input by considering its $K$ nearest neighbors in the training dataset.
Shen \textit{et al.}~\cite{shen2022towards} introduce an RFFI framework with KNN to address open-set scenarios. However, the computational cost is overlooked, as KNN requires comparing the inference input against all training samples, resulting in an overhead that scales linearly with both the size of the dataset and the number of classes~\cite{knn2021}. 
Score-based methods estimate the confidence that an input belongs to known classes. For example, OpenMax~\cite{openmax2016cvpr} extends the Softmax output from $N$ known classes to $N+1$ classes by modeling mean activation vectors (MAVs) with Weibull distributions, enabling the detection of unknown inputs.
Guo \textit{et al.}~\cite{guo2024tccn} leverage a generative adversarial network (GAN) and OpenMax to accomplish anomaly detection alongside closed-set classification. 
\mmj{More recently, extended logit normalization (ELogitNorm) was proposed in~\cite{ding2026enhancing} and has also demonstrated strong open-set recognition performance.}
Reconstruction-based methods, such as autoencoders (AEs)~\cite{hanna2021tccn}, are trained to reconstruct input signals, while rogue devices generally produce higher reconstruction errors, enabling outlier detection. 
However, an AE only provides detection capabilities and does not support classification~\cite{hanna2021tccn, wang2024design}. 
It is worth noting that, although several studies have investigated open-set scenarios in RFFI, the influence of real-world environmental variations is often neglected, and performance under varying signal-to-noise ratio (SNR) levels and deployment positions remains insufficiently evaluated.
% The reconstruction error serves as an anomaly indicator, with higher reconstruction values occurring when the rogue device is present.
% As the AE consists of an encoder–decoder architecture, the encoder maps the input to a lower-dimensional latent space, called the bottleneck. This bottleneck retains limited and prominent features for reconstruction. 
% The decoder aims to reconstruct the input data from the latent bottleneck representation.
% thereby capturing the underlying distribution of the normal training data

In order to address the aforementioned challenges,  this paper proposes a multi-task RFFI system that utilizes feature fusion and ViT.
% which integrates an AE and a fully connected layer to enable device identification and classification, respectively.
Specifically, two complementary signal representations are derived from the same I/Q signal but generated with different parameter configurations, resulting in spectrograms with distinct time-frequency resolutions.
The two representations are independently fed into two ViT models, and their resulting features are subsequently concatenated as a fused feature. 
The fused feature is fed to a multi-task module, where an AE is employed for device identification and a fully connected (FC) layer performs device classification.
We built a testbed consisting of ten legitimate LoRa devices and five rogue devices. A comprehensive experimental evaluation was conducted.
The contributions of this paper are summarized as follows.

\begin{itemize}
		\item Two channel-independent spectrograms (CISs) with complementary information are employed as the inputs. They are generated using different window lengths, which helps overcome the temporal-spectral resolution trade-off introduced during IQ-to-spectrogram transformation.
        They are independently processed by two ViT models to extract features, which are concatenated to form a fused feature.
\item A ViT adapted to CISs is proposed in this work. The proposed ViT architecture is specifically designed to accommodate the properties of time-frequency signals, enabling more effective extraction of features.
\item To handle open-set RFFI, a multi-task module is designed, combining an AE for identification and an FC layer for classification. The AE identifies devices as legitimate or rogue by evaluating reconstruction error, and the FC layer is used to predict the device's label. 
\item Extensive experiments are carried out with different simulated SNR levels, achieved by adding additive white Gaussian noise (AWGN) to the captured signals. The evaluation demonstrated the improvement of our approach. For instance, at an SNR of 15 dB, the fused feature achieves approximately a 10\% improvement in accuracy over the single-feature configuration. Compared to the traditional ViT and CNN models, the proposed approach demonstrates higher classification accuracy and improved overall rogue device detection performance.
\mmj{Finally, compared to the benchmark methods OpenMax, KNN, and ELogitNorm, the proposed approach achieved the best identification performance in the majority of evaluated cases.}
\item Extensive experiments are also conducted across diverse real-world scenarios to evaluate the effectiveness of the proposed approach. Four test datasets are collected, each corresponding to signal conditions with SNR ranges greater than 40 dB, 30–40 dB, 20-30 dB and 10–20 dB, respectively, to assess the effectiveness of the proposed approach. In a realistic wireless scenario with SNRs ranging from 20 dB to 30 dB, the proposed RFFI system achieves a classification accuracy of 91.1\%, while maintaining AUC values exceeding 0.72 across a variety of rogue devices.
\end{itemize}

The rest of the paper is organized as follows. 
Section~\ref{sec:selfsystem} provides a system overview.
Section~\ref{sec:dualinput_featureextractor} presents the fused feature extraction, while Section~\ref{sec:our_nn} introduces the multi-task authentication module, which jointly performs device identification and classification. 
Section~\ref{sec:setupandbench} describes the experimental setup and the open-set benchmark algorithms.
Sections~\ref{sec:experiment_simulated} and~\ref{sec:expreal} evaluate the identification and classification performance of the proposed methods under simulated SNR conditions and in real-world scenarios, respectively.
Section~\ref{sec:conclusion} concludes the paper.

\section{System Overview}\label{sec:selfsystem}
\begin{figure}[!t]
	\begin{center}
		\includegraphics[width = 3.4in]{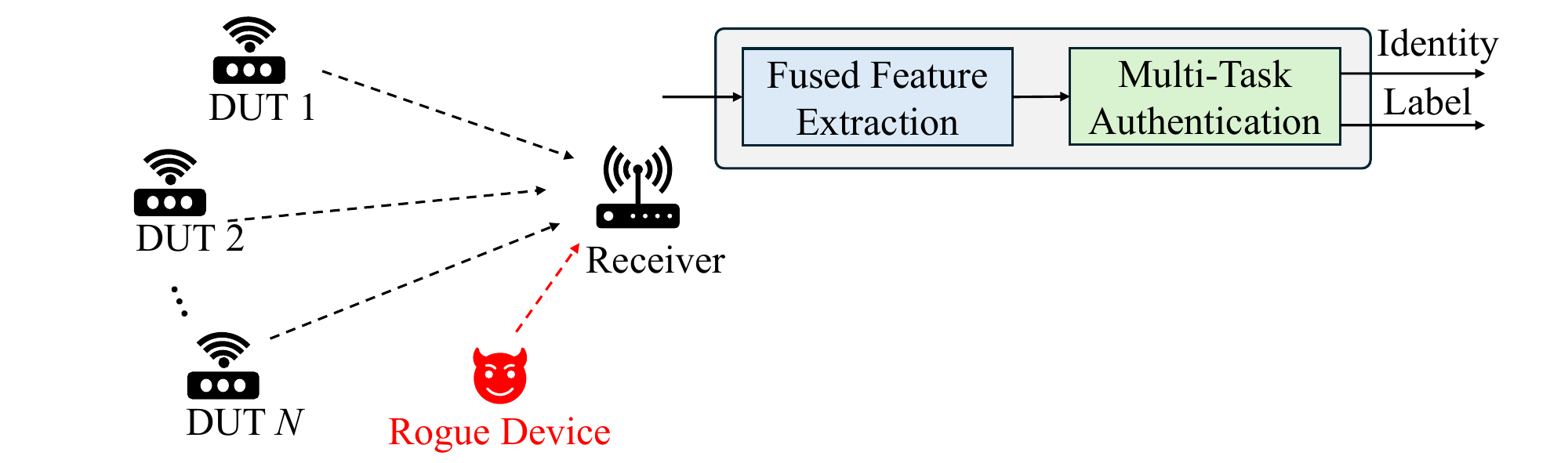}
		\caption{System model of an open-set RFFI system with feature fusion.}
		\label{fig:RFFIsys}
	\end{center}	
\end{figure}

As shown in Fig.~\ref{fig:RFFIsys}, this paper considers an RFFI system involving $N$ legitimate devices under test (DUTs) as well as rogue devices, constituting an open-set recognition scenario. Therefore, we designed a multi-task authentication module to perform joint identification and classification. In addition, we also implemented fused feature extraction to enhance the open-set recognition performance. The detailed algorithm design is described below, involving training and inference stages.

\begin{figure}[!t]
	\begin{center}
		\includegraphics[width = 3.4in]{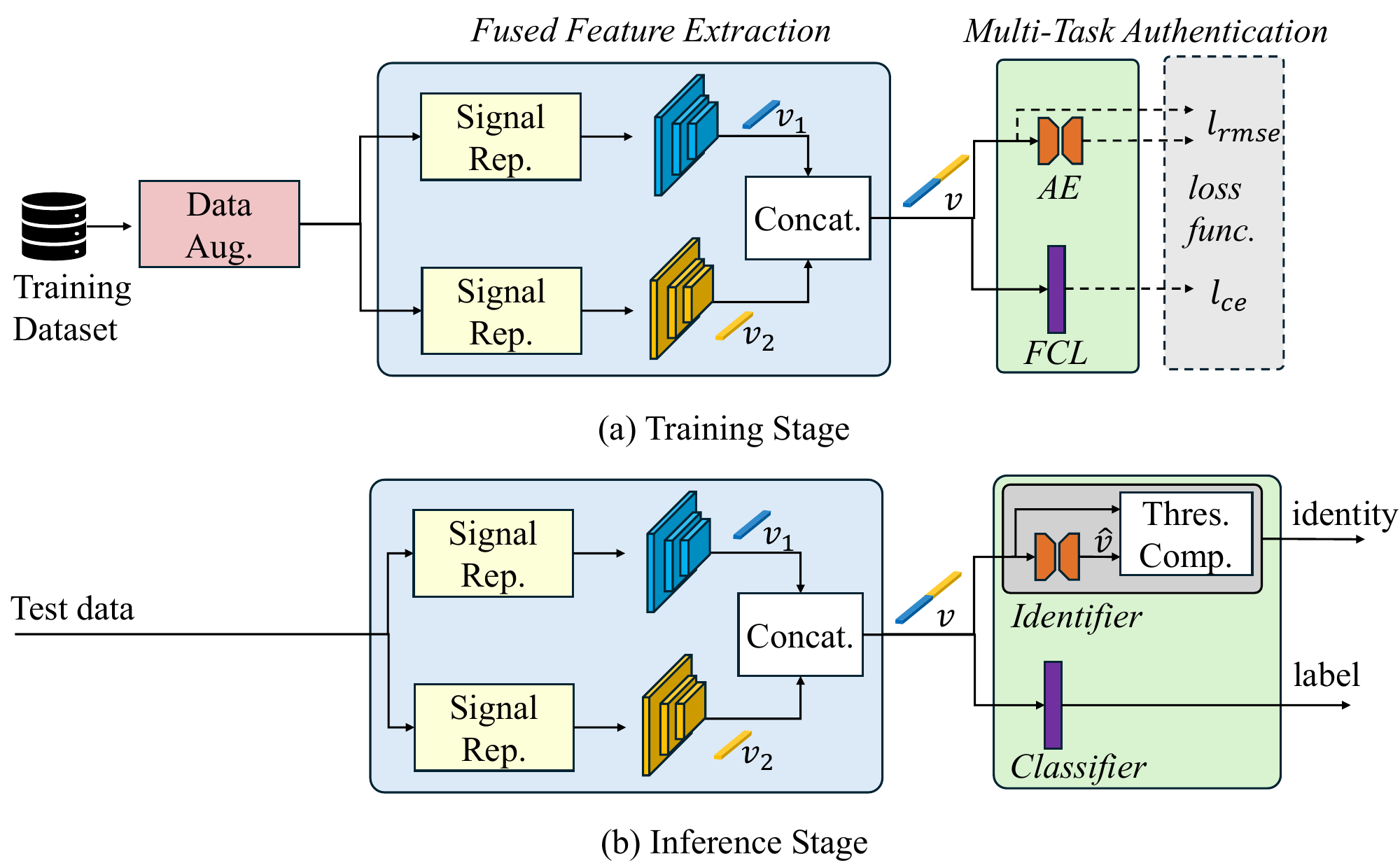}
		\caption{Overview of the proposed multi-task RFFI system. An AE and an FC layer are integrated with two feature extractors to construct the neural network. (a) Training stage. (b) Inference stage. }
		\label{fig:rffisystem}
	\end{center}	
\end{figure}

\subsection{Training}\label{sec:training}
As shown in Fig.~\ref{fig:rffisystem}(a), the training dataset is augmented, and each signal is transformed into two distinct signal representations that serve as inputs to the neural network.
The proposed network architecture employs two parallel feature extractors, each dedicated to a signal representation. 
The extracted features are subsequently concatenated and forwarded to a multi-task module, which consists of an AE for device identification and an FC layer for device classification.

\subsubsection{Fused Feature Extraction}\label{sec:signal}
The proposed system employs two distinct signal representations and corresponding feature extractors, from which a fused feature vector is derived.
Specifically, the \textbf{Signal Representation} module converts the I/Q samples to a different domain. The \textbf{Feature Extractor} is designed to capture discriminative features from the input, and its architecture is adapted to the specific representation. Details on the fused feature extraction are discussed in Section~\ref{sec:dualinput_featureextractor}.

\subsubsection{Multi-task Authentication}
The fused feature is subsequently utilized by a multi-task module comprising an AE and an FC layer, which serve as the identifier and classifier, respectively.
The \textbf{AE} is trained to minimize the reconstruction error, with the goal of accurately reconstructing the input fused feature.
The \textbf{FC Layer} aims to learn a mapping from the fused feature to its associated device label.
More details of the identifier and classifier will be provided in Section~\ref{sec:our_nn}.

\subsection{Inference}\label{sec:inference}
During the inference stage, as illustrated in Fig.~\ref{fig:rffisystem}(b), the test data are processed through the same fused feature extraction and multi-task authentication module used in training. 
The system will output the device identity and classification label.

\section{Fused Feature Extraction}\label{sec:dualinput_featureextractor}

\subsection{LoRa Signals}
LoRa is a widely adopted technology optimized for IoT applications.
This paper uses LoRa as a case study to demonstrate the implementation of the proposed RFFI system.

\begin{figure}[!t]
	\begin{center}
		\includegraphics[width = 3.1in]{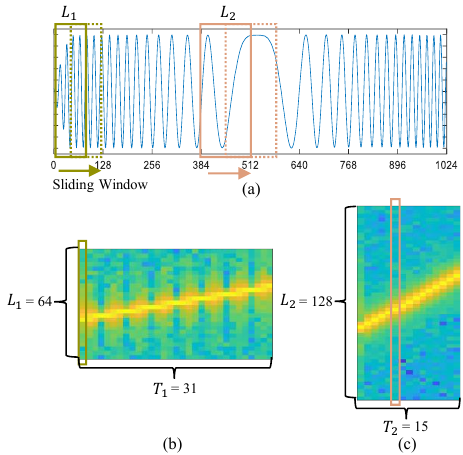}
		\caption{Signal representation of a preamble. (a) Real part (I) of the time-domain signal. (b) Spectrogram, $L_1 = 64$. (c) Spectrogram, $L_2 = 128$.}
		\label{fig:sp}
	\end{center}	
\end{figure}

LoRa employs chirp spread spectrum (CSS) modulation with continuously varying instantaneous frequency, resulting in non-stationary signals.
Each LoRa packet begins with a sequence of preambles, which are essential for facilitating synchronization at the receiver. 
These preambles are consistent and independent of the packet’s payload, making them particularly well-suited for RFFI. 
Due to this property, the proposed system leverages the preamble segment of the LoRa signal to train the DL model. 
The in-phase (I) component of the time-domain baseband waveform for a preamble is shown in Fig.~\ref{fig:sp}(a).

\subsection{Data Augmentation}\label{sec:dataaug}
Data augmentation is employed in the training stage, which enhances the diversity of the training dataset.
The tapped delay line (TDL) model and Doppler shift can be incorporated to simulate diverse propagation conditions. 
In addition, various levels of AWGN can be introduced to the original signals. The detailed design methodology can be found in~\cite{shen2022towards,length2023}.

\subsection{Signal Representation}

\subsubsection{Short-Time Fourier Transform (STFT)}\label{sec:STFT}
STFT is a widely used technique for time-frequency analysis. 
As shown in Fig.~\ref{fig:sp}(a), STFT applies a sliding window function $w[n]$ to signal $x[n]$ and performs a Fourier transform at each time step $t$. 
The window slides across the signal with a step size of $L \cdot (1-\alpha)$, where $L$ is the window length and $\alpha \in[0,1)$  denotes the overlap ratio between adjacent columns.
STFT is applied to the time-domain signal $x[n]$ to obtain its spectrogram $X$, given as 
\begin{equation}
\begin{aligned}
X_{l, t}=\sum_{n=0}^{N_{tot}-1} \ x[n] w[n-t \cdot L \cdot (1-\alpha)] e^{-j 2 \pi \frac{l}{L} n}
\\
\text { for } l=1,2, \ldots, L \text { and } t=1,2, \ldots, T,
\end{aligned}
\label{eq:stft}  
\end{equation}
where $X_{l, t}$ denotes the $(l,t)$-th element of the STFT complex matrix $X$, and $N_{tot}$ is the length of the signal.
When the time-domain signal $x[n]$ consists of $N_p$ LoRa preambles, the number of columns in matrix $X$ is calculated as
\begin{equation}
T= \left\lfloor\frac{N_p \cdot \frac{2^{SF}}{B} \cdot f_s- L}{L \cdot (1-\alpha)}\right\rfloor+1,
\label{eq:time}  
\end{equation}
where $SF$ is the spreading factor, $B$ is the bandwidth, and $f_s$ is the sampling frequency.

Based on (\ref{eq:time}), it is evident that a trade-off exists between time and frequency resolution~\cite{stft2021tradeoff}, which is directly influenced by the selection of window length, $L$.
% Specifically, a shorter window provides finer temporal resolution at the expense of frequency resolution, whereas a longer window yields better frequency resolution at the expense of temporal details.
Specifically, a shorter window spans a smaller number of signal samples, which enhances the ability to observe temporal changes. However, it leads to a broader frequency response, thereby degrading frequency resolution.
In contrast, a longer window increases the time-domain observation interval, resulting in finer frequency discrimination. 

Given this trade-off, instead of relying on a single window length, we adopt two different window lengths to extract complementary time–frequency characteristics of the signal, which provides a more comprehensive representation of the underlying signal characteristics.
To illustrate, Fig.~\ref{fig:sp}(b) and Fig.~\ref{fig:sp}(c) show the results of applying STFT to the IQ signal segment with window lengths of $L_1=64$ and $L_2=128$, respectively.
In our experimental setting, $SF$ is 7, $B$ is 125 kHz, $f_s$ is 1 MHz, and $\alpha$ is 0.5. 
Therefore, when $N_p$ is 1, according to (\ref{eq:time}), $T_1$ is 31 for $L_1=64$ and $T_2 =$ 15 for $L_2=128$.

\subsubsection{Channel-Independent Spectrogram}\label{sec:CIS}

\begin{figure}[!t]
	\begin{center}
		\includegraphics[width = 3.2in]{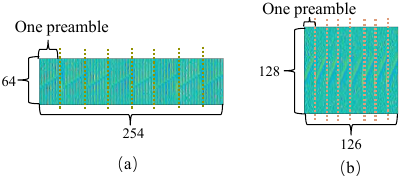}
		\caption{CIS generated from eight LoRa preambles. (a) $L$ = 64. (b) $L$ = 128.}
		\label{fig:signal}
	\end{center}	
\end{figure}

Channel variations are inevitable in RFFI and are reflected in the spectrogram.
The work in~\cite{shen2022towards} designed the CIS to eliminate the channel effects residing in the spectrogram. Specifically, denote the $t$-th column of the spectrogram as $\mathbf{X}_t=\left[X_{1, t}, X_{2, t} \cdots X_{L, t}\right]^\top$.
Given that the time interval between two adjacent signal segments is extremely short, at the microsecond level, it is reasonable to assume that the wireless channel remains constant. Hence, the channel effect can be eliminated by calculating $\mathbf{S}_t = \mathbf{X}_{t+1}/\mathbf{X}_{t}$. The CIS can be constructed as
\begin{align}
    \mathbf{S} = \{\mathbf{S}_{1}, ..., \mathbf{S}_{t},..., \mathbf{S}_{T-1} \}.
\label{eq:cis}  
\end{align}
It exhibits robust and consistent performance across both stationary and dynamic scenarios. 
The complete theoretical derivation and detailed performance evaluations of CIS can be found in~\cite{shen2022towards}.

The CIS is the signal representation employed in this paper. Specifically, every eight preambles are treated as one sample, transformed into a CIS, and then fed into the neural network for feature extraction. 
For a clearer illustration, Fig.~\ref{fig:signal} presents the CISs obtained with $L_1=64$ and $L_2=128$, where the number of columns can be calculated by (\ref{eq:time}) and (\ref{eq:cis}).

\subsection{ViT-Based Feature Extractor}\label{sec:extractor}
\begin{figure}[!t]
	\begin{center}
		\includegraphics[width = 3.4in]{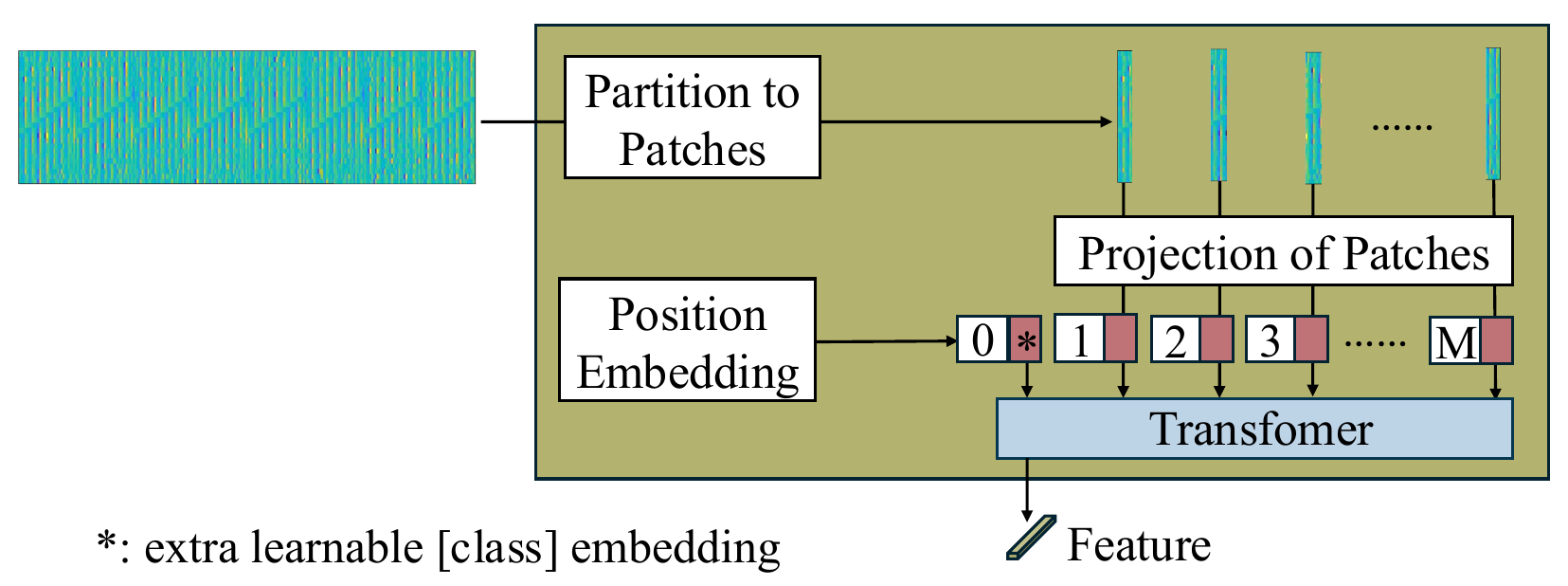}
		\caption{The architecture of ViT used in the proposed RFFI system. The CIS was generated with $L=64$.}
		\label{fig:vit}
	\end{center}	
\end{figure}

Since a CIS can be regarded as a 2D image and ViT performs well in image classification tasks, we adopted ViT as the feature extractor. 
The architecture of the proposed ViT model, designed for CIS, is illustrated in Fig.~\ref{fig:vit}.

\subsubsection{Partitioning into Patches}
The CIS is partitioned along the time axis. 
Inputting patches in this form, i.e., as the frequency components contained within specific temporal segments, facilitates the network's ability to extract temporal and frequency information from the CIS more effectively.

The CIS is denoted as $\mathbf{S} \in \mathbb{R}^{L \times T}$, and it is partitioned along the time axis into non-overlapping patches. 
Each patch consists of two consecutive columns and their corresponding frequency components, resulting in a patch tensor of shape $\mathbf{S_p} \in \mathbb{R}^{L \times 2}$.
Accordingly, the total number of patches of this CIS is given by $M=T/2$. 
This patch-based representation enables the model to effectively learn local temporal and spectral patterns from the spectrogram input.

% which also serves as the effective input sequence length for the transformer. 
\subsubsection{Projection of Patches}
Each 2D patch is first flattened and then projected into a $G$-dimensional embedding space through a trainable FC layer. In this work, we set $G$ to 1024.

\subsubsection{Position Embedding}

Positional embeddings are incorporated into the patch embeddings to retain positional information. Additionally, a learnable token ($\mathbf{z}_0^0=\mathbf{S}_{\text {class }}$) will be prepended to the sequence of patch embeddings. 
The resulting vectors can be expressed as
\begin{equation}
\begin{aligned}
\mathbf{z}_0 & =\left[\mathbf{S}_{\text {class }} ; \mathbf{S_p^1} \mathbf{E} ; \mathbf{S_p^2} \mathbf{E} ; \cdots ; \mathbf{S_p^M} \mathbf{E}\right]+\mathbf{E}_{p o s},\\
&\mathbf{E} \in \mathbb{R}^{\left(L \cdot 2\right) \times G}, \mathbf{E}_{p o s} \in \mathbb{R}^{(M+1) \times G}. 
\label{eq:imagepatch}
\end{aligned}
\end{equation}
The combined sequence of embedding vectors is then provided as input to the transformer.

\subsubsection{Transformer}
\begin{figure}[!t]
	\begin{center}
		\includegraphics[width = 3.4in]{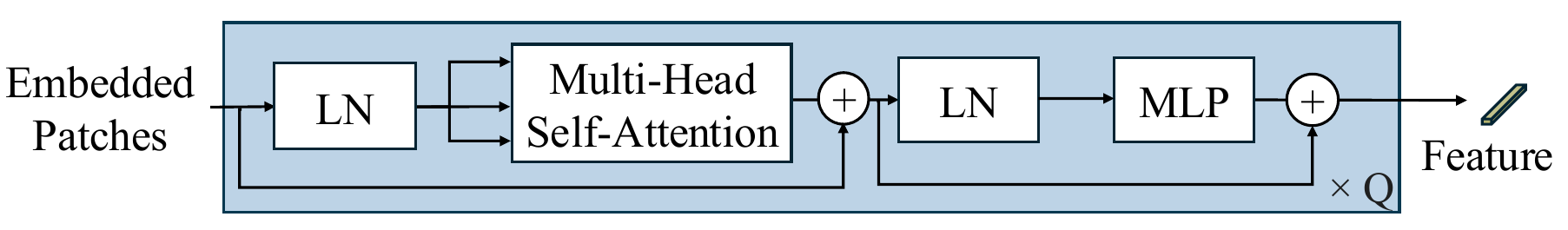}
		\caption{The architecture of the transformer used in ViT.}
		\label{fig:transformer}
	\end{center}	
\end{figure}

The transformer's architecture is shown in Fig.~\ref{fig:transformer}. The architecture comprises $Q$ layers of normalization, multi-head self-attention (MSA), and multilayer perceptron (MLP) blocks. The processes are given as
\begin{align}
\mathbf{z}_{q}^{\prime} &=\operatorname{MSA}\left(\operatorname{LN}\left(\mathbf{z}_{q-1}\right)\right)+\mathbf{z}_{q-1}, q=1 \ldots Q,
\label{eq:msa}\\    
\mathbf{z}_{q} & =\operatorname{MLP}\left(\operatorname{LN}\left(\mathbf{z}_{q}^{\prime}\right)\right)+\mathbf{z}_{q}^{\prime}, \; q=1 \ldots Q.
\label{eq:mlp}
\end{align}
Each block is preceded by layer normalization (LN) and followed by a residual connection. 
The MLP module comprises two FC layers with a Gaussian error linear unit (GELU) activation function applied between them. 

The output of the transformer corresponding to the classification token is denoted as $\mathbf{z}_Q^0$, which can serve as the feature of the input. 
$Q$ is set to 4 in our work.
This feature is then employed as a component in the subsequent fusion stage.

\subsection{Feature Fusion}
A fused feature is obtained from spectrograms generated using different window lengths, enabling complementary integration of rich temporal and frequency information.

The features extracted by the two ViTs are denoted as $v_1$ and $v_2$, respectively. 
The two extracted features from the ViTs are concatenated into a fused feature $v$, given as
\begin{align}
v = \text concat{(v_1, v_2)},
\end{align}
where $concat(\cdot)$ is the operation of appending the head of $v_2$ directly to the end of $v_1$.
In our implementation, the dimension of each individual vector is 1024, and the fused vector $v$ is of dimension 2048. 
The concatenation introduces no additional learnable parameters, ensuring computational efficiency while enhancing feature richness.

\section{Multi-task Authentication}\label{sec:our_nn}
The fused feature is processed by the multi-task module comprising an AE for identification and an FC layer for classification, as shown in Fig.~\ref{fig:rffisystem}.

\subsection{Training Stage}
\subsubsection{AE for Identification}\label{sec:ae}
The AE, composed of an encoder and a decoder, is employed as an identifier to determine whether a device is legitimate or rogue.
The encoder $\mathcal{E}(\cdot)$ compresses the fused feature into a bottleneck representation, and the decoder $\mathcal{D}(\cdot)$ then reconstructs it by expanding the bottleneck back to the original size, formulated as
\begin{equation}
\hat{v} = \mathcal{D}(\mathcal{E}(v)).
\end{equation}
The reconstruction error $e$ between the fused feature $v$ and its reconstruction $\hat{v}$ serves as the anomaly score for identification.
The root mean squared error (RMSE)~\cite{rmse2020} is adopted to calculate the reconstruction error $e$ between $v$ and $\hat{v}$, given as
\begin{equation}
e = \sqrt{(v - \hat{v})^2}.
\end{equation}

The fused feature with a size of 2048 is projected into a 256-dimensional bottleneck representation via two FC layers with an intermediate dimensionality of 1024. Then it is reconstructed to the original size of 2048 through a symmetric two-layer decoder, with an intermediate dimensionality of 1024.

\subsubsection{Fully Connected Layer for Classification}\label{sec:clf_linear}
Fig.~\ref{fig:rffisystem} shows that an FC layer $\mathcal{C}_{dense}(\cdot)$ is utilized as the classification module. This layer operates on the fused feature $v$ and projects it to a vector whose dimension corresponds to the number of legitimate devices. It can be given as
\begin{equation}
    % \hat{y} = \mathcal{N}(S_1,S_2;\theta),
% \hat{y} = \mathcal{N}_{\mathrm{dense}}(\mathbf{v})
\mathbf{\hat{p}} = \mathcal{C}_{dense}(v),
\end{equation}
where $\mathbf{\hat{p}}$ is a list of probabilities over all the classes.

\subsubsection{Loss Function}\label{sec:lossfunction}
The neural network is optimized using the reconstruction loss, $\ell_{\textit {rmse}}$, and the cross-entropy loss, $\ell_{\textit {ce}}$. 
The AE is trained using $\ell_{\textit {rmse}}$, given as
\begin{equation}
\ell_{\textit {rmse}} = \sqrt{\left(v-\hat{v}\right)^2},
\end{equation}
The classifier is updated based on $\ell_{\textit {ce}}$, expressed by
\begin{equation}
\ell_{\textit {ce}}=-\sum_{n=1}^N {y}_n \log \left(\hat{p}_n\right).
\end{equation}
$\hat{p}_n$ represents the $n$-th element of the predicted probabilities $\mathbf{\hat{p}}$, and $y_n$ is the corresponding ground truth.

The total loss function $\ell$ is given as
% \begin{equation}
% \mathcal{L} = \mathcal{L}_{\textit{rmse}} + \mathcal{L}_{\textit{ce}}.
% \end{equation}
\begin{equation}
\ell = \ell_{\textit{rmse}} + \ell_{\textit{ce}}.
\end{equation}
By jointly optimizing these losses, the neural network is trained in a multi-task manner to simultaneously perform identification and classification. %The two loss terms are optimized with equal weight, which was empirically found to provide the best overall performance.

\subsection{Inference Stage}
During the inference stage, the test data will also be transformed into CISs with $L_1 = 64$ and $L_2 = 128$, and subsequently fed into the trained model.

For device identification, following the AE framework, the reconstruction error $e$ is compared to a threshold $\tau$, defined as
\begin{equation}
\phi = \begin{cases}\text { legitimate, } & e \leq \tau \\ \text { rogue, } & e>\tau\end{cases}.
\end{equation}
$\phi$ represents the identification result.
% It $e$ a high reconstruction error indicates that the received signal originates from a rogue device, while a low reconstruction error suggests it is likely from a legitimate device. Therefore, 
% the $e$ will compare to the 

For classification, an FC layer is applied to produce a probability list $\mathbf{\hat{p}}$ over all legitimate classes, and the class corresponding to the highest probability is selected as the predicted label.

\section{Experimental Setup and Benchmarks}\label{sec:setupandbench}

\subsection{Experimental Setup}\label{sec:setup}
\subsubsection{Devices}\label{sec:setup_dev}
\begin{figure}[!t]
	\centering
	\subfloat[]{\includegraphics[width=1.1in]{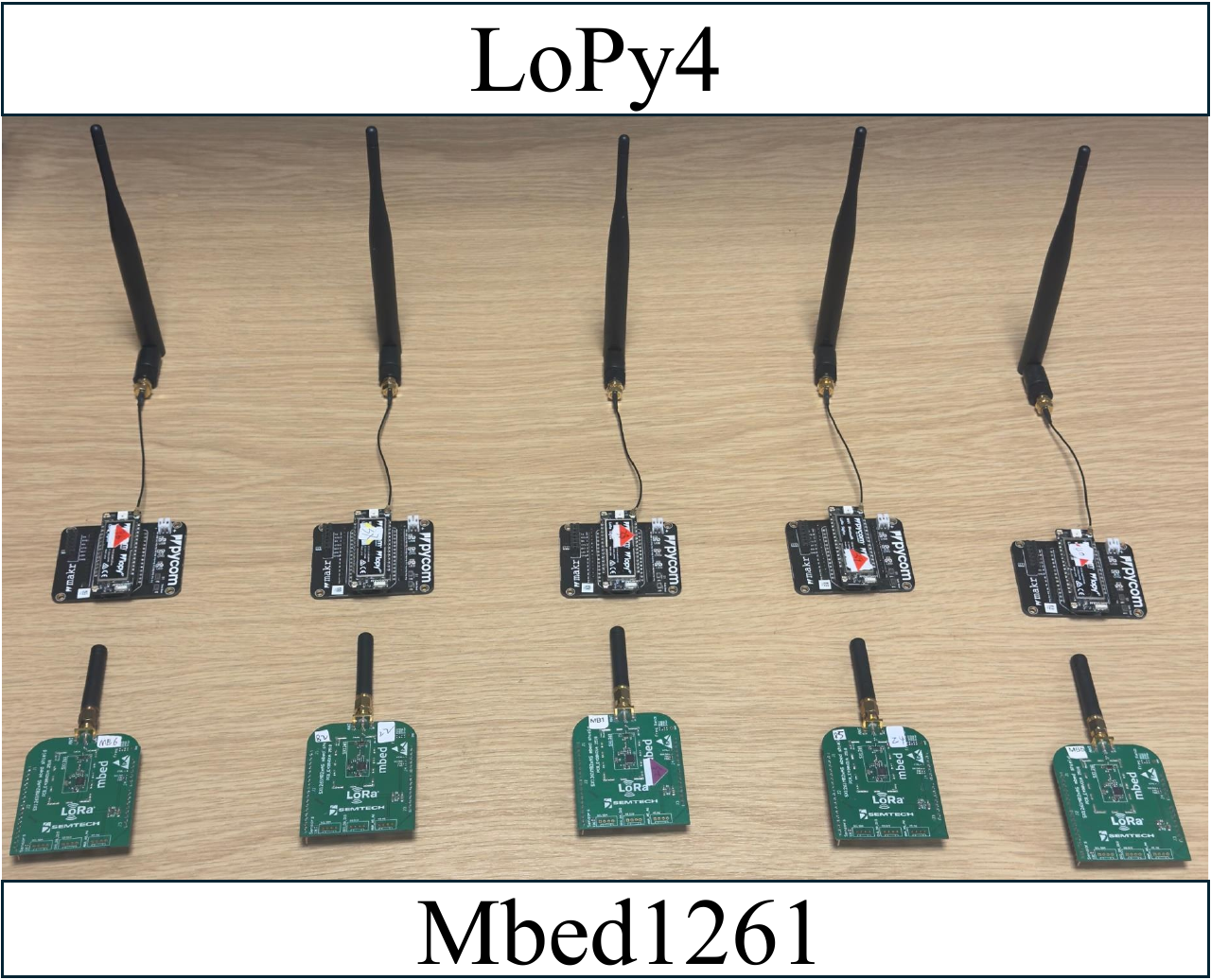}
		\label{fig:legit}}
	\subfloat[]{\includegraphics[width=1.1in]{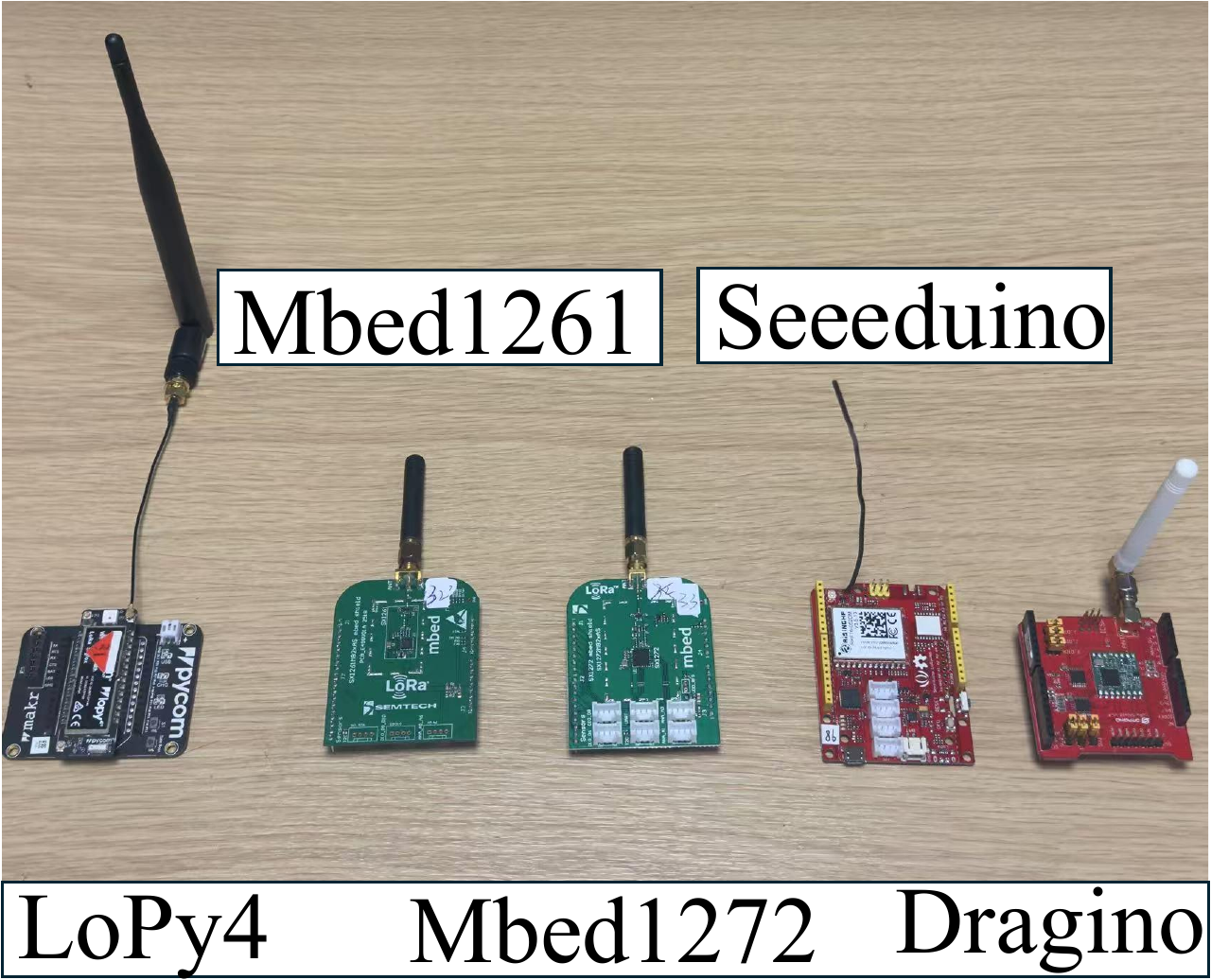}
		\label{fig:rog}}
        \subfloat[]{\includegraphics[width=1.1in]{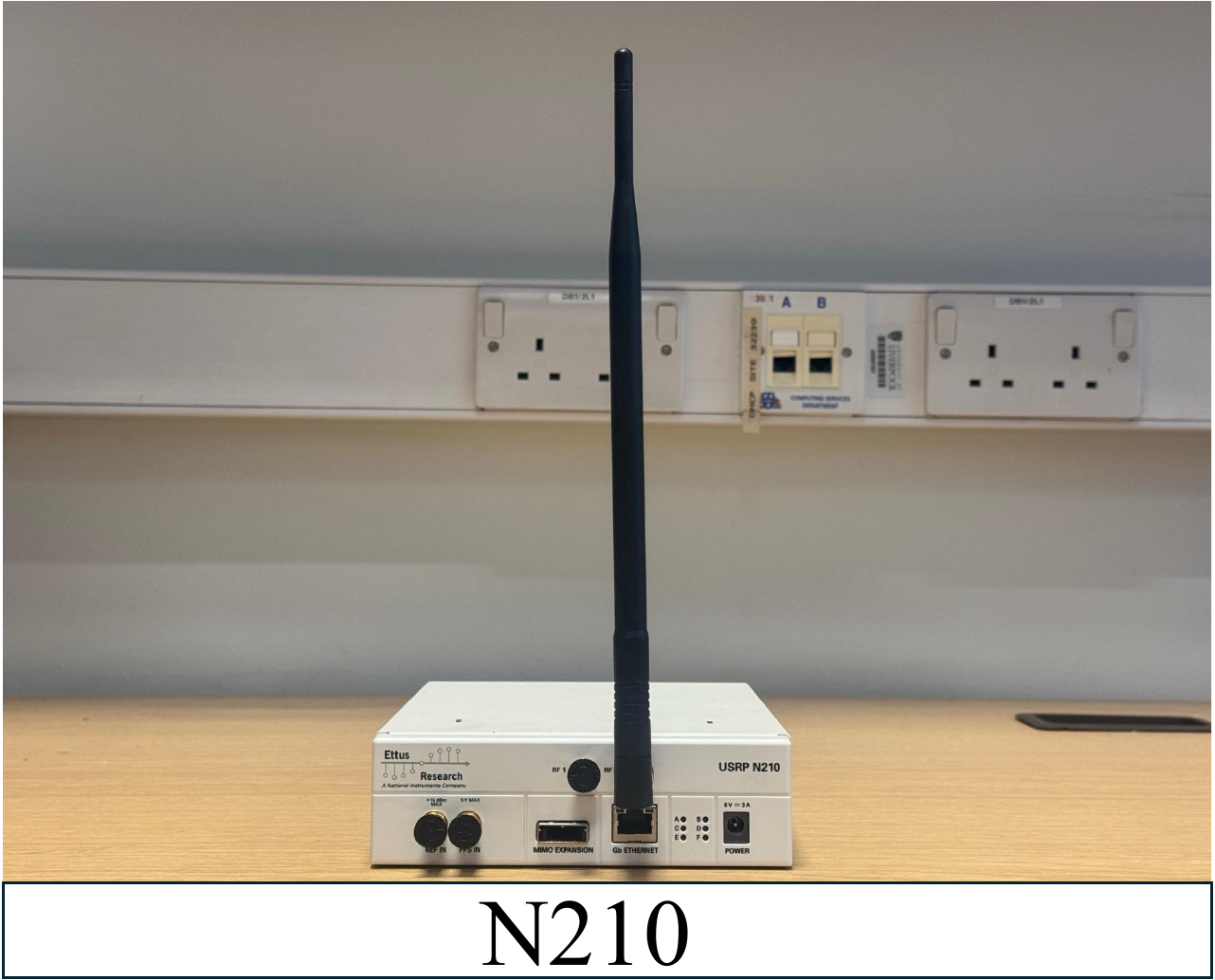}
		\label{fig:n210}}
	\caption{Experimental devices. (a) Legitimate devices: 5 LoPy4 and 5 Mbed1261.  (b) Rogue devices: LoPy4, Mbed1261, Mbed1272, Seeeduino, Dragino. (c) Receiver: N210 SDR platform. }
	\label{fig:devs}
\end{figure}

Ten LoRa devices were used as the legitimate devices, comprising five LoPy4 boards and five Mbed1261 shields, as shown in Fig.~\ref{fig:devs}(a). 
To evaluate rogue device detection, five types of rogue devices were considered, as illustrated in Fig.~\ref{fig:devs}(b). These included two device types observed during training (LoPy4 and Mbed1261 shield) and three previously unseen device types, namely the Mbed1272 shield, Seeeduino, and Dragino.

All the LoRa devices were configured with the same carrier frequency $f_c$ = 868.1~MHz, a spreading factor $SF$ of 7, and a bandwidth $B$ of 125~kHz.
Signal reception was performed using a USRP N210 SDR, shown in Fig.~\ref{fig:devs}(c), which was interfaced through the Matlab Communications Toolbox Support Package for USRP radio\footnote{https://www.mathworks.com/help/comm/usrpradio.html}.
The receiver sampling rate was set to 1 MS/s. All device and receiver parameters remained unchanged throughout the experiments.

\subsubsection{Signal Preprocessing}\label{sec:preprocess}
All the collected signals were preprocessed, including CFO compensation and normalization~\cite{shen2021radio}. 
The CFO compensation is carried out to mitigate the effects of temperature variations and oscillator frequency drift. 
Signal normalization is performed by dividing the signal by its root mean square (RMS) value, minimizing the impact of amplitude variations across different signals.

\subsection{RFFI Model Training}
As shown in Fig.~\ref{fig:rffisystem}(a), the RFFI training stage involves training dataset collection, data augmentation, signal representation generation, and DL model training. Their configurations are described below.

\subsubsection{Training Dataset Collection}
The LoRa transmitter and the USRP N210 receiver were positioned in a stationary, line-of-sight (LOS) configuration with a separation distance of one meter. This setup provided a high SNR while minimizing channel effects.
The training dataset consists of 1000 packets from each device.

\subsubsection{Data Augmentation Settings}\label{sec:data_aug_set}
The original high-SNR training dataset did not include various channel conditions and SNR levels. It was then augmented with various channel conditions and noise levels via the data augmentation method described in Section~\ref{sec:dataaug}, in order to simulate realistic wireless environments.

Specifically, the TDL channel model was implemented using the PyPhysim library\footnote{https://pyphysim.readthedocs.io/en/latest/} with the following parameters~\cite{shen2024federated}:
\begin{itemize}
	\item RMS delay spreads ranging from 5 to 300~ns;
	\item Doppler spread: Jakes model with Doppler frequencies between 0 and 5~Hz. 
\end{itemize}
In addition, synthetic AWGN was introduced to generate SNR conditions ranging from 10~dB to 40~dB.

\subsubsection{Signal Representation}
The augmented signal was then converted to spectrograms. Two spectrograms were generated from the same time-domain signal, with $L_1 = 64$ and $L_2 = 128$, respectively. Each spectrogram was further converted to a CIS, which served as an input to the DL model. 

\subsubsection{Training Configuration}
All the neural networks were trained on a PC equipped with an NVIDIA GeForce RTX 4090 GPU, and PyTorch was used. 
The DL models were trained with the same parameters. 
% We select adaptive moment estimation with weight decay (AdamW) as the optimizer, and the initial learning rate is set to 0.0003. The learning rate drops every 10 epochs with a drop factor of 0.5. The mini-batch size is set to 64, and the L2 regularization factor is 0.001. 
We adopt the adaptive moment estimation with weight decay (AdamW) optimizer, with an L2 regularization factor of 0.001. 
The initial learning rate is set to 0.0003 and is decayed by a factor of 0.5 if validation performance does not improve over 10 consecutive epochs.
The mini-batch size is fixed at 64.
A label smoothing factor of 0.2 is applied to enhance generalization. 
Training terminated when the validation loss did not change within 30 epochs.

\subsection{Open-Set Benchmarks: KNN, OpenMax, and ELogitNorm}\label{sec:openset_benchmark}
\mmj{We considered three open-set benchmarks in this paper, namely OpenMax, KNN, and ELogitNorm. All operate after the feature extractor, and decisions are made based on the extracted features.}

\begin{figure}[!t]
	\begin{center}
		\includegraphics[width = 3.4in]{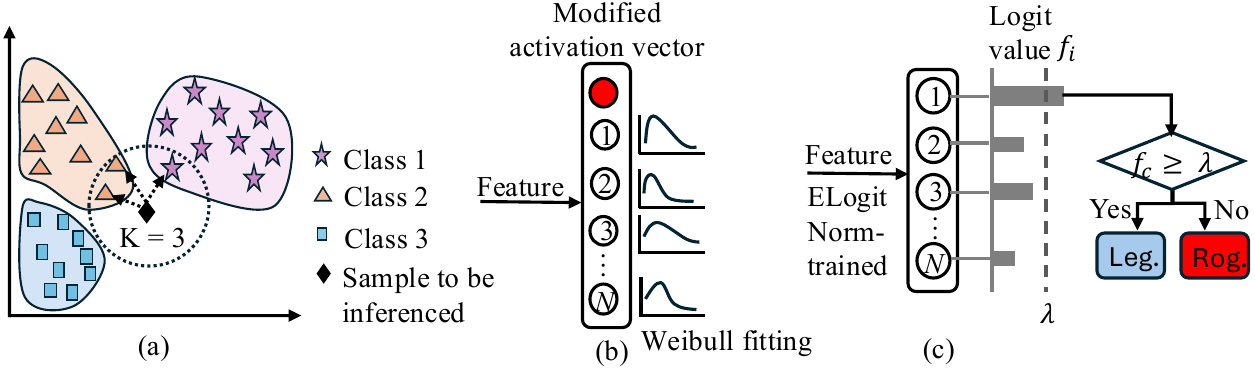}
		\caption{\mmj{Open-set recognition. (a) KNN. (b) OpenMax. (c) ELogitNorm. }}
		\label{fig:joint}
	\end{center}	
\end{figure}

\subsubsection{KNN}
KNN~\cite{cover1967nearest} predicts an input by comparing its features, extracted via a feature extractor, with those of the training dataset.
Given a test input, the algorithm computes the distance between the input and every sample in the training set, using a predefined distance metric such as Euclidean or cosine distance.
It then selects the $K$ training samples that are closest to the input, which are referred to as the ``$K$ nearest neighbors''.
After selecting these neighbors, KNN determines the predicted label of the input based on a majority vote among the neighbors’ labels, and analyzes the distances to assess how far the input deviates from typical training samples for outlier detection.

As demonstrated in Fig.~\ref{fig:joint}(a), when $K=3$, the sample to be inferred has two of its three nearest neighbors belonging to Class 2 and one to Class 1, so the sample is classified as Class 2.
Whether this sample is considered legitimate or rogue is determined based on the distances between the sample and its $K$ nearest neighbors.
If the distance exceeds a predefined threshold, the sample is regarded as rogue; otherwise, it is considered legitimate.

As KNN stores the entire training dataset and computes distances from the sample to be inferred to all training samples, it typically results in high memory usage and computational cost.

\subsubsection{OpenMax}
OpenMax~\cite{openmax2016cvpr} generates a modified activation vector $\textbf{a}^{\prime}$ by adjusting the activation vector $\textbf{a}$
\begin{equation}
a_n^{\prime}= \begin{cases}a_n \omega_n, & n \in\{1, \ldots,|N|\} \\ \sum_{n=1}^{|N|} a_n\left(1-\omega_n\right) & n=|N|+1\end{cases},
\end{equation}
where $\omega_n$ is the confidence in the membership of the given sample to class $n$~\cite{hanna2021tccn}.
To calculate $\omega_n$, the MAV for each known class is calculated. 
A Weibull distribution $\text{WeibullCDF}(\cdot)$ is fitted to the largest distances $d_n$ between activation vectors $\textbf{a}$ and their corresponding class means, and the number of such distances is determined by a predefined tail size $\lambda$.
These fitted distributions, $\text{WeibullCDF}(d_n(\textbf{a}),\cdot)$, are shown in Fig.~\ref{fig:joint}(b). 
$\omega_n$ is calculated using
\begin{equation}
\omega_n=1-\frac{\alpha_{top}-n}{\alpha_{top}} \times \operatorname{WeibullCDF}\left(d_n(\mathbf{a}),\cdot)\right..
\end{equation}
$\alpha_{top}$ denotes the number of top activation scores selected.
After obtaining $\textbf{a}^{\prime}$, the updated $N+1$ probabilities are computed to facilitate the prediction.
A sample is identified as originating from a rogue device if the unknown class has the highest probability. Otherwise, it is assigned to the known class with the highest probability.

The performance of OpenMax is highly dependent on the proper design of hyperparameters, such as the tail size $\lambda$ and the number of top-ranked classes $\alpha_{top}$. 
Due to the sensitivity of these parameters, suboptimal configurations can lead to noticeable performance degradation~\cite{openmax2023}, including reduced accuracy and misclassification of rogue devices.

\mmj{\subsubsection{ELogitNorm}
As illustrated in Fig.~\ref{fig:joint}(c), the proposed detector is built upon a classifier trained on legitimate devices with ELogitNorm~\cite{ding2026enhancing}.
Let $\mathbf{v}$ be the feature vector before the final layer, and let $c$ denote the index of the maximum entry in the logit vector $\mathbf{W}^{\top}\mathbf{v}+\mathbf{b}$. 
ELogitNorm scales the logits by the average distance from $\mathbf{v}$ to the decision boundaries separating the predicted class $c$ from the others:
\begin{equation}
\mathcal{D}(\mathbf{v}):=\frac{1}{N-1}\sum^N_{i\neq c }
\frac{\left|(\mathbf{w}_c-\mathbf{w}_{i})^{\top}\mathbf{v}+(\mathbf{b}_c-\mathbf{b}_{i})\right|}
{\left\|\mathbf{w}_c-\mathbf{w}_{i}\right\|_{2}}.
\label{eq:dist}
\end{equation}
% where $\mathbf{w}_{i}$ is the $i$-th column of $\mathbf{W}$. 
The network is trained by minimizing
\begin{equation}
\mathcal{L}=-\log\frac{\exp\!\left(f_{y}/\mathcal{D}(\mathbf{v})\right)}
{\sum_{i=1}^{N}\exp\!\left(f_{i}/\mathcal{D}(\mathbf{v})\right)},
\label{eq:loss}
\end{equation}
where $y$ is the ground-truth label and $f_i$ denotes the $i$-th entry of the logit vector.

During inference, the maximum logit $f_c$ of the received packet is
used for authentication.
% \begin{equation}
% S= f_c.
% \label{eq:score}
% \end{equation}
If $f_c\geq\lambda$, the packet is identified as legitimate and assigned to class $c$.
Otherwise, it is identified as originating from a rogue device. Here, $\lambda$ is a predefined decision threshold.}

\section{Experimental Evaluation Under Simulated SNR Conditions}\label{sec:experiment_simulated}
To validate the effectiveness of our approach, we conducted a detailed experimental evaluation under simulated SNR levels. The noise power can be calculated to emulate SNR levels, which allows us to check performance under a specific SNR.

First, we compared the performance of our approach with single-feature methods and the decision fusion technique.
Second, we compared the performance of the proposed ViT with conventional ViT and CNN-based models.
Third, we evaluated the performance of the joint identification and classification using our approach, comparing it with KNN, OpenMax, and ELogitNorm.

The identification performance was evaluated by introducing each of the five rogue devices into the inference stage alongside ten legitimate devices. 
The classification evaluation was conducted on the ten legitimate devices. 

\subsection{Test Dataset Collection}
The testing dataset was collected on a different date from the training dataset. As in the training setup, the DUT and the receiver were placed approximately one meter apart in a LOS configuration, resulting in a high SNR. 
AWGN at different levels was then introduced to emulate a range of SNR conditions, enabling controlled adjustment of the signal quality.
The testing dataset consists of 100 packets per device.

\subsection{Performance Evaluation of the Fused Feature} \label{sec:expdual}
In this section, we evaluated the performance of the fused feature approach against the single-feature and decision fusion (Fused Decision) methods.
Moreover, we conduct a sensitivity analysis of key parameters, including different window lengths and their fusion.
\subsubsection{Benchmark Design}\label{sec:dualinput_setup1}
To assess the effectiveness of the fusion strategy, two individual inputs used for fusion, CISs with window lengths of 64 and 128, are evaluated separately. 
A comparison is also made with the decision fusion approach, in which each ViT is independently connected to an AE and an FC layer. 
The identification result is obtained by fusing RMSE values from two AEs, and the classification result is obtained by averaging Softmax probabilities from two FC layers.
\subsubsection{Results of Identification}\label{sec:identify_dual}
\begin{table*}[!t]
\caption{Performance of fused and single features across different rogue devices and SNRs.}
\centering
\scriptsize
{%
  \begin{threeparttable}
\begin{tabular}{|c|c|llll|llll|llll|llll|llll|}
\hline
                        &                            & \multicolumn{4}{c|}{}                                                                                                                                                                                                                                                                                     & \multicolumn{4}{c|}{}                                                                                                                                                                                                                                                                                     & \multicolumn{4}{c|}{}                                                                                                                                                                                                                                                                                     & \multicolumn{4}{c|}{}                                                                                                                                                                                                                                                                                     & \multicolumn{4}{c|}{}                                                                                                                                                                                                                                                                                     \\
\multirow{-2}{*}{SNR}   & \multirow{-2}{*}{Metric}   & \multicolumn{4}{c|}{\multirow{-2}{*}{LoPy4}}                                                                                                                                                                                                                                                              & \multicolumn{4}{c|}{\multirow{-2}{*}{Mbed1261}}                                                                                                                                                                                                                                                           & \multicolumn{4}{c|}{\multirow{-2}{*}{Mbed1272}}                                                                                                                                                                                                                                                           & \multicolumn{4}{c|}{\multirow{-2}{*}{Seeduino}}                                                                                                                                                                                                                                                           & \multicolumn{4}{c|}{\multirow{-2}{*}{Dragino}}                                                                                                                                                                                                                                                            \\ \hline
                        & {\color[HTML]{0D0D0D} AUC} & \multicolumn{1}{l|}{\cellcolor[HTML]{E8DFBF}{\color[HTML]{0D0D0D} \textbf{0.566}}} & \multicolumn{1}{l|}{\cellcolor[HTML]{FBE2D5}{\color[HTML]{0D0D0D} 0.44}} & \multicolumn{1}{l|}{\cellcolor[HTML]{CAEDFB}{\color[HTML]{0D0D0D} 0.53}}  & \cellcolor[HTML]{D0D0D0}{\color[HTML]{0D0D0D} 0.559}          & \multicolumn{1}{l|}{\cellcolor[HTML]{E8DFBF}{\color[HTML]{0D0D0D} 0.774}}          & \multicolumn{1}{l|}{\cellcolor[HTML]{FBE2D5}{\color[HTML]{0D0D0D} \textbf{0.81}}} & \multicolumn{1}{l|}{\cellcolor[HTML]{CAEDFB}{\color[HTML]{0D0D0D} 0.651}} & \cellcolor[HTML]{D0D0D0}{\color[HTML]{0D0D0D} 0.604} & \multicolumn{1}{l|}{\cellcolor[HTML]{E8DFBF}{\color[HTML]{0D0D0D} 0.627}}          & \multicolumn{1}{l|}{\cellcolor[HTML]{FBE2D5}{\color[HTML]{0D0D0D} 0.52}} & \multicolumn{1}{l|}{\cellcolor[HTML]{CAEDFB}{\color[HTML]{0D0D0D} 0.496}} & \cellcolor[HTML]{D0D0D0}{\color[HTML]{0D0D0D} \textbf{0.677}} & \multicolumn{1}{l|}{\cellcolor[HTML]{E8DFBF}{\color[HTML]{0D0D0D} 0.516}}          & \multicolumn{1}{l|}{\cellcolor[HTML]{FBE2D5}{\color[HTML]{0D0D0D} 0.37}} & \multicolumn{1}{l|}{\cellcolor[HTML]{CAEDFB}{\color[HTML]{0D0D0D} 0.516}} & \cellcolor[HTML]{D0D0D0}{\color[HTML]{0D0D0D} \textbf{0.557}} & \multicolumn{1}{l|}{\cellcolor[HTML]{E8DFBF}{\color[HTML]{0D0D0D} 0.631}}          & \multicolumn{1}{l|}{\cellcolor[HTML]{FBE2D5}{\color[HTML]{0D0D0D} 0.55}} & \multicolumn{1}{l|}{\cellcolor[HTML]{CAEDFB}{\color[HTML]{0D0D0D} 0.447}} & \cellcolor[HTML]{D0D0D0}{\color[HTML]{0D0D0D} \textbf{0.675}} \\ \cline{2-22} 
\multirow{-2}{*}{10 dB} & {\color[HTML]{0D0D0D} EER} & \multicolumn{1}{l|}{\cellcolor[HTML]{E8DFBF}{\color[HTML]{0D0D0D} \textbf{0.421}}} & \multicolumn{1}{l|}{\cellcolor[HTML]{FBE2D5}{\color[HTML]{0D0D0D} 0.55}} & \multicolumn{1}{l|}{\cellcolor[HTML]{CAEDFB}{\color[HTML]{0D0D0D} 0.47}}  & \cellcolor[HTML]{D0D0D0}{\color[HTML]{0D0D0D} 0.447}          & \multicolumn{1}{l|}{\cellcolor[HTML]{E8DFBF}{\color[HTML]{0D0D0D} 0.29}}           & \multicolumn{1}{l|}{\cellcolor[HTML]{FBE2D5}{\color[HTML]{0D0D0D} \textbf{0.27}}} & \multicolumn{1}{l|}{\cellcolor[HTML]{CAEDFB}{\color[HTML]{0D0D0D} 0.381}} & \cellcolor[HTML]{D0D0D0}{\color[HTML]{0D0D0D} 0.439} & \multicolumn{1}{l|}{\cellcolor[HTML]{E8DFBF}{\color[HTML]{0D0D0D} 0.402}}          & \multicolumn{1}{l|}{\cellcolor[HTML]{FBE2D5}{\color[HTML]{0D0D0D} 0.49}} & \multicolumn{1}{l|}{\cellcolor[HTML]{CAEDFB}{\color[HTML]{0D0D0D} 0.499}} & \cellcolor[HTML]{D0D0D0}{\color[HTML]{0D0D0D} \textbf{0.39}}  & \multicolumn{1}{l|}{\cellcolor[HTML]{E8DFBF}{\color[HTML]{0D0D0D} 0.497}}          & \multicolumn{1}{l|}{\cellcolor[HTML]{FBE2D5}{\color[HTML]{0D0D0D} 0.64}} & \multicolumn{1}{l|}{\cellcolor[HTML]{CAEDFB}{\color[HTML]{0D0D0D} 0.49}}  & \cellcolor[HTML]{D0D0D0}{\color[HTML]{0D0D0D} \textbf{0.462}} & \multicolumn{1}{l|}{\cellcolor[HTML]{E8DFBF}{\color[HTML]{0D0D0D} 0.39}}           & \multicolumn{1}{l|}{\cellcolor[HTML]{FBE2D5}{\color[HTML]{0D0D0D} 0.47}} & \multicolumn{1}{l|}{\cellcolor[HTML]{CAEDFB}{\color[HTML]{0D0D0D} 0.548}} & \cellcolor[HTML]{D0D0D0}{\color[HTML]{0D0D0D} \textbf{0.381}} \\ \hline
                        & {\color[HTML]{0D0D0D} AUC} & \multicolumn{1}{l|}{\cellcolor[HTML]{E8DFBF}{\color[HTML]{0D0D0D} \textbf{0.842}}} & \multicolumn{1}{l|}{\cellcolor[HTML]{FBE2D5}{\color[HTML]{0D0D0D} 0.59}} & \multicolumn{1}{l|}{\cellcolor[HTML]{CAEDFB}{\color[HTML]{0D0D0D} 0.561}} & \cellcolor[HTML]{D0D0D0}{\color[HTML]{0D0D0D} 0.784}          & \multicolumn{1}{l|}{\cellcolor[HTML]{E8DFBF}{\color[HTML]{0D0D0D} 0.673}}          & \multicolumn{1}{l|}{\cellcolor[HTML]{FBE2D5}{\color[HTML]{0D0D0D} \textbf{0.79}}} & \multicolumn{1}{l|}{\cellcolor[HTML]{CAEDFB}{\color[HTML]{0D0D0D} 0.615}} & \cellcolor[HTML]{D0D0D0}{\color[HTML]{0D0D0D} 0.624} & \multicolumn{1}{l|}{\cellcolor[HTML]{E8DFBF}{\color[HTML]{0D0D0D} 0.897}}          & \multicolumn{1}{l|}{\cellcolor[HTML]{FBE2D5}{\color[HTML]{0D0D0D} 0.8}}  & \multicolumn{1}{l|}{\cellcolor[HTML]{CAEDFB}{\color[HTML]{0D0D0D} 0.693}} & \cellcolor[HTML]{D0D0D0}{\color[HTML]{0D0D0D} \textbf{0.927}} & \multicolumn{1}{l|}{\cellcolor[HTML]{E8DFBF}{\color[HTML]{0D0D0D} \textbf{0.789}}} & \multicolumn{1}{l|}{\cellcolor[HTML]{FBE2D5}{\color[HTML]{0D0D0D} 0.45}} & \multicolumn{1}{l|}{\cellcolor[HTML]{CAEDFB}{\color[HTML]{0D0D0D} 0.537}} & \cellcolor[HTML]{D0D0D0}{\color[HTML]{0D0D0D} 0.748}          & \multicolumn{1}{l|}{\cellcolor[HTML]{E8DFBF}{\color[HTML]{0D0D0D} 0.817}}          & \multicolumn{1}{l|}{\cellcolor[HTML]{FBE2D5}{\color[HTML]{0D0D0D} 0.61}} & \multicolumn{1}{l|}{\cellcolor[HTML]{CAEDFB}{\color[HTML]{0D0D0D} 0.668}} & \cellcolor[HTML]{D0D0D0}{\color[HTML]{0D0D0D} \textbf{0.861}} \\ \cline{2-22} 
\multirow{-2}{*}{20 dB} & {\color[HTML]{0D0D0D} EER} & \multicolumn{1}{l|}{\cellcolor[HTML]{E8DFBF}{\color[HTML]{0D0D0D} \textbf{0.239}}} & \multicolumn{1}{l|}{\cellcolor[HTML]{FBE2D5}{\color[HTML]{0D0D0D} 0.47}} & \multicolumn{1}{l|}{\cellcolor[HTML]{CAEDFB}{\color[HTML]{0D0D0D} 0.42}}  & \cellcolor[HTML]{D0D0D0}{\color[HTML]{0D0D0D} 0.282}          & \multicolumn{1}{l|}{\cellcolor[HTML]{E8DFBF}{\color[HTML]{0D0D0D} 0.39}}           & \multicolumn{1}{l|}{\cellcolor[HTML]{FBE2D5}{\color[HTML]{0D0D0D} \textbf{0.29}}} & \multicolumn{1}{l|}{\cellcolor[HTML]{CAEDFB}{\color[HTML]{0D0D0D} 0.439}} & \cellcolor[HTML]{D0D0D0}{\color[HTML]{0D0D0D} 0.458} & \multicolumn{1}{l|}{\cellcolor[HTML]{E8DFBF}{\color[HTML]{0D0D0D} 0.169}}          & \multicolumn{1}{l|}{\cellcolor[HTML]{FBE2D5}{\color[HTML]{0D0D0D} 0.26}} & \multicolumn{1}{l|}{\cellcolor[HTML]{CAEDFB}{\color[HTML]{0D0D0D} 0.352}} & \cellcolor[HTML]{D0D0D0}{\color[HTML]{0D0D0D} \textbf{0.13}}  & \multicolumn{1}{l|}{\cellcolor[HTML]{E8DFBF}{\color[HTML]{0D0D0D} \textbf{0.289}}} & \multicolumn{1}{l|}{\cellcolor[HTML]{FBE2D5}{\color[HTML]{0D0D0D} 0.56}} & \multicolumn{1}{l|}{\cellcolor[HTML]{CAEDFB}{\color[HTML]{0D0D0D} 0.489}} & \cellcolor[HTML]{D0D0D0}{\color[HTML]{0D0D0D} 0.302}          & \multicolumn{1}{l|}{\cellcolor[HTML]{E8DFBF}{\color[HTML]{0D0D0D} 0.252}}          & \multicolumn{1}{l|}{\cellcolor[HTML]{FBE2D5}{\color[HTML]{0D0D0D} 0.43}} & \multicolumn{1}{l|}{\cellcolor[HTML]{CAEDFB}{\color[HTML]{0D0D0D} 0.387}} & \cellcolor[HTML]{D0D0D0}{\color[HTML]{0D0D0D} \textbf{0.21}}  \\ \hline
                        & {\color[HTML]{0D0D0D} AUC} & \multicolumn{1}{l|}{\cellcolor[HTML]{E8DFBF}{\color[HTML]{0D0D0D} 0.93}}           & \multicolumn{1}{l|}{\cellcolor[HTML]{FBE2D5}{\color[HTML]{0D0D0D} 0.7}}  & \multicolumn{1}{l|}{\cellcolor[HTML]{CAEDFB}{\color[HTML]{0D0D0D} 0.804}} & \cellcolor[HTML]{D0D0D0}{\color[HTML]{0D0D0D} \textbf{0.947}} & \multicolumn{1}{l|}{\cellcolor[HTML]{E8DFBF}{\color[HTML]{0D0D0D} 0.929}}          & \multicolumn{1}{l|}{\cellcolor[HTML]{FBE2D5}{\color[HTML]{0D0D0D} \textbf{0.98}}} & \multicolumn{1}{l|}{\cellcolor[HTML]{CAEDFB}{\color[HTML]{0D0D0D} 0.574}} & \cellcolor[HTML]{D0D0D0}{\color[HTML]{0D0D0D} 0.682} & \multicolumn{1}{l|}{\cellcolor[HTML]{E8DFBF}{\color[HTML]{0D0D0D} 0.944}}          & \multicolumn{1}{l|}{\cellcolor[HTML]{FBE2D5}{\color[HTML]{0D0D0D} 0.88}} & \multicolumn{1}{l|}{\cellcolor[HTML]{CAEDFB}{\color[HTML]{0D0D0D} 0.718}} & \cellcolor[HTML]{D0D0D0}{\color[HTML]{0D0D0D} \textbf{0.983}} & \multicolumn{1}{l|}{\cellcolor[HTML]{E8DFBF}{\color[HTML]{0D0D0D} 0.979}}          & \multicolumn{1}{l|}{\cellcolor[HTML]{FBE2D5}{\color[HTML]{0D0D0D} 0.87}} & \multicolumn{1}{l|}{\cellcolor[HTML]{CAEDFB}{\color[HTML]{0D0D0D} 0.971}} & \cellcolor[HTML]{D0D0D0}{\color[HTML]{0D0D0D} \textbf{0.997}} & \multicolumn{1}{l|}{\cellcolor[HTML]{E8DFBF}{\color[HTML]{0D0D0D} \textbf{0.981}}} & \multicolumn{1}{l|}{\cellcolor[HTML]{FBE2D5}{\color[HTML]{0D0D0D} 0.92}} & \multicolumn{1}{l|}{\cellcolor[HTML]{CAEDFB}{\color[HTML]{0D0D0D} 0.809}} & \cellcolor[HTML]{D0D0D0}{\color[HTML]{0D0D0D} 0.962}          \\ \cline{2-22} 
\multirow{-2}{*}{30 dB} & {\color[HTML]{0D0D0D} EER} & \multicolumn{1}{l|}{\cellcolor[HTML]{E8DFBF}{\color[HTML]{0D0D0D} 0.131}}          & \multicolumn{1}{l|}{\cellcolor[HTML]{FBE2D5}{\color[HTML]{0D0D0D} 0.32}} & \multicolumn{1}{l|}{\cellcolor[HTML]{CAEDFB}{\color[HTML]{0D0D0D} 0.259}} & \cellcolor[HTML]{D0D0D0}{\color[HTML]{0D0D0D} \textbf{0.129}} & \multicolumn{1}{l|}{\cellcolor[HTML]{E8DFBF}{\color[HTML]{0D0D0D} 0.142}}          & \multicolumn{1}{l|}{\cellcolor[HTML]{FBE2D5}{\color[HTML]{0D0D0D} \textbf{0.07}}} & \multicolumn{1}{l|}{\cellcolor[HTML]{CAEDFB}{\color[HTML]{0D0D0D} 0.476}} & \cellcolor[HTML]{D0D0D0}{\color[HTML]{0D0D0D} 0.42}  & \multicolumn{1}{l|}{\cellcolor[HTML]{E8DFBF}{\color[HTML]{0D0D0D} 0.105}}          & \multicolumn{1}{l|}{\cellcolor[HTML]{FBE2D5}{\color[HTML]{0D0D0D} 0.18}} & \multicolumn{1}{l|}{\cellcolor[HTML]{CAEDFB}{\color[HTML]{0D0D0D} 0.34}}  & \cellcolor[HTML]{D0D0D0}{\color[HTML]{0D0D0D} \textbf{0.062}} & \multicolumn{1}{l|}{\cellcolor[HTML]{E8DFBF}{\color[HTML]{0D0D0D} 0.045}}          & \multicolumn{1}{l|}{\cellcolor[HTML]{FBE2D5}{\color[HTML]{0D0D0D} 0.19}} & \multicolumn{1}{l|}{\cellcolor[HTML]{CAEDFB}{\color[HTML]{0D0D0D} 0.064}} & \cellcolor[HTML]{D0D0D0}{\color[HTML]{0D0D0D} \textbf{0.019}} & \multicolumn{1}{l|}{\cellcolor[HTML]{E8DFBF}{\color[HTML]{0D0D0D} \textbf{0.061}}} & \multicolumn{1}{l|}{\cellcolor[HTML]{FBE2D5}{\color[HTML]{0D0D0D} 0.14}} & \multicolumn{1}{l|}{\cellcolor[HTML]{CAEDFB}{\color[HTML]{0D0D0D} 0.268}} & \cellcolor[HTML]{D0D0D0}{\color[HTML]{0D0D0D} 0.089}          \\ \hline
                        & {\color[HTML]{0D0D0D} AUC} & \multicolumn{1}{l|}{\cellcolor[HTML]{E8DFBF}{\color[HTML]{0D0D0D} \textbf{0.987}}} & \multicolumn{1}{l|}{\cellcolor[HTML]{FBE2D5}{\color[HTML]{0D0D0D} 0.71}} & \multicolumn{1}{l|}{\cellcolor[HTML]{CAEDFB}{\color[HTML]{0D0D0D} 0.966}} & \cellcolor[HTML]{D0D0D0}{\color[HTML]{0D0D0D} 0.965}          & \multicolumn{1}{l|}{\cellcolor[HTML]{E8DFBF}{\color[HTML]{0D0D0D} \textbf{0.989}}} & \multicolumn{1}{l|}{\cellcolor[HTML]{FBE2D5}{\color[HTML]{0D0D0D} 0.98}}          & \multicolumn{1}{l|}{\cellcolor[HTML]{CAEDFB}{\color[HTML]{0D0D0D} 0.656}} & \cellcolor[HTML]{D0D0D0}{\color[HTML]{0D0D0D} 0.738} & \multicolumn{1}{l|}{\cellcolor[HTML]{E8DFBF}{\color[HTML]{0D0D0D} \textbf{0.993}}} & \multicolumn{1}{l|}{\cellcolor[HTML]{FBE2D5}{\color[HTML]{0D0D0D} 0.9}}  & \multicolumn{1}{l|}{\cellcolor[HTML]{CAEDFB}{\color[HTML]{0D0D0D} 0.812}} & \cellcolor[HTML]{D0D0D0}{\color[HTML]{0D0D0D} 0.991}          & \multicolumn{1}{l|}{\cellcolor[HTML]{E8DFBF}{\color[HTML]{0D0D0D} 0.995}}          & \multicolumn{1}{l|}{\cellcolor[HTML]{FBE2D5}{\color[HTML]{0D0D0D} 0.93}} & \multicolumn{1}{l|}{\cellcolor[HTML]{CAEDFB}{\color[HTML]{0D0D0D} 0.998}} & \cellcolor[HTML]{D0D0D0}{\color[HTML]{0D0D0D} \textbf{1}}     & \multicolumn{1}{l|}{\cellcolor[HTML]{E8DFBF}{\color[HTML]{0D0D0D} \textbf{0.997}}} & \multicolumn{1}{l|}{\cellcolor[HTML]{FBE2D5}{\color[HTML]{0D0D0D} 0.96}} & \multicolumn{1}{l|}{\cellcolor[HTML]{CAEDFB}{\color[HTML]{0D0D0D} 0.956}} & \cellcolor[HTML]{D0D0D0}{\color[HTML]{0D0D0D} 0.977}          \\ \cline{2-22} 
\multirow{-2}{*}{40 dB} & {\color[HTML]{0D0D0D} EER} & \multicolumn{1}{l|}{\cellcolor[HTML]{E8DFBF}{\color[HTML]{0D0D0D} \textbf{0.031}}} & \multicolumn{1}{l|}{\cellcolor[HTML]{FBE2D5}{\color[HTML]{0D0D0D} 0.33}} & \multicolumn{1}{l|}{\cellcolor[HTML]{CAEDFB}{\color[HTML]{0D0D0D} 0.071}} & \cellcolor[HTML]{D0D0D0}{\color[HTML]{0D0D0D} 0.059}          & \multicolumn{1}{l|}{\cellcolor[HTML]{E8DFBF}{\color[HTML]{0D0D0D} \textbf{0.031}}} & \multicolumn{1}{l|}{\cellcolor[HTML]{FBE2D5}{\color[HTML]{0D0D0D} 0.05}}          & \multicolumn{1}{l|}{\cellcolor[HTML]{CAEDFB}{\color[HTML]{0D0D0D} 0.418}} & \cellcolor[HTML]{D0D0D0}{\color[HTML]{0D0D0D} 0.382} & \multicolumn{1}{l|}{\cellcolor[HTML]{E8DFBF}{\color[HTML]{0D0D0D} \textbf{0.02}}}  & \multicolumn{1}{l|}{\cellcolor[HTML]{FBE2D5}{\color[HTML]{0D0D0D} 0.16}} & \multicolumn{1}{l|}{\cellcolor[HTML]{CAEDFB}{\color[HTML]{0D0D0D} 0.266}} & \cellcolor[HTML]{D0D0D0}{\color[HTML]{0D0D0D} 0.041}          & \multicolumn{1}{l|}{\cellcolor[HTML]{E8DFBF}{\color[HTML]{0D0D0D} 0.011}}          & \multicolumn{1}{l|}{\cellcolor[HTML]{FBE2D5}{\color[HTML]{0D0D0D} 0.1}}  & \multicolumn{1}{l|}{\cellcolor[HTML]{CAEDFB}{\color[HTML]{0D0D0D} 0.008}} & \cellcolor[HTML]{D0D0D0}{\color[HTML]{0D0D0D} \textbf{0.001}} & \multicolumn{1}{l|}{\cellcolor[HTML]{E8DFBF}{\color[HTML]{0D0D0D} \textbf{0.002}}} & \multicolumn{1}{l|}{\cellcolor[HTML]{FBE2D5}{\color[HTML]{0D0D0D} 0.1}}  & \multicolumn{1}{l|}{\cellcolor[HTML]{CAEDFB}{\color[HTML]{0D0D0D} 0.093}} & \cellcolor[HTML]{D0D0D0}{\color[HTML]{0D0D0D} 0.058}          \\ \hline
\end{tabular}
\begin{tablenotes}
 \item[] 
\begin{tikzpicture}[baseline=0.5ex]
\fill[brown!50] (0,0) rectangle (0.6,0.2);
\end{tikzpicture}
\hspace{0.2em}: results of our approach (fused feature);
\begin{tikzpicture}[baseline=0.5ex]
\fill[red!20!white] (0,0) rectangle (0.6,0.2);
\end{tikzpicture}
\hspace{0.2em}: results of fused decision;
\begin{tikzpicture}[baseline=0.5ex]
\fill[cyan!30!white] (0,0) rectangle (0.6,0.2);
\end{tikzpicture}
\hspace{0.2em}: results of single feature (\emph{L}=64);
\begin{tikzpicture}[baseline=0.5ex]
\fill[gray!50] (0,0) rectangle (0.6,0.2);
\end{tikzpicture}
\hspace{0.2em}: results of single feature (\emph{L}=128).
 \item[] 
Entries highlighted in bold denote the best detection performance for each dataset, characterized by the highest AUC and the lowest EER.
\end{tablenotes}
\end{threeparttable}
}
\label{tab:eva_identify_dual}
\end{table*}

The identification results are summarized in Table~\ref{tab:eva_identify_dual}.
Our fused feature approach exhibits the best performance, with the majority of cases achieving the highest AUC and lowest EER. 
Although alternative methods achieve better performance in certain cases, our approach exhibits enhanced overall robustness, indicating that it is well suited for a broad range of application scenarios.

It can be observed that the AUC decreases by more than 0.3, from nearly 1.0 to approximately 0.65, as the SNR is reduced from 40 dB to 10 dB. This degradation is primarily attributed to the increased impact of noise at low SNRs. Under such conditions, noise-induced fluctuations become more pronounced, violating the assumption that the wireless channel remains relatively stable over short temporal intervals~\cite{shen2022towards}. Consequently, the effectiveness of the CIS is diminished, leading to degraded authentication performance.

\subsubsection{Results of Classification}\label{sec:clfdual}
\begin{figure}[!t]
	\begin{center}
		\includegraphics[width = 3.4in]{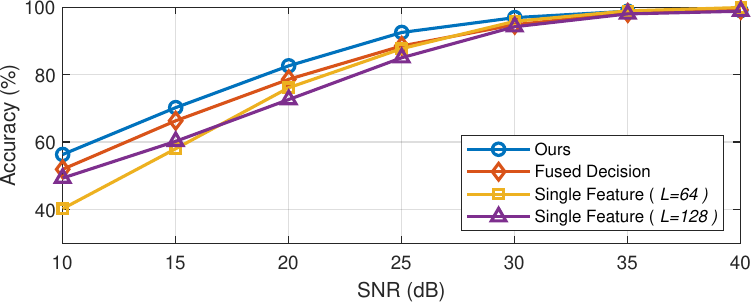}
		\caption{Comparison of classification performance using fused features, fused decisions, and single features.}
		\label{fig:clf_dual}
	\end{center}	
\end{figure}

Fig.~\ref{fig:clf_dual} compares the classification performance.
The fused approaches consistently outperform the single-feature configurations across all SNR levels. 
This can be attributed to the complementary information from the two inputs, which enables the DL models to learn from more comprehensive representations. 
Our approach also demonstrates better performance compared to ``Fused Decision'', because feature-level fusion enables more effective exploitation of the input before classification. 
In contrast, decision fusion may suffer from information loss due to the compression effects introduced by softmax outputs.

As the SNR decreases, the superiority of our approach becomes increasingly evident. 
At an SNR of 15 dB, the fused feature achieves approximately a 10\% improvement in accuracy over the single-feature configuration and outperforms ``Fused Decision'' by around 4\%. 
The consistently superior performance of our approach across varying SNRs demonstrates its enhanced robustness and noise resilience.

\subsubsection{Sensitivity Analysis}\label{sec:sensitive}
\begin{figure}[!t]
	\begin{center}
		\includegraphics[width = 3.4in]{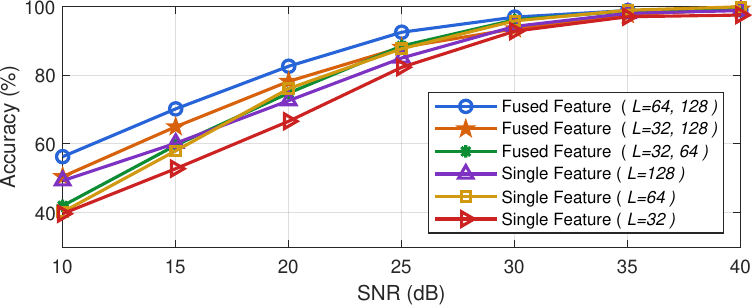}
		\caption{Sensitivity analysis of the effects of window length and fusion strategy on the proposed method.}
		\label{fig:sensitive}
	\end{center}	
\end{figure}

Fig.~\ref{fig:sensitive} presents a sensitivity analysis of the proposed method under different window lengths and fusion configurations for classification.
It can be observed that the ``Fused Feature" methods consistently outperform ``Single Feature" approaches. 
The performance differences among fusion combinations are relatively small when the SNR exceeds 25 dB, typically within 5\%.
However, at lower SNR levels, the advantages of the fused feature with $L=64$ and $L=128$ become more pronounced.
This aligns with the single-feature results, where $L=64$ and $L=128$ show slightly superior performance to $L=32$. 
It suggests that fused features derived from features with stronger single-feature performance tend to achieve better results.

% We investigate key parameters in the proposed method, in particular the impact of different window lengths and fusion choices.

\subsection{Performance Evaluation of the Deep Learning Backbone} \label{sec:evavit}
In this section, we compare the proposed ViT architecture against several baseline models, including a ViT with quasi-square patches, the lightweight MobileViT model~\cite{mehta2022mobilevit}, and a CNN-based model. 
All models are trained under identical configurations and using the same datasets, with the neural network architecture being the only varying factor.

\subsubsection{Benchmark Design}\label{sec:dualinput_setup}
The baseline ViT is adapted from the architecture proposed in~\cite{dosovitskiy2020vit}, which was configured with an input size of $256 \times 256$ and a patch size of $16 \times 16$.
However, the CISs used in our study, with dimensions of $26 \times 254$ and $52 \times 126$, are not directly divisible by the above patch size. 
To ensure compatibility with square patch-based processing in~\cite{dosovitskiy2020vit}, we apply zero-padding along the time dimension, resulting in adjusted input sizes of $26 \times 256$ and $52 \times 128$, respectively. 
These adjusted inputs are subsequently partitioned into patches of size $13 \times 16 $, following a patching strategy consistent with the baseline method, which employs square patches.
The MobileViT is adapted from MobileViT-S in~\cite{mehta2022mobilevit}, and the CNN architecture follows the design in~\cite{shen2022towards}.

\subsubsection{Results of Identification}\label{sec:identinet}

Table~\ref{tab:eva_identify_arch} shows the identification results of different DL models under various SNRs, and the proposed model and ViT with quasi-square patches generally outperform the CNN and MobileViT architectures.
As noted in~\cite{pooling2022}, pooling and downsampling operations in CNNs lead to the loss of fine-grained information, which can negatively impact the identification performance of the AE.
MobileViT's design combines CNN- and ViT-like properties~\cite{mehta2022mobilevit}, and its relatively lower identification performance may therefore be partially attributed to the limitations of CNNs in preserving fine-grained information.
In AE-based anomaly detection, a more informative latent space is beneficial for more effective anomaly detection, especially in distinguishing subtle deviations in RFFI.

The proposed method achieves the best detection performance in the majority of cases, particularly at lower SNR levels of 10 dB and 20 dB.
ViT with quasi-square patches performs competitively when the rogue device is Mbed1272 or Seeeduino, with its results differing from those of our proposed method by no more than 0.031. However, our approach consistently outperforms it in other cases, particularly when the rogue device is Mbed1261. 
For instance, at an SNR of 30 dB, the proposed approach achieves AUC gains of 0.267 and 0.509 over the ViT with quasi-square patches and the CNN, respectively, highlighting a substantial performance improvement. 
These findings suggest that the proposed approach offers a more reliable and effective identification solution.

\begin{table*}[!t]
\caption{Performance of the proposed method, CNN, ViT with quasi-square patches, and MobileViT across different rogue devices and SNRs.}
\centering
% \scriptsize
%\setlength{\tabcolsep}{5pt}
%\renewcommand{\arraystretch}{1.0}
{%
  \scalebox{0.9}
{%
\begin{threeparttable}
\begin{tabular}{|c|c|llll|llll|llll|llll|llll|}
\hline
                        &                          & \multicolumn{4}{c|}{}                                     & \multicolumn{4}{c|}{}                                     & \multicolumn{4}{c|}{}                                     & \multicolumn{4}{c|}{}                                     & \multicolumn{4}{c|}{}                                     \\
\multirow{-2}{*}{SNR}   & \multirow{-2}{*}{Metric} & \multicolumn{4}{c|}{\multirow{-2}{*}{LoPy4}}              & \multicolumn{4}{c|}{\multirow{-2}{*}{Mbed1261}}           & \multicolumn{4}{c|}{\multirow{-2}{*}{Mbed1272}}           & \multicolumn{4}{c|}{\multirow{-2}{*}{Seeeduino}}          & \multicolumn{4}{c|}{\multirow{-2}{*}{Dragino}}            \\ \hline
% 10dB AUC
% LoPy4:    bold0.566, 0.44,   0.478,      0.265
% Mbed1261: 0.774,    0.408,   0.689,      bold0.775
% Mbed1272: bold0.627,0.454,   0.565,      0.343
% Seeeduino: 0.516,  bold0.539,0.459,      0.232
% Dragino:  bold0.631,0.517,   0.591,      0.327
                        & AUC                      & \multicolumn{1}{l|}{\cellcolor[HTML]{E8DFBF}{\textbf{0.566}}} & \multicolumn{1}{l|}{\cellcolor[HTML]{FBE2D5}0.44}   & \multicolumn{1}{l|}{\cellcolor[HTML]{CAEDFB}0.478}            & \cellcolor[HTML]{C0C0C0}0.265 & \multicolumn{1}{l|}{\cellcolor[HTML]{E8DFBF}0.774}            & \multicolumn{1}{l|}{\cellcolor[HTML]{FBE2D5}0.408}  & \multicolumn{1}{l|}{\cellcolor[HTML]{CAEDFB}0.689}            & \cellcolor[HTML]{C0C0C0}{\textbf{0.775}} & \multicolumn{1}{l|}{\cellcolor[HTML]{E8DFBF}{\textbf{0.627}}} & \multicolumn{1}{l|}{\cellcolor[HTML]{FBE2D5}0.454}  & \multicolumn{1}{l|}{\cellcolor[HTML]{CAEDFB}0.565}            & \cellcolor[HTML]{C0C0C0}0.343 & \multicolumn{1}{l|}{\cellcolor[HTML]{E8DFBF}0.516}            & \multicolumn{1}{l|}{\cellcolor[HTML]{FBE2D5}{\textbf{0.539}}} & \multicolumn{1}{l|}{\cellcolor[HTML]{CAEDFB}0.459}            & \cellcolor[HTML]{C0C0C0}0.232 & \multicolumn{1}{l|}{\cellcolor[HTML]{E8DFBF}{\textbf{0.631}}} & \multicolumn{1}{l|}{\cellcolor[HTML]{FBE2D5}0.517}  & \multicolumn{1}{l|}{\cellcolor[HTML]{CAEDFB}0.591}            & \cellcolor[HTML]{C0C0C0}0.327 \\ \cline{2-22}
% 10dB EER
% LoPy4:    bold0.421,0.558,   0.509,      0.67
% Mbed1261: bold0.29, 0.581,   0.35,       0.319
% Mbed1272: bold0.402,0.54,    0.47,       0.61
% Seeeduino: 0.497,  bold0.46, 0.532,      0.69
% Dragino:  bold0.39, 0.5,     0.432,      0.639
\multirow{-2}{*}{10 dB} & EER                      & \multicolumn{1}{l|}{\cellcolor[HTML]{E8DFBF}{\textbf{0.421}}} & \multicolumn{1}{l|}{\cellcolor[HTML]{FBE2D5}0.558}  & \multicolumn{1}{l|}{\cellcolor[HTML]{CAEDFB}0.509}            & \cellcolor[HTML]{C0C0C0}0.67  & \multicolumn{1}{l|}{\cellcolor[HTML]{E8DFBF}{\textbf{0.29}}}  & \multicolumn{1}{l|}{\cellcolor[HTML]{FBE2D5}0.581}  & \multicolumn{1}{l|}{\cellcolor[HTML]{CAEDFB}0.35}             & \cellcolor[HTML]{C0C0C0}0.319 & \multicolumn{1}{l|}{\cellcolor[HTML]{E8DFBF}{\textbf{0.402}}} & \multicolumn{1}{l|}{\cellcolor[HTML]{FBE2D5}0.54}   & \multicolumn{1}{l|}{\cellcolor[HTML]{CAEDFB}0.47}             & \cellcolor[HTML]{C0C0C0}0.61  & \multicolumn{1}{l|}{\cellcolor[HTML]{E8DFBF}0.497}            & \multicolumn{1}{l|}{\cellcolor[HTML]{FBE2D5}{\textbf{0.46}}}  & \multicolumn{1}{l|}{\cellcolor[HTML]{CAEDFB}0.532}            & \cellcolor[HTML]{C0C0C0}0.69  & \multicolumn{1}{l|}{\cellcolor[HTML]{E8DFBF}{\textbf{0.39}}}  & \multicolumn{1}{l|}{\cellcolor[HTML]{FBE2D5}0.5}    & \multicolumn{1}{l|}{\cellcolor[HTML]{CAEDFB}0.432}            & \cellcolor[HTML]{C0C0C0}0.639 \\ \hline
% 20dB AUC
% LoPy4:    bold0.842,0.608,   0.771,      0.591
% Mbed1261: 0.673,    0.362,   0.519,      bold0.789
% Mbed1272: 0.897,    0.449,   bold0.912,  0.653
% Seeeduino: 0.789,  bold0.852,0.73,       0.702
% Dragino:  bold0.817,0.634,   0.755,      0.479
                        & AUC                      & \multicolumn{1}{l|}{\cellcolor[HTML]{E8DFBF}{\textbf{0.842}}} & \multicolumn{1}{l|}{\cellcolor[HTML]{FBE2D5}0.608}  & \multicolumn{1}{l|}{\cellcolor[HTML]{CAEDFB}0.771}            & \cellcolor[HTML]{C0C0C0}0.591 & \multicolumn{1}{l|}{\cellcolor[HTML]{E8DFBF}0.673}            & \multicolumn{1}{l|}{\cellcolor[HTML]{FBE2D5}0.362}  & \multicolumn{1}{l|}{\cellcolor[HTML]{CAEDFB}0.519}            & \cellcolor[HTML]{C0C0C0}{\textbf{0.789}} & \multicolumn{1}{l|}{\cellcolor[HTML]{E8DFBF}0.897}            & \multicolumn{1}{l|}{\cellcolor[HTML]{FBE2D5}0.449}  & \multicolumn{1}{l|}{\cellcolor[HTML]{CAEDFB}{\textbf{0.912}}} & \cellcolor[HTML]{C0C0C0}0.653 & \multicolumn{1}{l|}{\cellcolor[HTML]{E8DFBF}0.789}            & \multicolumn{1}{l|}{\cellcolor[HTML]{FBE2D5}{\textbf{0.852}}} & \multicolumn{1}{l|}{\cellcolor[HTML]{CAEDFB}0.73}             & \cellcolor[HTML]{C0C0C0}0.702 & \multicolumn{1}{l|}{\cellcolor[HTML]{E8DFBF}{\textbf{0.817}}} & \multicolumn{1}{l|}{\cellcolor[HTML]{FBE2D5}0.634}  & \multicolumn{1}{l|}{\cellcolor[HTML]{CAEDFB}0.755}            & \cellcolor[HTML]{C0C0C0}0.479 \\ \cline{2-22}
% 20dB EER
% LoPy4:    bold0.239,0.408,   0.308,      0.38
% Mbed1261: 0.39,     0.59,    0.479,      bold0.239
% Mbed1272: 0.169,    0.53,    bold0.166,  0.344
% Seeeduino: 0.289,  bold0.201,0.323,      0.305
% Dragino:  bold0.252,0.38,    0.318,      0.498
\multirow{-2}{*}{20 dB} & EER                      & \multicolumn{1}{l|}{\cellcolor[HTML]{E8DFBF}{\textbf{0.239}}} & \multicolumn{1}{l|}{\cellcolor[HTML]{FBE2D5}0.408}  & \multicolumn{1}{l|}{\cellcolor[HTML]{CAEDFB}0.308}            & \cellcolor[HTML]{C0C0C0}0.38  & \multicolumn{1}{l|}{\cellcolor[HTML]{E8DFBF}0.39}             & \multicolumn{1}{l|}{\cellcolor[HTML]{FBE2D5}0.59}   & \multicolumn{1}{l|}{\cellcolor[HTML]{CAEDFB}0.479}            & \cellcolor[HTML]{C0C0C0}{\textbf{0.239}} & \multicolumn{1}{l|}{\cellcolor[HTML]{E8DFBF}0.169}            & \multicolumn{1}{l|}{\cellcolor[HTML]{FBE2D5}0.53}   & \multicolumn{1}{l|}{\cellcolor[HTML]{CAEDFB}{\textbf{0.166}}} & \cellcolor[HTML]{C0C0C0}0.344 & \multicolumn{1}{l|}{\cellcolor[HTML]{E8DFBF}0.289}            & \multicolumn{1}{l|}{\cellcolor[HTML]{FBE2D5}{\textbf{0.201}}} & \multicolumn{1}{l|}{\cellcolor[HTML]{CAEDFB}0.323}            & \cellcolor[HTML]{C0C0C0}0.305 & \multicolumn{1}{l|}{\cellcolor[HTML]{E8DFBF}{\textbf{0.252}}} & \multicolumn{1}{l|}{\cellcolor[HTML]{FBE2D5}0.38}   & \multicolumn{1}{l|}{\cellcolor[HTML]{CAEDFB}0.318}            & \cellcolor[HTML]{C0C0C0}0.498 \\ \hline
% 30dB AUC
% LoPy4:    0.93,     0.697,   bold0.946,  0.746
% Mbed1261: bold0.929,0.42,    0.662,      0.776
% Mbed1272: 0.944,    0.612,   bold0.975,  0.748
% Seeeduino: 0.979,   0.983,   bold0.995,  0.798
% Dragino:  bold0.981,0.79,    0.878,      0.621
                        & AUC                      & \multicolumn{1}{l|}{\cellcolor[HTML]{E8DFBF}0.93}             & \multicolumn{1}{l|}{\cellcolor[HTML]{FBE2D5}0.697}  & \multicolumn{1}{l|}{\cellcolor[HTML]{CAEDFB}{\textbf{0.946}}} & \cellcolor[HTML]{C0C0C0}0.746 & \multicolumn{1}{l|}{\cellcolor[HTML]{E8DFBF}{\textbf{0.929}}} & \multicolumn{1}{l|}{\cellcolor[HTML]{FBE2D5}0.42}   & \multicolumn{1}{l|}{\cellcolor[HTML]{CAEDFB}0.662}            & \cellcolor[HTML]{C0C0C0}0.776 & \multicolumn{1}{l|}{\cellcolor[HTML]{E8DFBF}0.944}            & \multicolumn{1}{l|}{\cellcolor[HTML]{FBE2D5}0.612}  & \multicolumn{1}{l|}{\cellcolor[HTML]{CAEDFB}{\textbf{0.975}}} & \cellcolor[HTML]{C0C0C0}0.748 & \multicolumn{1}{l|}{\cellcolor[HTML]{E8DFBF}0.979}            & \multicolumn{1}{l|}{\cellcolor[HTML]{FBE2D5}0.983}  & \multicolumn{1}{l|}{\cellcolor[HTML]{CAEDFB}{\textbf{0.995}}} & \cellcolor[HTML]{C0C0C0}0.798 & \multicolumn{1}{l|}{\cellcolor[HTML]{E8DFBF}{\textbf{0.981}}} & \multicolumn{1}{l|}{\cellcolor[HTML]{FBE2D5}0.79}   & \multicolumn{1}{l|}{\cellcolor[HTML]{CAEDFB}0.878}            & \cellcolor[HTML]{C0C0C0}0.621 \\ \cline{2-22}
% 30dB EER
% LoPy4:    0.131,    0.357,   bold0.13,   0.26
% Mbed1261: bold0.142,0.58,    0.361,      0.239
% Mbed1272: 0.105,    0.423,   bold0.08,   0.252
% Seeeduino: 0.045,   0.068,   bold0.014,  0.213
% Dragino:  bold0.061,0.29,    0.213,      0.37
\multirow{-2}{*}{30 dB} & EER                      & \multicolumn{1}{l|}{\cellcolor[HTML]{E8DFBF}0.131}            & \multicolumn{1}{l|}{\cellcolor[HTML]{FBE2D5}0.357}  & \multicolumn{1}{l|}{\cellcolor[HTML]{CAEDFB}{\textbf{0.13}}}  & \cellcolor[HTML]{C0C0C0}0.26  & \multicolumn{1}{l|}{\cellcolor[HTML]{E8DFBF}{\textbf{0.142}}} & \multicolumn{1}{l|}{\cellcolor[HTML]{FBE2D5}0.58}   & \multicolumn{1}{l|}{\cellcolor[HTML]{CAEDFB}0.361}            & \cellcolor[HTML]{C0C0C0}0.239 & \multicolumn{1}{l|}{\cellcolor[HTML]{E8DFBF}0.105}            & \multicolumn{1}{l|}{\cellcolor[HTML]{FBE2D5}0.423}  & \multicolumn{1}{l|}{\cellcolor[HTML]{CAEDFB}{\textbf{0.08}}}  & \cellcolor[HTML]{C0C0C0}0.252 & \multicolumn{1}{l|}{\cellcolor[HTML]{E8DFBF}0.045}            & \multicolumn{1}{l|}{\cellcolor[HTML]{FBE2D5}0.068}  & \multicolumn{1}{l|}{\cellcolor[HTML]{CAEDFB}{\textbf{0.014}}} & \cellcolor[HTML]{C0C0C0}0.213 & \multicolumn{1}{l|}{\cellcolor[HTML]{E8DFBF}{\textbf{0.061}}} & \multicolumn{1}{l|}{\cellcolor[HTML]{FBE2D5}0.29}   & \multicolumn{1}{l|}{\cellcolor[HTML]{CAEDFB}0.213}            & \cellcolor[HTML]{C0C0C0}0.37  \\ \hline
% 40dB AUC
% LoPy4:    0.987,    0.776,   bold0.995,  0.778
% Mbed1261: bold0.989,0.521,   0.778,      0.789
% Mbed1272: 0.993,    0.436,   bold0.995,  0.775
% Seeeduino: 0.995,   0.982,   bold1,      0.799
% Dragino:  bold0.997,0.909,   0.918,      0.641
                        & AUC                      & \multicolumn{1}{l|}{\cellcolor[HTML]{E8DFBF}0.987}            & \multicolumn{1}{l|}{\cellcolor[HTML]{FBE2D5}0.776}  & \multicolumn{1}{l|}{\cellcolor[HTML]{CAEDFB}{\textbf{0.995}}} & \cellcolor[HTML]{C0C0C0}0.778 & \multicolumn{1}{l|}{\cellcolor[HTML]{E8DFBF}{\textbf{0.989}}} & \multicolumn{1}{l|}{\cellcolor[HTML]{FBE2D5}0.521}  & \multicolumn{1}{l|}{\cellcolor[HTML]{CAEDFB}0.778}            & \cellcolor[HTML]{C0C0C0}0.789 & \multicolumn{1}{l|}{\cellcolor[HTML]{E8DFBF}0.993}            & \multicolumn{1}{l|}{\cellcolor[HTML]{FBE2D5}0.436}  & \multicolumn{1}{l|}{\cellcolor[HTML]{CAEDFB}{\textbf{0.995}}} & \cellcolor[HTML]{C0C0C0}0.775 & \multicolumn{1}{l|}{\cellcolor[HTML]{E8DFBF}0.995}            & \multicolumn{1}{l|}{\cellcolor[HTML]{FBE2D5}0.982}  & \multicolumn{1}{l|}{\cellcolor[HTML]{CAEDFB}{\textbf{1}}}     & \cellcolor[HTML]{C0C0C0}0.799 & \multicolumn{1}{l|}{\cellcolor[HTML]{E8DFBF}{\textbf{0.997}}} & \multicolumn{1}{l|}{\cellcolor[HTML]{FBE2D5}0.909}  & \multicolumn{1}{l|}{\cellcolor[HTML]{CAEDFB}0.918}            & \cellcolor[HTML]{C0C0C0}0.641 \\ \cline{2-22}
% 40dB EER
% LoPy4:    0.031,    0.28,    bold0.037,  0.117
% Mbed1261: bold0.031,0.451,   0.251,      0.267
% Mbed1272: bold0.02, 0.543,   0.042,      0.237
% Seeeduino: 0.011,   0.078,   bold0,      0.222
% Dragino:  bold0.002,0.158,   0.18,       0.342
\multirow{-2}{*}{40 dB} & EER                      & \multicolumn{1}{l|}{\cellcolor[HTML]{E8DFBF}0.031}            & \multicolumn{1}{l|}{\cellcolor[HTML]{FBE2D5}0.28}   & \multicolumn{1}{l|}{\cellcolor[HTML]{CAEDFB}{\textbf{0.037}}} & \cellcolor[HTML]{C0C0C0}0.117 & \multicolumn{1}{l|}{\cellcolor[HTML]{E8DFBF}{\textbf{0.031}}} & \multicolumn{1}{l|}{\cellcolor[HTML]{FBE2D5}0.451}  & \multicolumn{1}{l|}{\cellcolor[HTML]{CAEDFB}0.251}            & \cellcolor[HTML]{C0C0C0}0.267 & \multicolumn{1}{l|}{\cellcolor[HTML]{E8DFBF}{\textbf{0.02}}}  & \multicolumn{1}{l|}{\cellcolor[HTML]{FBE2D5}0.543}  & \multicolumn{1}{l|}{\cellcolor[HTML]{CAEDFB}0.042}            & \cellcolor[HTML]{C0C0C0}0.237 & \multicolumn{1}{l|}{\cellcolor[HTML]{E8DFBF}0.011}            & \multicolumn{1}{l|}{\cellcolor[HTML]{FBE2D5}0.078}  & \multicolumn{1}{l|}{\cellcolor[HTML]{CAEDFB}{\textbf{0}}}     & \cellcolor[HTML]{C0C0C0}0.222 & \multicolumn{1}{l|}{\cellcolor[HTML]{E8DFBF}{\textbf{0.002}}} & \multicolumn{1}{l|}{\cellcolor[HTML]{FBE2D5}0.158}  & \multicolumn{1}{l|}{\cellcolor[HTML]{CAEDFB}0.18}             & \cellcolor[HTML]{C0C0C0}0.342 \\ \hline
\end{tabular}
\begin{tablenotes}
 \item[] 
\begin{tikzpicture}[baseline=0.5ex]
\fill[brown!50] (0,0) rectangle (0.6,0.2);
\end{tikzpicture}
\hspace{0.2em}: results of our approach;
\begin{tikzpicture}[baseline=0.5ex]
\fill[red!20!white] (0,0) rectangle (0.6,0.2);
\end{tikzpicture}
\hspace{0.2em}: results of CNN;
\begin{tikzpicture}[baseline=0.5ex]
\fill[cyan!30!white] (0,0) rectangle (0.6,0.2);
\end{tikzpicture}
\hspace{0.2em}: results of ViT with quasi-square patches;
\begin{tikzpicture}[baseline=0.5ex]
\fill[gray!50] (0,0) rectangle (0.6,0.2);
\end{tikzpicture}
\hspace{0.2em}: results of MobileViT.
 \item[] 
Entries highlighted in bold denote the best detection performance for each dataset, characterized by the highest AUC and the lowest EER.
\end{tablenotes}
\end{threeparttable}%
}
}
\label{tab:eva_identify_arch}
\end{table*}

\subsubsection{Results of Classification}\label{sec:clfnet}
\begin{figure}[!t]
	\begin{center}
		\includegraphics[width = 3.4in]{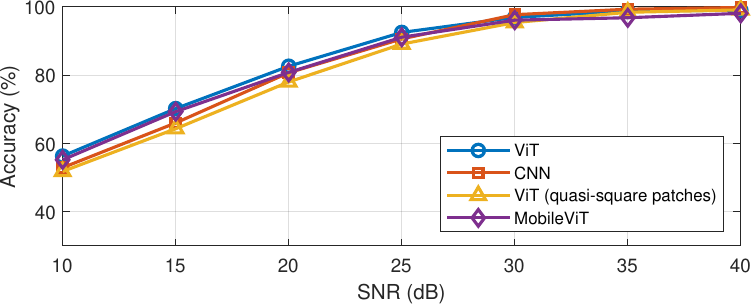}
		\caption{Classification performance comparison among the proposed approach, CNN, ViT with quasi-square patches, and MobileViT.}
		\label{fig:clf_vitcnn}
	\end{center}	
\end{figure}

Fig.~\ref{fig:clf_vitcnn} demonstrates the classification accuracy of the proposed method, CNN, ViT with quasi-square patches, and MobileViT across SNRs ranging from 10 dB to 40 dB. 
Our approach consistently outperforms the baseline models across all SNR conditions.
Specifically, at an SNR of 15~dB, the proposed method achieves a 5.9\% accuracy improvement over the ViT with quasi-square patches. 
This result demonstrates that tailoring the patch partition strategy to the intrinsic time-frequency structure of spectrograms is more effective than directly adopting the quasi-square patch design used in image-based tasks.

\subsection{Performance Evaluation of Multi-Task Authentication} \label{sec:clfnid}
\mmj{KNN, OpenMax, and ELogitNorm also take the fused feature from the ViT extractors as input, consistent with our approach.
}

\subsubsection{Benchmark Design}\label{sec:knnopenmax_setup}
\begin{itemize}
\item \textbf{KNN:}
% Following the parameter setting in~\cite{shen2023towards}, a neighborhood size $K = 15$ is employed for both identification and classification tasks, wherein neighbors are determined based on the Euclidean distance metric.
With reference to the parameter settings in~\cite{shen2023towards} and empirical evaluation of alternative values for the neighborhood size $K$, we finally set $K = 15$ for both identification and classification tasks, wherein neighbors are determined based on the Euclidean distance metric, as it yields favorable overall performance.
\item \textbf{OpenMax:}
% Following the configuration in~\cite{openmax2016cvpr}, a hybrid distance metric combining Euclidean and cosine distances is employed. The tail size is set to 40, and the number of top-score predictions to recalibrate is set to 1. 
Based on the configuration in~\cite{openmax2016cvpr}, we empirically evaluated different parameter combinations (e.g., tail sizes of 20 or 40, and recalibration of the top 1, 2, or 3 score predictions). 
Given its favorable performance, we apply a hybrid distance metric combining Euclidean and cosine distances, with the tail size set to 40 and the number of top-score predictions to recalibrate set to 1.
\mmj{\item \textbf{ELogitNorm:}
Following the configuration in~\cite{ding2026enhancing}, we train
the classifier with the ELogitNorm loss in~\eqref{eq:loss}. 
Regarding the parameter settings, we use $\ell_2$  feature normalization and the maximum logit as the detection score. 
% Specifically, the feature is
% $\ell_2$-normalized before being fed into the
% classifier, and the maximum logit is used as the authentication score at inference.
}
\end{itemize}
\subsubsection{\mmj{Results of Identification}}\label{sec:identify_joint}

\begin{figure}[!t]
	\begin{center}
		\includegraphics[width = 3.4in]{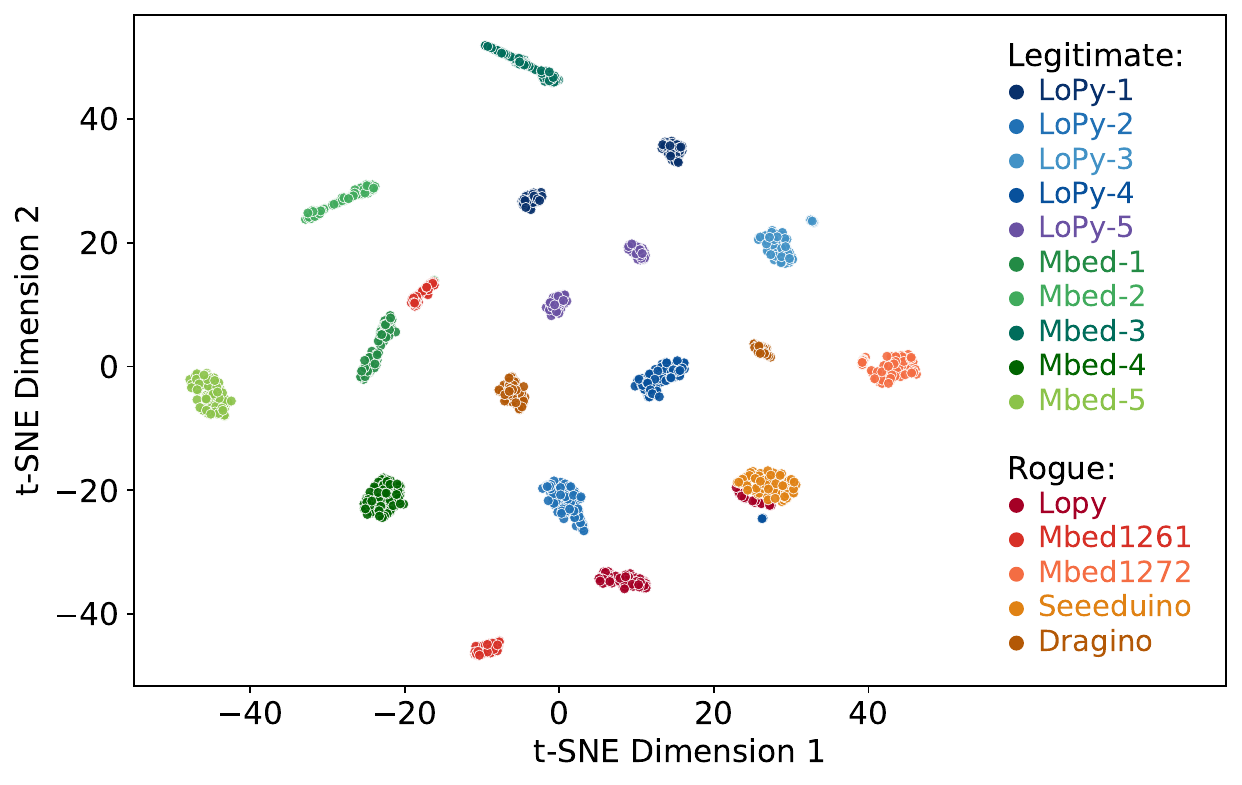}
		\caption{\mmj{t-SNE visualisation of the extracted features for legitimate and rogue devices.}}
		\label{fig:tsne}
	\end{center}	
\end{figure}

\begin{table*}[!t]
\caption{\mmj{Performance of our proposed method, OpenMax, KNN, and ELogitNorm across different rogue devices and SNRs.}}
\centering
{%
  \scalebox{0.9}{
\begin{threeparttable}
\begin{tabular}{|c|c|llll|llll|llll|llll|llll|}
\hline
                        &                          & \multicolumn{4}{c|}{}                                                                                                                                     & \multicolumn{4}{c|}{}                                                                                                                                     & \multicolumn{4}{c|}{}                                                                                                                                     & \multicolumn{4}{c|}{}                                                                                                                                     & \multicolumn{4}{c|}{}                                                                                                                                     \\
\multirow{-2}{*}{SNR}   & \multirow{-2}{*}{Metric} & \multicolumn{4}{c|}{\multirow{-2}{*}{LoPy4}}                                                                                                              & \multicolumn{4}{c|}{\multirow{-2}{*}{Mbed1261}}                                                                                                           & \multicolumn{4}{c|}{\multirow{-2}{*}{Mbed1272}}                                                                                                           & \multicolumn{4}{c|}{\multirow{-2}{*}{Seeeduino}}                                                                                                           & \multicolumn{4}{c|}{\multirow{-2}{*}{Dragino}}                                                                                                            \\ \hline
                         &  AUC                       & \multicolumn{1}{l|}{\cellcolor[HTML]{E8DFBF}0.566} & \multicolumn{1}{l|}{\cellcolor[HTML]{FBE2D5}0.33} & \multicolumn{1}{l|}{\cellcolor[HTML]{CAEDFB}\textbf{0.647}} & \cellcolor[HTML]{D9D9D9}0.498 & \multicolumn{1}{l|}{\cellcolor[HTML]{E8DFBF}\textbf{0.774}} & \multicolumn{1}{l|}{\cellcolor[HTML]{FBE2D5}0.759} & \multicolumn{1}{l|}{\cellcolor[HTML]{CAEDFB}0.598} & \cellcolor[HTML]{D9D9D9}0.426 & \multicolumn{1}{l|}{\cellcolor[HTML]{E8DFBF}0.627} & \multicolumn{1}{l|}{\cellcolor[HTML]{FBE2D5}0.434} & \multicolumn{1}{l|}{\cellcolor[HTML]{CAEDFB}\textbf{0.683}} & \cellcolor[HTML]{D9D9D9}0.594 & \multicolumn{1}{l|}{\cellcolor[HTML]{E8DFBF}0.516} & \multicolumn{1}{l|}{\cellcolor[HTML]{FBE2D5}0.32} & \multicolumn{1}{l|}{\cellcolor[HTML]{CAEDFB}\textbf{0.563}} & \cellcolor[HTML]{D9D9D9}0.449 & \multicolumn{1}{l|}{\cellcolor[HTML]{E8DFBF}0.631} & \multicolumn{1}{l|}{\cellcolor[HTML]{FBE2D5}0.435} & \multicolumn{1}{l|}{\cellcolor[HTML]{CAEDFB}\textbf{0.683}} & \cellcolor[HTML]{D9D9D9}0.564 \\ \cline{2-22} 
\multirow{-2}{*}{10 dB}  &  EER                       & \multicolumn{1}{l|}{\cellcolor[HTML]{E8DFBF}0.421} & \multicolumn{1}{l|}{\cellcolor[HTML]{FBE2D5}0.63} & \multicolumn{1}{l|}{\cellcolor[HTML]{CAEDFB}\textbf{0.419}} & \cellcolor[HTML]{D9D9D9}0.500 & \multicolumn{1}{l|}{\cellcolor[HTML]{E8DFBF}\textbf{0.29}} & \multicolumn{1}{l|}{\cellcolor[HTML]{FBE2D5}0.318} & \multicolumn{1}{l|}{\cellcolor[HTML]{CAEDFB}0.462} & \cellcolor[HTML]{D9D9D9}0.515 & \multicolumn{1}{l|}{\cellcolor[HTML]{E8DFBF}0.402} & \multicolumn{1}{l|}{\cellcolor[HTML]{FBE2D5}0.554} & \multicolumn{1}{l|}{\cellcolor[HTML]{CAEDFB}\textbf{0.372}} & \cellcolor[HTML]{D9D9D9}0.440 & \multicolumn{1}{l|}{\cellcolor[HTML]{E8DFBF}0.497} & \multicolumn{1}{l|}{\cellcolor[HTML]{FBE2D5}0.647} & \multicolumn{1}{l|}{\cellcolor[HTML]{CAEDFB}\textbf{0.463}} & \cellcolor[HTML]{D9D9D9}0.533 & \multicolumn{1}{l|}{\cellcolor[HTML]{E8DFBF}0.39} & \multicolumn{1}{l|}{\cellcolor[HTML]{FBE2D5}0.526} & \multicolumn{1}{l|}{\cellcolor[HTML]{CAEDFB}\textbf{0.381}} & \cellcolor[HTML]{D9D9D9}0.453 \\ \hline
                         &  AUC                       & \multicolumn{1}{l|}{\cellcolor[HTML]{E8DFBF}\textbf{0.842}} & \multicolumn{1}{l|}{\cellcolor[HTML]{FBE2D5}0.702} & \multicolumn{1}{l|}{\cellcolor[HTML]{CAEDFB}0.75} & \cellcolor[HTML]{D9D9D9}0.707 & \multicolumn{1}{l|}{\cellcolor[HTML]{E8DFBF}0.673} & \multicolumn{1}{l|}{\cellcolor[HTML]{FBE2D5}\textbf{0.87}} & \multicolumn{1}{l|}{\cellcolor[HTML]{CAEDFB}0.585} & \cellcolor[HTML]{D9D9D9}0.396 & \multicolumn{1}{l|}{\cellcolor[HTML]{E8DFBF}\textbf{0.897}} & \multicolumn{1}{l|}{\cellcolor[HTML]{FBE2D5}0.87} & \multicolumn{1}{l|}{\cellcolor[HTML]{CAEDFB}0.701} & \cellcolor[HTML]{D9D9D9}0.889 & \multicolumn{1}{l|}{\cellcolor[HTML]{E8DFBF}0.789} & \multicolumn{1}{l|}{\cellcolor[HTML]{FBE2D5}0.521} & \multicolumn{1}{l|}{\cellcolor[HTML]{CAEDFB}0.796} & \cellcolor[HTML]{D9D9D9}\textbf{0.842} & \multicolumn{1}{l|}{\cellcolor[HTML]{E8DFBF}\textbf{0.817}} & \multicolumn{1}{l|}{\cellcolor[HTML]{FBE2D5}0.667} & \multicolumn{1}{l|}{\cellcolor[HTML]{CAEDFB}0.769} & \cellcolor[HTML]{D9D9D9}0.678 \\ \cline{2-22} 
\multirow{-2}{*}{20 dB}  &  EER                       & \multicolumn{1}{l|}{\cellcolor[HTML]{E8DFBF}\textbf{0.239}} & \multicolumn{1}{l|}{\cellcolor[HTML]{FBE2D5}0.364} & \multicolumn{1}{l|}{\cellcolor[HTML]{CAEDFB}0.307} & \cellcolor[HTML]{D9D9D9}0.343 & \multicolumn{1}{l|}{\cellcolor[HTML]{E8DFBF}0.39} & \multicolumn{1}{l|}{\cellcolor[HTML]{FBE2D5}\textbf{0.234}} & \multicolumn{1}{l|}{\cellcolor[HTML]{CAEDFB}0.529} & \cellcolor[HTML]{D9D9D9}0.605 & \multicolumn{1}{l|}{\cellcolor[HTML]{E8DFBF}\textbf{0.169}} & \multicolumn{1}{l|}{\cellcolor[HTML]{FBE2D5}0.19} & \multicolumn{1}{l|}{\cellcolor[HTML]{CAEDFB}0.36} & \cellcolor[HTML]{D9D9D9}0.201 & \multicolumn{1}{l|}{\cellcolor[HTML]{E8DFBF}0.289} & \multicolumn{1}{l|}{\cellcolor[HTML]{FBE2D5}0.488} & \multicolumn{1}{l|}{\cellcolor[HTML]{CAEDFB}0.26} & \cellcolor[HTML]{D9D9D9}\textbf{0.225} & \multicolumn{1}{l|}{\cellcolor[HTML]{E8DFBF}\textbf{0.252}} & \multicolumn{1}{l|}{\cellcolor[HTML]{FBE2D5}0.384} & \multicolumn{1}{l|}{\cellcolor[HTML]{CAEDFB}0.299} & \cellcolor[HTML]{D9D9D9}0.362 \\ \hline
                         &  AUC                       & \multicolumn{1}{l|}{\cellcolor[HTML]{E8DFBF}0.93} & \multicolumn{1}{l|}{\cellcolor[HTML]{FBE2D5}0.893} & \multicolumn{1}{l|}{\cellcolor[HTML]{CAEDFB}0.853} & \cellcolor[HTML]{D9D9D9}\textbf{0.955} & \multicolumn{1}{l|}{\cellcolor[HTML]{E8DFBF}0.929} & \multicolumn{1}{l|}{\cellcolor[HTML]{FBE2D5}\textbf{0.989}} & \multicolumn{1}{l|}{\cellcolor[HTML]{CAEDFB}0.925} & \cellcolor[HTML]{D9D9D9}0.249 & \multicolumn{1}{l|}{\cellcolor[HTML]{E8DFBF}0.944} & \multicolumn{1}{l|}{\cellcolor[HTML]{FBE2D5}0.739} & \multicolumn{1}{l|}{\cellcolor[HTML]{CAEDFB}0.72} & \cellcolor[HTML]{D9D9D9}\textbf{0.999} & \multicolumn{1}{l|}{\cellcolor[HTML]{E8DFBF}0.979} & \multicolumn{1}{l|}{\cellcolor[HTML]{FBE2D5}0.907} & \multicolumn{1}{l|}{\cellcolor[HTML]{CAEDFB}0.805} & \cellcolor[HTML]{D9D9D9}\textbf{0.997} & \multicolumn{1}{l|}{\cellcolor[HTML]{E8DFBF}\textbf{0.981}} & \multicolumn{1}{l|}{\cellcolor[HTML]{FBE2D5}0.878} & \multicolumn{1}{l|}{\cellcolor[HTML]{CAEDFB}0.834} & \cellcolor[HTML]{D9D9D9}0.794 \\ \cline{2-22} 
\multirow{-2}{*}{30 dB}  &  EER                       & \multicolumn{1}{l|}{\cellcolor[HTML]{E8DFBF}0.131} & \multicolumn{1}{l|}{\cellcolor[HTML]{FBE2D5}0.181} & \multicolumn{1}{l|}{\cellcolor[HTML]{CAEDFB}0.199} & \cellcolor[HTML]{D9D9D9}\textbf{0.121} & \multicolumn{1}{l|}{\cellcolor[HTML]{E8DFBF}0.142} & \multicolumn{1}{l|}{\cellcolor[HTML]{FBE2D5}\textbf{0.043}} & \multicolumn{1}{l|}{\cellcolor[HTML]{CAEDFB}0.207} & \cellcolor[HTML]{D9D9D9}0.732 & \multicolumn{1}{l|}{\cellcolor[HTML]{E8DFBF}0.105} & \multicolumn{1}{l|}{\cellcolor[HTML]{FBE2D5}0.308} & \multicolumn{1}{l|}{\cellcolor[HTML]{CAEDFB}0.303} & \cellcolor[HTML]{D9D9D9}\textbf{0.008} & \multicolumn{1}{l|}{\cellcolor[HTML]{E8DFBF}0.045} & \multicolumn{1}{l|}{\cellcolor[HTML]{FBE2D5}0.136} & \multicolumn{1}{l|}{\cellcolor[HTML]{CAEDFB}0.217} & \cellcolor[HTML]{D9D9D9}\textbf{0.014} & \multicolumn{1}{l|}{\cellcolor[HTML]{E8DFBF}\textbf{0.061}} & \multicolumn{1}{l|}{\cellcolor[HTML]{FBE2D5}0.206} & \multicolumn{1}{l|}{\cellcolor[HTML]{CAEDFB}0.236} & \cellcolor[HTML]{D9D9D9}0.306 \\ \hline
                         &  AUC                       & \multicolumn{1}{l|}{\cellcolor[HTML]{E8DFBF}\textbf{0.987}} & \multicolumn{1}{l|}{\cellcolor[HTML]{FBE2D5}0.955} & \multicolumn{1}{l|}{\cellcolor[HTML]{CAEDFB}0.879} & \cellcolor[HTML]{D9D9D9}0.962 & \multicolumn{1}{l|}{\cellcolor[HTML]{E8DFBF}0.989} & \multicolumn{1}{l|}{\cellcolor[HTML]{FBE2D5}\textbf{0.997}} & \multicolumn{1}{l|}{\cellcolor[HTML]{CAEDFB}0.923} & \cellcolor[HTML]{D9D9D9}0.121 & \multicolumn{1}{l|}{\cellcolor[HTML]{E8DFBF}0.993} & \multicolumn{1}{l|}{\cellcolor[HTML]{FBE2D5}0.885} & \multicolumn{1}{l|}{\cellcolor[HTML]{CAEDFB}0.777} & \cellcolor[HTML]{D9D9D9}\textbf{1} & \multicolumn{1}{l|}{\cellcolor[HTML]{E8DFBF}0.995} & \multicolumn{1}{l|}{\cellcolor[HTML]{FBE2D5}0.924} & \multicolumn{1}{l|}{\cellcolor[HTML]{CAEDFB}0.892} & \cellcolor[HTML]{D9D9D9}\textbf{1} & \multicolumn{1}{l|}{\cellcolor[HTML]{E8DFBF}\textbf{0.997}} & \multicolumn{1}{l|}{\cellcolor[HTML]{FBE2D5}0.909} & \multicolumn{1}{l|}{\cellcolor[HTML]{CAEDFB}0.87} & \cellcolor[HTML]{D9D9D9}0.801 \\ \cline{2-22} 
\multirow{-2}{*}{40 dB}  &  EER                       & \multicolumn{1}{l|}{\cellcolor[HTML]{E8DFBF}\textbf{0.031}} & \multicolumn{1}{l|}{\cellcolor[HTML]{FBE2D5}0.098} & \multicolumn{1}{l|}{\cellcolor[HTML]{CAEDFB}0.154} & \cellcolor[HTML]{D9D9D9}0.120 & \multicolumn{1}{l|}{\cellcolor[HTML]{E8DFBF}0.031} & \multicolumn{1}{l|}{\cellcolor[HTML]{FBE2D5}\textbf{0.012}} & \multicolumn{1}{l|}{\cellcolor[HTML]{CAEDFB}0.161} & \cellcolor[HTML]{D9D9D9}0.831 & \multicolumn{1}{l|}{\cellcolor[HTML]{E8DFBF}0.02} & \multicolumn{1}{l|}{\cellcolor[HTML]{FBE2D5}0.158} & \multicolumn{1}{l|}{\cellcolor[HTML]{CAEDFB}0.243} & \cellcolor[HTML]{D9D9D9}\textbf{0} & \multicolumn{1}{l|}{\cellcolor[HTML]{E8DFBF}0.011} & \multicolumn{1}{l|}{\cellcolor[HTML]{FBE2D5}0.093} & \multicolumn{1}{l|}{\cellcolor[HTML]{CAEDFB}0.141} & \cellcolor[HTML]{D9D9D9}\textbf{0.005} & \multicolumn{1}{l|}{\cellcolor[HTML]{E8DFBF}\textbf{0.002}} & \multicolumn{1}{l|}{\cellcolor[HTML]{FBE2D5}0.191} & \multicolumn{1}{l|}{\cellcolor[HTML]{CAEDFB}0.199} & \cellcolor[HTML]{D9D9D9}0.340 \\ \hline
\end{tabular}
\begin{tablenotes}
 \item[] 
\begin{tikzpicture}[baseline=0.5ex]
\fill[brown!50] (0,0) rectangle (0.6,0.2);
\end{tikzpicture}
\hspace{0.2em}: results of our approach;
\begin{tikzpicture}[baseline=0.5ex]
\fill[red!20!white] (0,0) rectangle (0.6,0.2);
\end{tikzpicture}
\hspace{0.2em}: results of KNN;
\begin{tikzpicture}[baseline=0.5ex]
\fill[cyan!30!white] (0,0) rectangle (0.6,0.2);
\end{tikzpicture}
\hspace{0.2em}: results of OpenMax.
\begin{tikzpicture}[baseline=0.5ex]
\fill[gray!50] (0,0) rectangle (0.6,0.2);
\end{tikzpicture}
\hspace{0.2em}: results of ELogitNorm.
 \item[] 
Entries highlighted in bold denote the best detection performance for each dataset, characterized by the highest AUC and the lowest EER.
\end{tablenotes}
\end{threeparttable}}
}
\label{tab:eva_identify_knn}
\end{table*}

\begin{figure}[!t]
	\begin{center}
		\includegraphics[width = 3.4in]{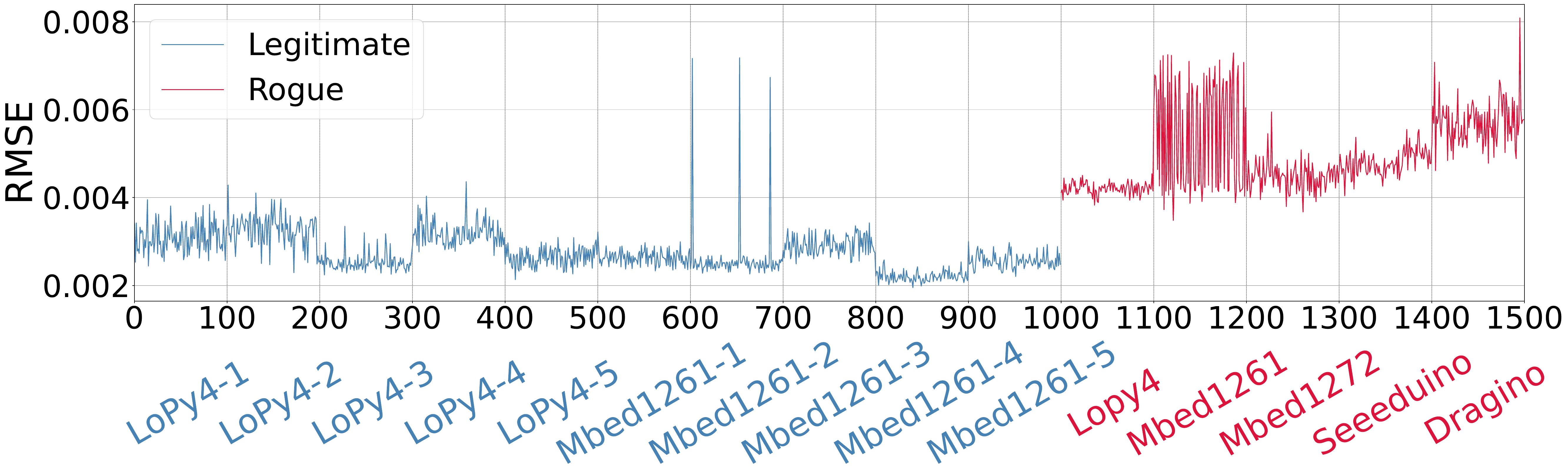}
		\caption{\mmj{Reconstruction error of packets from legitimate and rogue devices.}}
		\label{fig:rmse}
	\end{center}	
\end{figure}

\mmj{
To provide an intuitive view of the feature distributions, Fig.~\ref{fig:tsne} presents a t-SNE visualization of the features extracted from legitimate and rogue devices. The features generally form separate clusters, suggesting that the extracted features are discriminative.
As presented in Table~\ref{tab:eva_identify_knn}, 20 comparative experiments were performed at different SNRs and with different rogue devices.
Among the compared methods, the proposed approach achieves the best performance in the majority of cases.
Fig.~\ref{fig:rmse} shows the reconstruction errors of packets from legitimate and rogue devices obtained using our proposed approach at an SNR of 40 dB. 
Rogue devices exhibit higher reconstruction errors than legitimate devices, demonstrating the effectiveness of the proposed approach.

Although benchmark methods exhibit competitive performance in certain scenarios, such as KNN on Mbed1261 within 20 to 40 dB, OpenMax on Dragino at 10 dB, and ELogitNorm on Mbed1272 at 30 dB, their overall performance remains inferior to that of the proposed approach.}
KNN directly applies the distance metric to measure the similarity between data, without requiring an understanding of the training data, which can lead to incorrect results during inference~\cite{zhao2023}.
\mmj{Furthermore, KNN, OpenMax, and ELogitNorm yielded AUC values below 0.5 in some cases.
This is particularly evident for ELogitNorm when the rogue device is Mbed1261.}
An AUC score lower than 0.5 indicates that the model performs worse than random guessing, rendering it unreliable for identification.

\subsubsection{Results of Classification}\label{sec:clf}

\begin{figure}[!t]
	\begin{center}
		\includegraphics[width = 3.4in]{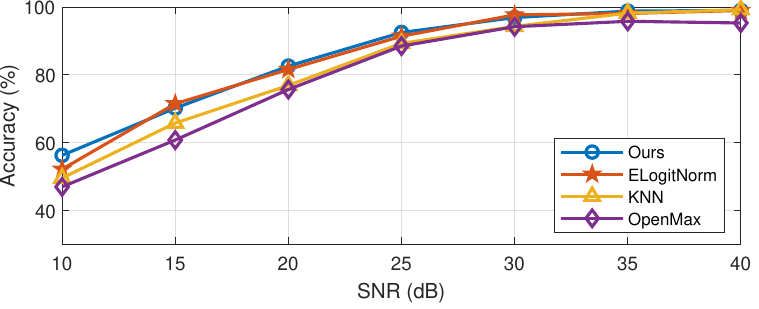}
		\caption{\mmj{Classification performance comparison among the proposed approach, KNN, OpenMax, and ELogitNorm.}}
		\label{fig:clf_vitknnopenmax}
	\end{center}	
\end{figure}

\mmj{Fig.~\ref{fig:clf_vitknnopenmax} illustrates the classification accuracy of our proposed method, KNN, OpenMax, and ELogitNorm under various SNRs.}
The four methods show comparable performance, with accuracies ranging from 95\% to 100\%, when the SNR is approximately 35 to 40 dB.
% Both KNN and OpenMax show significant performance degradation under noisy conditions.
\mmj{However, at SNRs below 30 dB, a more evident performance improvement is observed for the proposed method and ELogitNorm.}
Noisy conditions degrade the quality of the signal by introducing random noise that distorts discriminative features. 
OpenMax, a post-processing method, heavily relies on the features from the extractor~\cite{bisgin2024large}, demonstrating reduced effectiveness in noisy environments owing to its constrained adaptability.
The operation of KNN also depends on the provided features and demonstrates similar behavior.
\mmj{In contrast, the parameters of the FC layer and ELogitNorm are optimized via gradient-based iterative training, enabling improved adaptability to diverse environments.}

\section{Experimental Evaluation in Real Wireless Scenarios}\label{sec:expreal}
In the previous evaluation section, varying SNR conditions were emulated by injecting noise into high-SNR signals, rather than measured during actual wireless transmission.
Although such a setup is useful for preliminary analysis and controlled comparisons, it does not fully reflect the complexity and variability of real-world wireless environments. 
To further validate the real-world applicability of the proposed approach, this section presents evaluations under various realistic channel conditions with different SNR levels.
The same training dataset and trained model were used.

\subsection{Test Dataset Collection}\label{realsetup}
\begin{figure}[!t]
	\begin{center}
		\includegraphics[width = 3.4in]{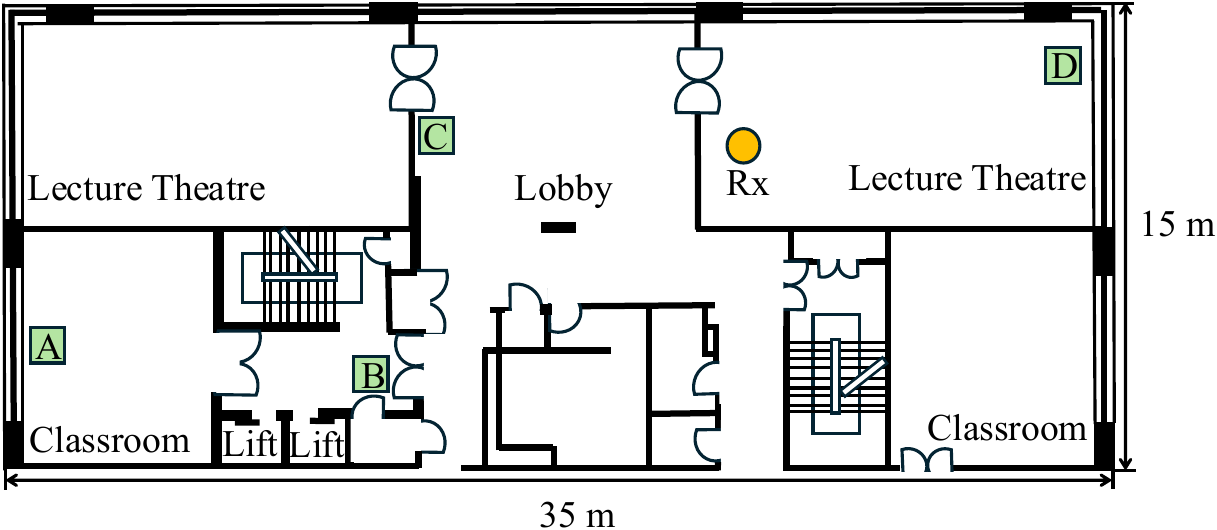}
		\caption{The floor plan of the experimental environment.}
		\label{fig:floorplan_vit}
	\end{center}	
\end{figure}

A new set of test data was collected under a range of realistic wireless channel conditions to evaluate the proposed approach.
Fig.~\ref{fig:floorplan_vit} illustrates the layout of the experimental environment used for data collection. During the experiments, the USRP N210 receiver was fixed at a designated location in the lecture theatre, while the legitimate and rogue transmitters were placed at four different locations (Locations A–D) in turn. These locations corresponded to SNR ranges of 10–20 dB, 20–30 dB, 30–40 dB, and above 40 dB, respectively. Both the receiver and the transmitters remained stationary during the collection of each dataset. For each test dataset, 100 packets were captured from every transmitter.

\subsection{Results of Identification}\label{sec:identify_real}
\subsubsection{\mmj{Performance}}
\begin{table*}[!t]
\caption{\mmj{Performance of our proposed method, KNN, OpenMax, and ELogitNorm, across different rogue devices and different locations.}}
\centering
{%
  
\scalebox{0.9}
{\begin{threeparttable}
\begin{tabular}{|c|c|llll|llll|llll|llll|llll|}
\hline
                         &                          & \multicolumn{4}{c|}{}                                                                                                                                     & \multicolumn{4}{c|}{}                                                                                                                                     & \multicolumn{4}{c|}{}                                                                                                                                     & \multicolumn{4}{c|}{}                                                                                                                                     & \multicolumn{4}{c|}{}                                                                                                                            \\
\multirow{-2}{*}{Location}    & \multirow{-2}{*}{Metric} & \multicolumn{4}{c|}{\multirow{-2}{*}{LoPy4}}                                                                                                              & \multicolumn{4}{c|}{\multirow{-2}{*}{Mbed1261}}                                                                                                           & \multicolumn{4}{c|}{\multirow{-2}{*}{Mbed1272}}                                                                                                           & \multicolumn{4}{c|}{\multirow{-2}{*}{Seeeduino}}                                                                                                           & \multicolumn{4}{c|}{\multirow{-2}{*}{Dragino}}                                                                                                   \\ \hline
                          &  AUC                       & \multicolumn{1}{l|}{\cellcolor[HTML]{E8DFBF}0.581} & \multicolumn{1}{l|}{\cellcolor[HTML]{FBE2D5}0.42} & \multicolumn{1}{l|}{\cellcolor[HTML]{CAEDFB}\textbf{0.656}} & \cellcolor[HTML]{D9D9D9}0.605 & \multicolumn{1}{l|}{\cellcolor[HTML]{E8DFBF}0.557} & \multicolumn{1}{l|}{\cellcolor[HTML]{FBE2D5}\textbf{0.77}} & \multicolumn{1}{l|}{\cellcolor[HTML]{CAEDFB}0.424} & \cellcolor[HTML]{D9D9D9}0.465 & \multicolumn{1}{l|}{\cellcolor[HTML]{E8DFBF}0.696} & \multicolumn{1}{l|}{\cellcolor[HTML]{FBE2D5}0.544} & \multicolumn{1}{l|}{\cellcolor[HTML]{CAEDFB}\textbf{0.734}} & \cellcolor[HTML]{D9D9D9}0.655 & \multicolumn{1}{l|}{\cellcolor[HTML]{E8DFBF}0.682} & \multicolumn{1}{l|}{\cellcolor[HTML]{FBE2D5}0.26} & \multicolumn{1}{l|}{\cellcolor[HTML]{CAEDFB}\textbf{0.832}} & \cellcolor[HTML]{D9D9D9}0.527 & \multicolumn{1}{l|}{\cellcolor[HTML]{E8DFBF}\textbf{0.701}} & \multicolumn{1}{l|}{\cellcolor[HTML]{FBE2D5}0.635} & \multicolumn{1}{l|}{\cellcolor[HTML]{CAEDFB}0.541} & \cellcolor[HTML]{D9D9D9}0.645 \\ \cline{2-22} 
\multirow{-2}{*}{A}  &  EER                       & \multicolumn{1}{l|}{\cellcolor[HTML]{E8DFBF}0.429} & \multicolumn{1}{l|}{\cellcolor[HTML]{FBE2D5}0.591} & \multicolumn{1}{l|}{\cellcolor[HTML]{CAEDFB}\textbf{0.399}} & \cellcolor[HTML]{D9D9D9}0.422 & \multicolumn{1}{l|}{\cellcolor[HTML]{E8DFBF}0.508} & \multicolumn{1}{l|}{\cellcolor[HTML]{FBE2D5}\textbf{0.333}} & \multicolumn{1}{l|}{\cellcolor[HTML]{CAEDFB}0.608} & \cellcolor[HTML]{D9D9D9}0.550 & \multicolumn{1}{l|}{\cellcolor[HTML]{E8DFBF}0.364} & \multicolumn{1}{l|}{\cellcolor[HTML]{FBE2D5}0.473} & \multicolumn{1}{l|}{\cellcolor[HTML]{CAEDFB}\textbf{0.349}} & \cellcolor[HTML]{D9D9D9}0.372 & \multicolumn{1}{l|}{\cellcolor[HTML]{E8DFBF}0.388} & \multicolumn{1}{l|}{\cellcolor[HTML]{FBE2D5}0.669} & \multicolumn{1}{l|}{\cellcolor[HTML]{CAEDFB}\textbf{0.231}} & \cellcolor[HTML]{D9D9D9}0.483 & \multicolumn{1}{l|}{\cellcolor[HTML]{E8DFBF}\textbf{0.351}} & \multicolumn{1}{l|}{\cellcolor[HTML]{FBE2D5}0.419} & \multicolumn{1}{l|}{\cellcolor[HTML]{CAEDFB}0.441} & \cellcolor[HTML]{D9D9D9}0.379 \\ \hline
                          &  AUC                       & \multicolumn{1}{l|}{\cellcolor[HTML]{E8DFBF}0.735} & \multicolumn{1}{l|}{\cellcolor[HTML]{FBE2D5}0.759} & \multicolumn{1}{l|}{\cellcolor[HTML]{CAEDFB}0.778} & \cellcolor[HTML]{D9D9D9}\textbf{0.874} & \multicolumn{1}{l|}{\cellcolor[HTML]{E8DFBF}0.725} & \multicolumn{1}{l|}{\cellcolor[HTML]{FBE2D5}\textbf{0.801}} & \multicolumn{1}{l|}{\cellcolor[HTML]{CAEDFB}0.648} & \cellcolor[HTML]{D9D9D9}0.382 & \multicolumn{1}{l|}{\cellcolor[HTML]{E8DFBF}0.84} & \multicolumn{1}{l|}{\cellcolor[HTML]{FBE2D5}0.717} & \multicolumn{1}{l|}{\cellcolor[HTML]{CAEDFB}0.569} & \cellcolor[HTML]{D9D9D9}\textbf{0.997} & \multicolumn{1}{l|}{\cellcolor[HTML]{E8DFBF}0.844} & \multicolumn{1}{l|}{\cellcolor[HTML]{FBE2D5}0.801} & \multicolumn{1}{l|}{\cellcolor[HTML]{CAEDFB}\textbf{0.858}} & \cellcolor[HTML]{D9D9D9}0.832 & \multicolumn{1}{l|}{\cellcolor[HTML]{E8DFBF}\textbf{0.915}} & \multicolumn{1}{l|}{\cellcolor[HTML]{FBE2D5}0.752} & \multicolumn{1}{l|}{\cellcolor[HTML]{CAEDFB}0.873} & \cellcolor[HTML]{D9D9D9}0.586 \\ \cline{2-22} 
\multirow{-2}{*}{B}  &  EER                       & \multicolumn{1}{l|}{\cellcolor[HTML]{E8DFBF}0.328} & \multicolumn{1}{l|}{\cellcolor[HTML]{FBE2D5}0.301} & \multicolumn{1}{l|}{\cellcolor[HTML]{CAEDFB}0.306} & \cellcolor[HTML]{D9D9D9}\textbf{0.211} & \multicolumn{1}{l|}{\cellcolor[HTML]{E8DFBF}0.386} & \multicolumn{1}{l|}{\cellcolor[HTML]{FBE2D5}\textbf{0.309}} & \multicolumn{1}{l|}{\cellcolor[HTML]{CAEDFB}0.459} & \cellcolor[HTML]{D9D9D9}0.595 & \multicolumn{1}{l|}{\cellcolor[HTML]{E8DFBF}0.242} & \multicolumn{1}{l|}{\cellcolor[HTML]{FBE2D5}0.36} & \multicolumn{1}{l|}{\cellcolor[HTML]{CAEDFB}0.45} & \cellcolor[HTML]{D9D9D9}\textbf{0.028} & \multicolumn{1}{l|}{\cellcolor[HTML]{E8DFBF}0.222} & \multicolumn{1}{l|}{\cellcolor[HTML]{FBE2D5}0.279} & \multicolumn{1}{l|}{\cellcolor[HTML]{CAEDFB}\textbf{0.209}} & \cellcolor[HTML]{D9D9D9}0.243 & \multicolumn{1}{l|}{\cellcolor[HTML]{E8DFBF}\textbf{0.17}} & \multicolumn{1}{l|}{\cellcolor[HTML]{FBE2D5}0.318} & \multicolumn{1}{l|}{\cellcolor[HTML]{CAEDFB}0.211} & \cellcolor[HTML]{D9D9D9}0.441 \\ \hline
                          &  AUC                       & \multicolumn{1}{l|}{\cellcolor[HTML]{E8DFBF}\textbf{0.958}} & \multicolumn{1}{l|}{\cellcolor[HTML]{FBE2D5}0.903} & \multicolumn{1}{l|}{\cellcolor[HTML]{CAEDFB}0.896} & \cellcolor[HTML]{D9D9D9}0.848 & \multicolumn{1}{l|}{\cellcolor[HTML]{E8DFBF}0.812} & \multicolumn{1}{l|}{\cellcolor[HTML]{FBE2D5}\textbf{0.874}} & \multicolumn{1}{l|}{\cellcolor[HTML]{CAEDFB}0.731} & \cellcolor[HTML]{D9D9D9}0.166 & \multicolumn{1}{l|}{\cellcolor[HTML]{E8DFBF}0.988} & \multicolumn{1}{l|}{\cellcolor[HTML]{FBE2D5}0.821} & \multicolumn{1}{l|}{\cellcolor[HTML]{CAEDFB}0.752} & \cellcolor[HTML]{D9D9D9}\textbf{1} & \multicolumn{1}{l|}{\cellcolor[HTML]{E8DFBF}0.975} & \multicolumn{1}{l|}{\cellcolor[HTML]{FBE2D5}0.848} & \multicolumn{1}{l|}{\cellcolor[HTML]{CAEDFB}0.88} & \cellcolor[HTML]{D9D9D9}\textbf{0.983} & \multicolumn{1}{l|}{\cellcolor[HTML]{E8DFBF}\textbf{0.991}} & \multicolumn{1}{l|}{\cellcolor[HTML]{FBE2D5}0.821} & \multicolumn{1}{l|}{\cellcolor[HTML]{CAEDFB}0.837} & \cellcolor[HTML]{D9D9D9}0.83 \\ \cline{2-22} 
\multirow{-2}{*}{C}  &  EER                       & \multicolumn{1}{l|}{\cellcolor[HTML]{E8DFBF}\textbf{0.1}} & \multicolumn{1}{l|}{\cellcolor[HTML]{FBE2D5}0.177} & \multicolumn{1}{l|}{\cellcolor[HTML]{CAEDFB}0.2} & \cellcolor[HTML]{D9D9D9}0.270 & \multicolumn{1}{l|}{\cellcolor[HTML]{E8DFBF}0.34} & \multicolumn{1}{l|}{\cellcolor[HTML]{FBE2D5}\textbf{0.252}} & \multicolumn{1}{l|}{\cellcolor[HTML]{CAEDFB}0.488} & \cellcolor[HTML]{D9D9D9}0.675 & \multicolumn{1}{l|}{\cellcolor[HTML]{E8DFBF}0.052} & \multicolumn{1}{l|}{\cellcolor[HTML]{FBE2D5}0.202} & \multicolumn{1}{l|}{\cellcolor[HTML]{CAEDFB}0.271} & \cellcolor[HTML]{D9D9D9}\textbf{0.008} & \multicolumn{1}{l|}{\cellcolor[HTML]{E8DFBF}0.089} & \multicolumn{1}{l|}{\cellcolor[HTML]{FBE2D5}0.18} & \multicolumn{1}{l|}{\cellcolor[HTML]{CAEDFB}0.212} & \cellcolor[HTML]{D9D9D9}\textbf{0.085} & \multicolumn{1}{l|}{\cellcolor[HTML]{E8DFBF}\textbf{0.052}} & \multicolumn{1}{l|}{\cellcolor[HTML]{FBE2D5}0.258} & \multicolumn{1}{l|}{\cellcolor[HTML]{CAEDFB}0.262} & \cellcolor[HTML]{D9D9D9}0.301 \\ \hline
                          &  AUC                       & \multicolumn{1}{l|}{\cellcolor[HTML]{E8DFBF}\textbf{0.996}} & \multicolumn{1}{l|}{\cellcolor[HTML]{FBE2D5}0.987} & \multicolumn{1}{l|}{\cellcolor[HTML]{CAEDFB}0.953} & \cellcolor[HTML]{D9D9D9}0.893 & \multicolumn{1}{l|}{\cellcolor[HTML]{E8DFBF}\textbf{0.985}} & \multicolumn{1}{l|}{\cellcolor[HTML]{FBE2D5}0.984} & \multicolumn{1}{l|}{\cellcolor[HTML]{CAEDFB}0.709} & \cellcolor[HTML]{D9D9D9}0.152 & \multicolumn{1}{l|}{\cellcolor[HTML]{E8DFBF}\textbf{1}} & \multicolumn{1}{l|}{\cellcolor[HTML]{FBE2D5}0.989} & \multicolumn{1}{l|}{\cellcolor[HTML]{CAEDFB}0.794} & \cellcolor[HTML]{D9D9D9}\textbf{1} & \multicolumn{1}{l|}{\cellcolor[HTML]{E8DFBF}\textbf{0.997}} & \multicolumn{1}{l|}{\cellcolor[HTML]{FBE2D5}0.991} & \multicolumn{1}{l|}{\cellcolor[HTML]{CAEDFB}0.923} & \cellcolor[HTML]{D9D9D9}0.931 & \multicolumn{1}{l|}{\cellcolor[HTML]{E8DFBF}\textbf{0.999}} & \multicolumn{1}{l|}{\cellcolor[HTML]{FBE2D5}0.954} & \multicolumn{1}{l|}{\cellcolor[HTML]{CAEDFB}0.865} & \cellcolor[HTML]{D9D9D9}0.885 \\ \cline{2-22} 
\multirow{-2}{*}{D}  &  EER                       & \multicolumn{1}{l|}{\cellcolor[HTML]{E8DFBF}\textbf{0.011}} & \multicolumn{1}{l|}{\cellcolor[HTML]{FBE2D5}0.06} & \multicolumn{1}{l|}{\cellcolor[HTML]{CAEDFB}0.078} & \cellcolor[HTML]{D9D9D9}0.228 & \multicolumn{1}{l|}{\cellcolor[HTML]{E8DFBF}\textbf{0.048}} & \multicolumn{1}{l|}{\cellcolor[HTML]{FBE2D5}0.07} & \multicolumn{1}{l|}{\cellcolor[HTML]{CAEDFB}0.501} & \cellcolor[HTML]{D9D9D9}0.712 & \multicolumn{1}{l|}{\cellcolor[HTML]{E8DFBF}0.002} & \multicolumn{1}{l|}{\cellcolor[HTML]{FBE2D5}0.026} & \multicolumn{1}{l|}{\cellcolor[HTML]{CAEDFB}0.224} & \cellcolor[HTML]{D9D9D9}\textbf{0} & \multicolumn{1}{l|}{\cellcolor[HTML]{E8DFBF}\textbf{0.011}} & \multicolumn{1}{l|}{\cellcolor[HTML]{FBE2D5}0.031} & \multicolumn{1}{l|}{\cellcolor[HTML]{CAEDFB}0.114} & \cellcolor[HTML]{D9D9D9}0.131 & \multicolumn{1}{l|}{\cellcolor[HTML]{E8DFBF}\textbf{0.002}} & \multicolumn{1}{l|}{\cellcolor[HTML]{FBE2D5}0.12} & \multicolumn{1}{l|}{\cellcolor[HTML]{CAEDFB}0.222} & \cellcolor[HTML]{D9D9D9}0.185 \\ \hline
\end{tabular}
\begin{tablenotes}
 \item[] 
\begin{tikzpicture}[baseline=0.5ex]
\fill[brown!50] (0,0) rectangle (0.6,0.2);
\end{tikzpicture}
\hspace{0.2em}: results of our approach;
\begin{tikzpicture}[baseline=0.5ex]
\fill[red!20!white] (0,0) rectangle (0.6,0.2);
\end{tikzpicture}
\hspace{0.2em}: results of KNN;
\begin{tikzpicture}[baseline=0.5ex]
\fill[cyan!30!white] (0,0) rectangle (0.6,0.2);
\end{tikzpicture}
\hspace{0.2em}: results of OpenMax.
\begin{tikzpicture}[baseline=0.5ex]
\fill[gray!50] (0,0) rectangle (0.6,0.2);
\end{tikzpicture}
\hspace{0.2em}: results of ELogitNorm.
 \item[] 
Entries highlighted in bold denote the best detection performance for each dataset, characterized by the highest AUC and the lowest EER.
\end{tablenotes}
\end{threeparttable}}
}
\label{tab:eva_identify_real}
\end{table*}

\begin{figure}[!t]
\centering
\subfloat[]{\includegraphics[width=1.75in]{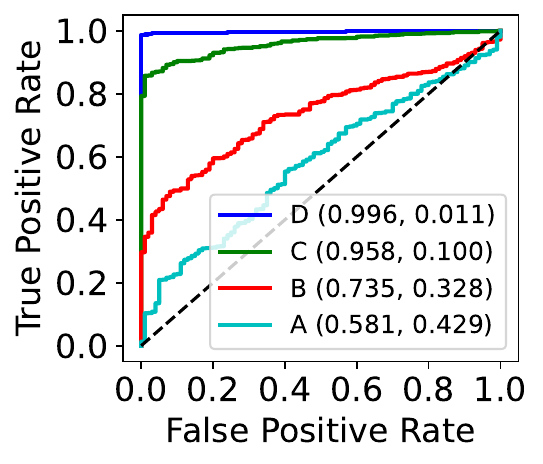}\label{fig:spec_wo_aa}}
% \hspace{0.5cm}
\subfloat[]{\includegraphics[width=1.75in]{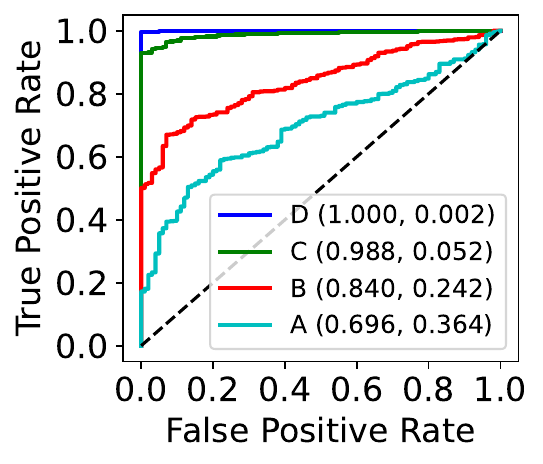}\label{fig:pgdv}}
\caption{ROC curves of the proposed open-set RFFI system for different rogue devices: (a) LoPy4 and (b) Mbed1272. 
Legend entries follow the format: Location (AUC, EER).}
\label{fig:roc}
\end{figure}

\mmj{Table~\ref{tab:eva_identify_real} compares our method with KNN, OpenMax, and ELogitNorm across various deployment locations and five rogue device scenarios.}
Across all test scenarios, the proposed method demonstrates its best performance at Location D under high-SNR conditions (over 40 dB), with AUC values ranging from 0.985 to 1 and correspondingly low EERs between 0.002 and 0.048.
% At Locations A and B, a noticeable degradation in performance is observed regardless of the specific rogue device involved or approach.

At Locations A and B, the performance is significantly inferior to that at Locations C and D for all evaluated rogue devices and approaches.
To facilitate a more detailed analysis, the ROC curves are presented in Fig.~\ref{fig:roc}(a) with the LoPy4 device as the rogue transmitter.
It can be found that the AUC of the proposed approach is only 0.581 and 0.735 at Locations A and B, respectively, whereas it reaches 0.958 and 0.996 at Locations C and D.
A similar trend is also observed when Mbed1272 acts as the rogue device, as shown in Fig.~\ref{fig:roc}(b).
The inferior performance at Locations A and B are attributed to the lower SNRs at these locations, which indicate increased distortion in the signals.

Compared to other rogue devices, detection is less effective when the rogue device is LoPy4 or Mbed1261.
For instance, at Location B, the AUCs drop to around 0.72 with these two rogue devices, whereas for the remaining rogue devices, the AUCs consistently exceed 0.84.
The phenomenon can be attributed to the composition of the training dataset, which consists of legitimate LoPy4 and Mbed1261 devices. 
Devices of the same type tend to exhibit similar hardware features. 
The AE performs anomaly detection by accurately reconstructing legitimate devices while producing higher reconstruction errors for rogue devices. 
\mmj{However, when rogue devices closely resemble legitimate ones, they might also be reconstructed with low error, making them more difficult to identify as anomalies.
% As shown in Fig.~\ref{fig:rmse}, we plot the reconstruction error of packets from legitimate and rogue devices at
% Location B
}

% \begin{figure}[!t]
% 	\begin{center}
% 		\includegraphics[width = 3.4in]{Pictures/RMSE_B.pdf}
% 		\caption{\mmj{Reconstruction error of packets from legitimate and rogue devices at Location B.}}
% 		\label{fig:rmse}
% 	\end{center}	
% \end{figure}

\subsubsection{Threshold Selection}\label{threshold}

In RFFI systems, the relative importance of the false positive rate (FPR) and true positive rate (TPR) depends on the application scenario, and the decision threshold should be adapted to it.
For security-critical applications, minimizing the FPR is of primary importance.
According to Fig.~\ref{fig:roc}(a), with LoPy4 as the rogue device, when FPR is set to 0.05 for Location~C, the corresponding threshold is 0.0041. 
On the other hand, in user-experience-focused scenarios, e.g., consumer electronics pairing and smart-home IoT ecosystems, maximizing TPR is often prioritized to provide a smooth experience for legitimate users and ensure that legitimate devices are not missed. 
As shown in Fig.~\ref{fig:roc}(a), at Location C, a TPR of 0.95 is achieved when the threshold is set to 0.0037.
In summary, selecting an appropriate threshold is crucial for achieving an optimal trade-off between security and usability in authentication systems.

\subsection{Results of Classification}\label{clf_real}

\begin{figure}[!t]
	\begin{center}
		\includegraphics[width = 3.4in]{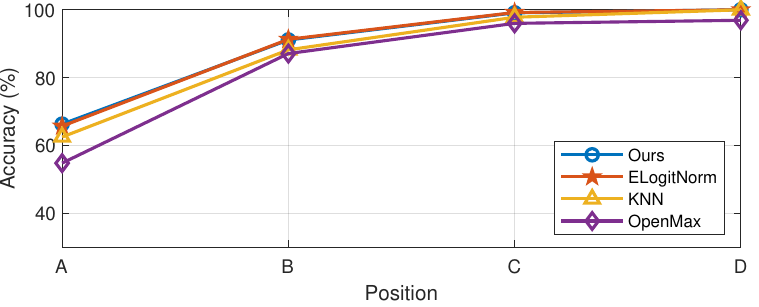}
		\caption{\mmj{Classification performance comparison among the proposed approach, KNN, OpenMax, and ELogitNorm in a real environment.}}
		\label{fig:clf_real}
	\end{center}	
\end{figure}

\mmj{Fig.~\ref{fig:clf_real} demonstrates the classification accuracy of our proposed method, KNN, OpenMax, and ELogitNorm at different locations. 
This observation is consistent with the trend shown in Fig.~\ref{fig:clf_vitknnopenmax}, where the proposed approach and ELogitNorm tend to perform best, especially at low SNRs.}
% All methods experience a performance drop of approximately 25\% relative to the results observed from Location B (91.1\%) to Location A (66.3\%).
All methods exhibit an approximately 25\% relative performance degradation when moving from Location B to Location A. Taking our method as an example, the accuracy decreases from 91.1\% at Location B to 66.3\% at Location A.
This performance gap can be attributed to more severe channel effects at lower SNRs, where increased noise makes it more challenging to extract discriminative features for classification.

\begin{figure}[!t]
	\centering
	\subfloat[]{\includegraphics[width=1.7in]{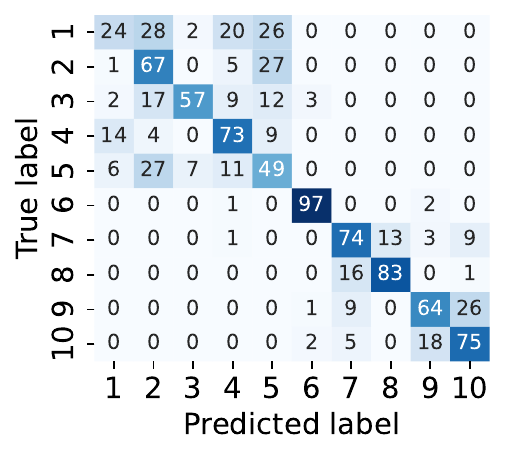}
		\label{fig:clf2010}}
    \subfloat[]{\includegraphics[width=1.7in]{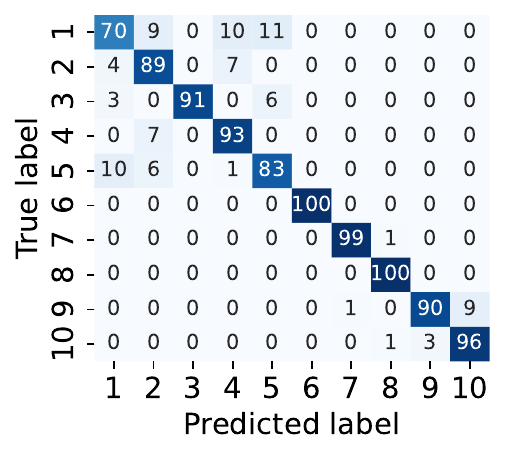}
    \label{fig:clf3020}}

    \subfloat[]{\includegraphics[width=1.7in]{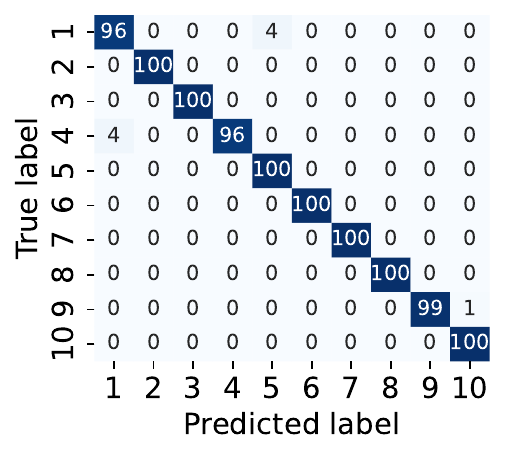}
    \label{fig:clf4030}}
    \subfloat[]{\includegraphics[width=1.7in]{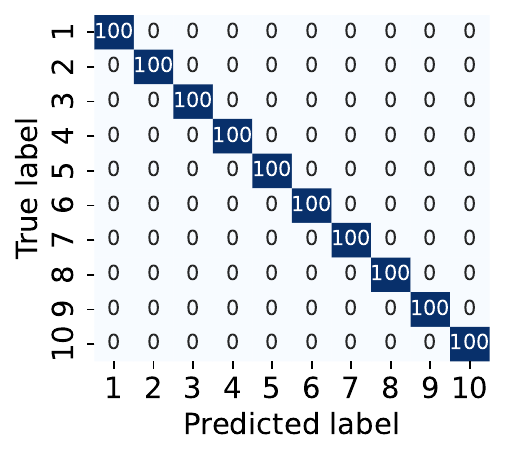}
    \label{fig:clfover40}}
        
	\caption{
    Classification performance at different transmitter locations. (a) Location A (10 to 20 dB), Acc = 66.3\%. (b) Location B (20 to 30 dB), Acc = 91.1\%. (c) Location C (30 to 40 dB), Acc = 99.1\%. (d) Location D (over 40 dB), Acc = 100\%.
    }
	\label{fig:clfabcd}
\end{figure}

Fig.~\ref{fig:clfabcd} presents the confusion matrices of the proposed method evaluated on datasets collected at Locations A through D.
The results indicate that the proposed model exhibits consistency with the simulation outcomes in Section~\ref{sec:experiment_simulated}. 
Specifically, the classification accuracy reaches 100\% for SNR values above 40 dB and 99.1\% within the 30–40 dB range, closely matching the performance illustrated earlier.
The consistency highlights the robustness and reliability of the proposed method under various wireless conditions and validates its effectiveness in realistic deployment scenarios.

Fig.~\ref{fig:clfabcd}(a) shows that most misclassifications occur among devices of the same series, such as within the groups LoPy4 1–5 and Mbed1261 6–10. 
This is likely due to the high similarity in manufacturing processes among devices from the same series, which increases the difficulty of distinguishing them.

\subsection{Overall Performance Comparison}\label{overall_comp}

As presented in Table~\ref{tab:inferencetime}, the proposed model achieves the shortest inference time while maintaining a moderate storage requirement.
\mmj{The proposed approach achieves an inference time of 0.33 ms, outperforming OpenMax and KNN. 
KNN has the highest inference cost because it computes distances between the features of the received packet and the stored training features during nearest-neighbour searches~\cite{knn2021disadvantage}. 
Despite incorporating an AE for reconstruction-based detection, the proposed approach achieves an inference time comparable to that of ELogitNorm.
% The KNN algorithm exhibits significant storage and inference time, as it necessitates retaining the entire training dataset and performing distance computations against all data points during the inference~\cite{knn2021disadvantage}.
% OpenMax exhibits minimal storage requirements, as it primarily focuses on probability recalibration. It maintains only a limited set of parameters, such as MAVs and Weibull distribution parameters, making it an efficient and lightweight post-processing module that can be seamlessly integrated with the feature extractor.
Regarding storage, all methods employ the same feature extractor, which occupies 193.75 MB. The differences arise from their identification and classification components. KNN has the largest storage requirement of 78.2 MB because it retains training features as references for inference. 
The proposed approach requires 18.18~MB, primarily due to the AE parameters. 
OpenMax and ELogitNorm require only 0.16 MB and 0.1 MB, respectively, as OpenMax uses a small set of parameters for probability recalibration, and ELogitNorm makes decisions by comparing the maximum logit with a predefined threshold.}
% OpenMax performs probability recalibration using class mean activation vectors and fitted Weibull parameters, without retaining individual training features. 
% ELogitNorm modifies the classifier training objective and uses the resulting logits for detection, requiring no additional detection network. Thus, the proposed approach introduces a moderate storage overhead while maintaining competitive inference efficiency.
% OpenMax focuses on probability recalibration maintain only a limited set of parameters, making it a lightweight post-processing module, while ELogitNorm requires no additional detection network.
% OpenMax exhibits relatively low storage requirements, as they primarily focus on probability recalibration. They maintain only a limited set of parameters, making them efficient and lightweight post-processing modules that can be seamlessly integrated with the feature extractor.

\begin{table}[!t]
\caption{\mmj{Comparison of inference time and storage requirements for the proposed approach, KNN, OpenMax, and ELogitNorm.}}
\centering
\begin{tabular}{|c|c|c|c|}
\hline
\multirow{2}{*}{Method} &
\multirow{2}{*}{Inference Time (ms)} &
\multicolumn{2}{c|}{Storage (MB)} \\ \cline{3-4}
& &
Feature Extractor &
Identifier/Classifier \\ \hline

Ours &
\textbf{0.33} &
\multirow{4}{*}{193.75} &
18.18 \\ \cline{1-2} \cline{4-4}

KNN &
0.64 &
&
78.2 \\ \cline{1-2} \cline{4-4}

OpenMax &
0.4 &
&
0.16 \\ \cline{1-2} \cline{4-4}

ELogitNorm &
\textbf{0.33} &
&
\textbf{0.1} \\ \hline

\end{tabular}
\label{tab:inferencetime}
\end{table}

The above metrics, inference time and storage, are important for RFFI systems. Moreover, identification and classification play a critical role in open-set scenarios, as they fundamentally determine the system’s ability to detect rogue devices and classify legitimate devices.
Our proposed method demonstrates strong performance on the four aforementioned metrics, particularly in identification and classification, thereby offering a promising approach to improving the reliability and security of device authentication.
\section{Discussion}

\subsection{Outdoor and Industrial Deployment}
The performance of RFFI systems in outdoor and industrial deployment scenarios remains insufficiently investigated. These environments differ substantially from indoor settings in terms of propagation characteristics and electromagnetic interference. Specifically, outdoor deployments introduce additional channel variability, including large-scale fading, shadowing, and pronounced Doppler effects caused by user mobility. Industrial environments are characterized by rich multipath propagation due to machinery and structures, as well as significant electromagnetic interference from existing industrial communication systems. Despite their practical importance, studies of RFFI under these challenging deployment scenarios remain limited.

\subsection{Low-SNR Scenarios}
The performance of RFFI systems under low-SNR conditions requires further investigation. 
Denoising techniques and DL-based methods have been explored to improve performance in low-SNR scenarios~\cite{zhang2025tifs}.
Shi \textit{et al.}~\cite{shi2025snr} employed the squared cross-power spectral density (SCPSD) as an SNR-aware feature to reconstruct low-SNR signals towards the feature distribution of high-SNR signals, thereby improving performance under low-SNR conditions.
Jiang \textit{et al.}~\cite{jiang2023rf} proposed a method in which bivariate variational mode decomposition (VMD) enables adaptive signal decomposition for effective denoising. 
For DL-based methods, Yin \textit{et al.}~\cite{yin2025wcnc} leveraged a diffusion model to effectively restore the RFF under low-SNR scenarios.
Wu \textit{et al.}~\cite{wu2022iotj} proposed a dynamic shrinkage learning network (DSLN) with a dynamic shrinkage threshold for improving the classification accuracy under low-SNR conditions.

In the context of LoRa-based RFFI,~\cite{shen2023towards} proposed online augmentation and multi-packet inference, which improved identification accuracy from 50\% to 90\% at 10 dB.
Moreover, low-SNR conditions pose additional challenges to LoRa-based RFFI systems, as the SF is adaptively adjusted under poor channel conditions. 
According to the LoRa protocol, the SF can vary from 6 to 12 depending on channel conditions, resulting in signals with variable lengths and introducing additional challenges for DL model design~\cite{length2023}.

\section{Conclusion}\label{sec:conclusion}
In this paper, we proposed an open-set RFFI system that employed ViT models as feature extractors and integrated them with a multi-task module.
Two complementary CISs were employed as the inputs, from which two features were extracted using ViT models.
The features were concatenated into a fused feature that comprised more comprehensive device-specific information, which was subsequently fed to the multi-task module.
The multi-task module consisted of an AE for identification and an FC layer for classification. 
Extensive experiments were conducted, involving ten legitimate transmitters for classification assessment and five additional devices designated as rogue transmitters.
Firstly, the fused feature was evaluated against its two components, demonstrating an accuracy improvement of about 10\%.
Secondly, the proposed ViT was compared with the traditional ViT model and CNN, demonstrating an AUC improvement of more than 0.267 over the baseline methods.
\mmj{Thirdly, the proposed method was evaluated against KNN, OpenMax, and ELogitNorm, demonstrating the best overall performance in both identification and classification.}
Finally, experiments were conducted in various realistic wireless environments with different SNR levels, demonstrating that the proposed RFFI system achieved a classification accuracy of 91.1\% and AUC values consistently exceeding 0.72 in the 20 to 30~dB SNR scenario.
While the proposed method demonstrates promising performance, improving its lightweight design remains an important direction for future research. 
In particular, several model compression techniques~\cite{tong2021}, such as pruning, weight quantization, low-rank decomposition, and knowledge distillation, can be explored. 

\bibliographystyle{IEEEtran}
\bibliography{IEEEabrv,mybibfile}

@STRING{IEEE_J_IFS = "{IEEE} Trans. Inf. Forensics Security"}

@STRING{IEEE_J_MC = "{IEEE} Trans. Mobile Comput."}

@STRING{IEEE_J_IOT = "{IEEE} Internet Things J."}

@STRING{IEEE_J_CCN = "{IEEE} Trans. Cogn. Commun. Netw."}

@STRING{IEEE_J_JSAC = "{IEEE} J. Sel. Areas Commun."}

@STRING{IEEE_J_VT = "{IEEE} Trans. Veh. Technol."}

@STRING{IEEE_J_WCOM = "{IEEE} Trans. Wireless Commun."}

@article{shen2024federated,
  title={Federated Radio Frequency Fingerprint Identification Powered by Unsupervised Contrastive Learning},
  author={Shen, Guanxiong and Zhang, Junqing and Wang, Xuyu and Mao, Shiwen},
  journal=IEEE_J_IFS,
  volume = {19},
  year={2024},
  pages ={9204 -- 9215}
}

@ARTICLE{tong2021,
  author={Jian, Tong and Gong, Yifan and Zhan, Zheng and Shi, Runbin and Soltani, Nasim and Wang, Zifeng and Dy, Jennifer and Chowdhury, Kaushik and Wang, Yanzhi and Ioannidis, Stratis},
  journal=IEEE_J_MC,
  title={Radio Frequency Fingerprinting on the Edge}, 
  year={2022},
  volume={21},
  number={11},
  pages={4078-4093}}

@inproceedings{ding2026enhancing,
  title={Enhancing out-of-distribution detection with extended logit normalization},
  author={Ding, Yifan and Liu, Xixi and Unger, Jonas and Eilertsen, Gabriel}, 
  booktitle={Proc. IEEE Conf. Comput. Vis. Pattern Recognit. Workshop (CVPR)},
  pages={24823--24832},
  year={2026},
  month={Jun.},
  address={Denver, Colorado, USA}
}

@ARTICLE{zeng2024,
  author={Zeng, Yuan and Gong, Yi and Liu, Jiawei and Lin, Shangao and Han, Zidong and Cao, Ruoxiao and Huang, Kaibin and Letaief, Khaled B.},
  journal={IEEE Trans. Wireless Commun.},
  title={Multi-Channel Attentive Feature Fusion for Radio Frequency Fingerprinting}, 
  year={2024},
  volume={23},
  number={5},
  pages={4243-4254}}

@inproceedings{yang2023slnet,
  title={{SLNet}: A spectrogram learning neural network for deep wireless sensing},
  author={Yang, Zheng and Zhang, Yi and Qian, Kun and Wu, Chenshu},
  booktitle={Proc. 20th USENIX Symp. Netw. Syst. Design Implement. (NSDI)},
  pages={1221--1236},
  month={Apr.},
  year={2023},
address = {Boston, MA, USA}
}

@ARTICLE{yang2025tmc,
  author={Yang, Jian and Feng, Shuai and Wang, Yatong and Wu, Xinghang and Yan, Mu},
  journal={IEEE Trans. Mobile Comput.}, 
  title={{OpenRFI}: Open-Set Radio Frequency Fingerprint Identification via Test-Time Fine-Tuning}, 
  year={2025},
  volume={24},
  number={11},
  pages={12297-12313}}

@article{ding2025enhanced,
  title={An Enhanced Radar Sensing Algorithm With Time-Frequency Spectrum Augmentation and Activity Recognition},
  author={Ding, Minhao and Zhang, Zhixiang and Tang, Bowen and Liu, Runjin and Ding, Yipeng},
  journal=IEEE_J_IOT, 
  year={2025},
  volume={12},
  number={13},
  pages={24996-25007},
  publisher={IEEE}
}

@article{wang2025detection,
  title={Detection and recognition of {UAV} radio frequency signals based on time--frequency processing and transfer learning with multi-channel input},
  author={Wang, Shuoyang and Luo, Yongqiang and Zheng, Yifan and Sun, Zhongsen and Zheng, Yu and Liu, Tian and Jin, Zhejun},
  journal={Signal, Image and Video Processing},
  volume={19},
  number={10},
  pages={868},
  year={2025},
  publisher={Springer}
}

@ARTICLE{tang2021sensor,
  author={Tang, Longzhen and Jia, Yong and Qian, Yujia and Yi, Shi and Yuan, Peng},
  journal={IEEE Sensors Journal}, 
  title={Human Activity Recognition Based on Mixed {CNN} With Radar Multi-Spectrogram}, 
  year={2021},
  volume={21},
  number={22},
  pages={25950-25962}}

@ARTICLE{aerosp2026,
  author={J, Swarubini P and Ganapathy, Nagarajan},
  journal = {IEEE Trans. Aerosp. Electron. Syst.},
  title={Automated Human Activity Recognition Using Multiple Contactless Ultra-Wideband Radar Signals and Time–Frequency Representations}, 
  year={2026},
  volume={62},
  number={},
  pages={9213-9225}}

@INPROCEEDINGS{openmax2016cvpr,
  author={Bendale, Abhijit and Boult, Terrance E.},
  title={Towards Open Set Deep Networks}, 
  pages={1563-1572},
  booktitle={Proc. IEEE Conf. Comput. Vis. Pattern Recognit. Workshop (CVPR)},
  year={2016},
  month={Jun.},
  address={Las Vegas, Nevada, USA}
}

@article{cover1967nearest,
  title={Nearest neighbor pattern classification},
  author={Cover, Thomas and Hart, Peter},
  journal={IEEE Trans. Inf. Theory},
  volume={13},
  number={1},
  pages={21--27},
  year={1967},
  publisher={IEEE}
}

@ARTICLE{knn2021,
  author={Younes, Hamoud and Ibrahim, Ali and Rizk, Mostafa and Valle, Maurizio},
  journal={IEEE Open J. Circuits Syst.}, 
  title={An Efficient Selection-Based {kNN} Architecture for Smart Embedded Hardware Accelerators}, 
  year={2021},
  volume={2},
  number={},
  pages={534-545}}

@ARTICLE{openmax2023,
  author={Oveis, Amir Hossein and Giusti, Elisa and Ghio, Selenia and Martorella, Marco},
  journal={IEEE Trans. Aerosp. Electron.}, 
  title={Extended Openmax Approach for the Classification of Radar Images With a Rejection Option}, 
  year={2023},
  volume={59},
  number={1},
  pages={196-208}}

@article{bisgin2024large,
  title={Large-scale evaluation of open-set image classification techniques},
  author={Bisgin, Halil and Palechor, Andres and Suter, Mike and G{\"u}nther, Manuel},
  journal={arXiv preprint arXiv:2406.09112},
  year={2024}
}

@ARTICLE{zhao2023,
  author={Zhang, Shichao and Li, Jiaye and Li, Yangding},
  journal={IEEE Trans. Knowl. Data Eng.}, 
  title={Reachable Distance Function for {KNN} Classification}, 
  year={2023},
  volume={35},
  number={7},
  pages={7382-7396}}

@ARTICLE{guo2024tccn,
  author={Guo, Lantu and Liu, Chao and Liu, Yuchao and Lin, Yun and Gui, Guan},
  journal=IEEE_J_CCN,
  title={Toward Open-Set Specific Emitter Identification Using Auxiliary Classifier Generative Adversarial Network and OpenMax}, 
  year={2024},
  volume={10},
  number={6},
  pages={2019-2028}}

@ARTICLE{stft2021tradeoff,
  author={Marafioti, Andrés and Holighaus, Nicki and Majdak, Piotr},
  journal={IEEE Trans. Signal Process.}, 
  title={Time-Frequency Phase Retrieval for Audio—The Effect of Transform Parameters}, 
  year={2021},
  volume={69},
  number={},
  pages={3585-3596}}

@inproceedings{wang2024design,
  title={Design of noise robust open-set radio frequency fingerprint identification method},
  author={Wang, Min and Peng, Linning and Xie, Lingnan and Zhang, Junqing and Liu, Ming and Fu, Hua},
  booktitle={Proc. IEEE Int. Conf. Comput. Commun. (INFOCOM)},
  pages={1--6},
  year={2024},
  month={Jul.},
  address={Vancouver, Canada}
}

@INPROCEEDINGS{yin2025wcnc,
  author={Yin, Guolin and Zhang, Junqing and Ding, Yuan and Cotton, Simon},
  booktitle={Proc. IEEE Wirel. Commun. and Netw. Conf. (WCNC)},
  address={Milan, Italy},
  month=mar,
  title={Noise-Robust Radio Frequency Fingerprint Identification Using Denoise Diffusion Model}, 
  year={2025},
  pages={1-6}}

@article{shi2025snr,
  title={An {SNR-Aware} Feature Reconstruction Method in Radio Frequency Fingerprint Identification},
  author={Shi, Junxian and Peng, Linning and Xie, Lingnan and Fu, Hua and Hu, Aiqun},
  journal=IEEE_J_IFS,
  year={2025},
  publisher={IEEE}
}

@INPROCEEDINGS{zhao2024explanation,
  author={Zhao, Tianya and Wang, Xuyu and Zhang, Junqing and Mao, Shiwen},
  booktitle={Proc. IEEE Int. Conf. Comput. Commun. (INFOCOM)}, 
  title={Explanation-Guided Backdoor Attacks on Model-Agnostic {RF} Fingerprinting}, 
  year={2024},
  pages={221-230},
  month={May.},
  address={Vancouver, Canada}}

@inproceedings{xie2025infocom,
  title={Towards Robust {RF} Fingerprint Identification Using Spectral Regrowth and Carrier Frequency Offset},
  author={Xie, Lingnan and Peng, Linning and Zhang, Junqing},
  booktitle={Proc. IEEE Int. Conf. Comput. Commun. (INFOCOM)},
  pages={1--6},
  year={2025},
  month={May.},
  address={London, UK}
}

@inproceedings{knn2021disadvantage,
author = {Ogden, Samuel S. and Kong, Xiangnan and Guo, Tian},
title = {{PieSlicer}: Dynamically Improving Response Time for Cloud-based {CNN} Inference},
year = {2021},
address = {New York, NY, USA},
booktitle = {Proc. ACM/SPEC Int. Conf. on Perf. Eng.},
pages = {249--256},
numpages = {8},
location = {Virtual Event, France}
}

@article{shen2021radio,
  title={Radio frequency fingerprint identification for {LoRa} using deep learning},
  author={Shen, Guanxiong and Zhang, Junqing and Marshall, Alan and Peng, Linning and Wang, Xianbin},
  journal = IEEE_J_JSAC,
  volume={39},
  number={8},
  pages={2604--2616},
  year={2021},
  publisher={IEEE}
}

@article{he2020cooperative,
  title={Cooperative specific emitter identification via multiple distorted receivers},
  author={He, Boxiang and Wang, Fanggang},
  journal=IEEE_J_IFS,
  volume={15},
  pages={3791--3806},
  year={2020},
  publisher={IEEE}
}

@inproceedings{fusion2019,
author = {Andrews, Seth and Gerdes, Ryan M. and Li, Ming},
title = {Crowdsourced measurements for device fingerprinting},
year = {2019},
address = {New York, USA},
booktitle = {Proc. 12th Conf. Security Privacy Wireless Mobile Netw.},
pages = {72--82},
numpages = {11},
location = {Miami, Florida}
}

@INPROCEEDINGS{yang2023infocom,
  author={Yang, Yanni and Cao, Jiannong and An, Zhenlin and Wang, Yanwen and Hu, Pengfei and Zhang, Guoming},
  booktitle={Proc. IEEE Int. Conf. Comput. Commun. (INFOCOM)}, 
  title={{NFChain}: A Practical Fingerprinting Scheme for {NFC} Tag Authentication}, 
  year={2023},
  pages={1-10},
  month={May.},
  address={New York, USA}}

@INPROCEEDINGS{al2020infocom,
  author={Al-Shawabka, Amani and Restuccia, Francesco and D’Oro, Salvatore and Jian, Tong and Costa Rendon, Bruno and Soltani, Nasim and Dy, Jennifer and Ioannidis, Stratis and Chowdhury, Kaushik and Melodia, Tommaso},
  booktitle={Proc. IEEE Int. Conf. Comput. Commun. (INFOCOM)}, 
  title={Exposing the Fingerprint: Dissecting the Impact of the Wireless Channel on Radio Fingerprinting}, 
  year={2020},
  pages={646-655},
  month={July.},
  address={virtual}}

@ARTICLE{fusion2022iotj,
  author={Zhang, Weifeng and Zhao, Wenhong and Tan, Xiaoheng and Shao, Liudong and Ran, Chuan},
  journal=IEEE_J_IOT, 
  title={Adaptive {RF} Fingerprints Fusion via Dual Attention Convolutions}, 
  year={2022},
  volume={9},
  number={24},
  pages={25181-25195}}

@article{he2025adaptive,
  title={Adaptive Channel-Robust Signal Fusion Method for Collaborative Device Fingerprint Identification},
  author={He, Jiashuo and Jiang, Weiwei and Tang, Qinxue and Wang, Yumeng and Chang, Shuo and Huang, Sai and Hao, Caiyong and Feng, Zhiyong},
  journal={Electronics Letters},
  volume={61},
  number={1},
  pages={e70319},
  year={2025},
  publisher={Wiley Online Library}
}

@article{zhang2024multisource,
  title={Multisource heterogeneous specific emitter identification using attention mechanism-based {RFF} fusion method},
  author={Zhang, Yibin and Zhang, Qianyun and Zhao, Haitao and Lin, Yun and Gui, Guan and Sari, Hikmet},
  journal=IEEE_J_IFS,
  volume={19},
  pages={2639--2650},
  year={2024},
  publisher={IEEE}
}

@ARTICLE{gan2024tim,
  author={Gan, Jiayan and Li, Qiang and Shao, Huaizong and Wen, Zhongyi and Yang, Tao and Pan, Ye and Sun, Guomin},
  journal={IEEE Trans. Instrum. Meas.}, 
  title={A {Z}ynq-Based Platform With Conditional-Reconfigurable Complex-Valued Neural Network for Specific Emitter Identification}, 
  year={2024},
  volume={73},
  number={},
  pages={1-11}}

@article{zhang2023real,
  title={Real-world aircraft recognition based on {RF} fingerprinting with few Labeled {ADS-B} signals},
  author={Zhang, Zechen and Li, Guyue and Shi, Jitong and Li, Haobo and Hu, Aiqun},
  journal=IEEE_J_VT,
  volume={73},
  number={2},
  pages={2866--2871},
  year={2023},
  publisher={IEEE}
}

@ARTICLE{zhang2025tifs,
  author={Zhang, Junqing and Ardizzon, Francesco and Piana, Mattia and Shen, Guanxiong and Tomasin, Stefano},
  journal=IEEE_J_IFS,
  title={Physical Layer-Based Device Fingerprinting For Wireless Security: From Theory To Practice}, 
  year={2025},
  volume={20},
  pages={5296 - 5325}}

@inproceedings{mehta2022mobilevit,
     title={{MobileViT}: Light-weight, General-purpose, and Mobile-friendly Vision Transformer},
     author={Sachin Mehta and Mohammad Rastegari},
  booktitle={Proc. Int. Conf. Lrn. Repr. (ICLR)},
     year={2022},
  address={virtual},
  month={Apr.},
}

@ARTICLE{xi2025,
  author={Xi, Bobo and Zhang, Wenjie and Li, Jiaojiao and Song, Rui and Li, Yunsong},
  journal={IEEE Trans. Image Process.}, 
  title={{HyperCASR}: Spectral-Spatial Open-Set Recognition With Category-Aware Semantic Reconstruction for Hyperspectral Imagery}, 
  year={2025},
  volume={34},
  number={},
  pages={7642-7655}}

@article{jiang2023rf,
  title={{RF} fingerprinting identification in low {SNR} scenarios for automatic identification system},
  author={Jiang, Qi and Sha, Jin},
  journal= IEEE_J_WCOM,
  volume={23},
  number={3},
  pages={2070--2081},
  year={2023},
  publisher={IEEE}
}

@article{shen2023towards,
  title={Towards receiver-agnostic and collaborative radio frequency fingerprint identification},
  author={Shen, Guanxiong and Zhang, Junqing and Marshall, Alan and Woods, Roger and Cavallaro, Joseph and Chen, Liquan},
  journal=IEEE_J_MC,
  year={2024},
  volume={23},
  number={7},
  pages={7618-7634},
  publisher={IEEE}
}

@article{shen2022towards,
  title={Towards scalable and channel-robust radio frequency fingerprint identification for {LoRa}},
  author={Shen, Guanxiong and Zhang, Junqing and Marshall, Alan and Cavallaro, Joseph R},
  journal=IEEE_J_IFS,
  volume={17},
  pages={774--787},
  year={2022},
  publisher={IEEE}
}

@ARTICLE{li2025jsac,
  author={Li, Kunling and Bao, Jiazhong and Xie, Xin and Hong, Jianan and Hua, Cunqing},
  journal=IEEE_J_JSAC,
  title={Receiver-Agnostic Radio Frequency Fingerprint Identification for Zero-Trust Wireless Networks}, 
  year={2025},
  volume={},
  number={},
  pages={1-1}}

@ARTICLE{hanna2021tccn,
  author={Hanna, Samer and Karunaratne, Samurdhi and Cabric, Danijela},
  journal=IEEE_J_CCN,
  title={Open Set Wireless Transmitter Authorization: Deep Learning Approaches and Dataset Considerations}, 
  year={2021},
  volume={7},
  number={1},
  pages={59-72}}

@inproceedings{dosovitskiy2020vit,
  title={An Image is Worth 16x16 Words: Transformers for Image Recognition at Scale},
  author={Dosovitskiy, Alexey and Beyer, Lucas and Kolesnikov, Alexander and Weissenborn, Dirk and Zhai, Xiaohua and Unterthiner, Thomas and  Dehghani, Mostafa and Minderer, Matthias and Heigold, Georg and Gelly, Sylvain and Uszkoreit, Jakob and Houlsby, Neil},
  booktitle={Proc. Int. Conf. Lrn. Repr. (ICLR)},
  address={virtual},
  month={May.},
  year={2021}
}

@article{deng2023lightweight,
  title={A lightweight transformer-based approach of specific emitter identification for the automatic identification system},
  author={Deng, Pengfei and Hong, Shaohua and Qi, Jie and Wang, Lin and Sun, Haixin},
  journal=IEEE_J_IFS,
  volume={18},
  pages={2303--2317},
  year={2023},
  publisher={IEEE}
}

@inproceedings{pooling2022,
title = {No More Strided Convolutions or Pooling: A New {CNN} Building Block for Low-Resolution Images and Small Objects},
author = {Sunkara, Raja and Luo, Tie},
booktitle={Proc. Eur. Conf. Mach. Learn. Princip. Pract. Knowl. Discov. Databases (ECML PKDD)},
month = {Sept.},
year = {2022},
address = {Grenoble, France},
pages = {443–459},
numpages = {17},
}

@ARTICLE{wu2022iotj,
  author={Wu, Weiwei and Hu, Su and Lin, Di and Liu, Zilong},
  journal=IEEE_J_IOT, 
  title={{DSLN}: Securing {Internet} of {Things} Through {RF} Fingerprint Recognition in Low-{SNR} Settings}, 
  year={2022},
  volume={9},
  number={5},
  pages={3838-3849}}

@ARTICLE{zhang2021tifs,
  author={Zhang, Junqing and Woods, Roger and Sandell, Magnus and Valkama, Mikko and Marshall, Alan and Cavallaro, Joseph},
  journal=IEEE_J_IFS, 
  title={Radio Frequency Fingerprint Identification for Narrowband Systems, Modelling and Classification}, 
  year={2021},
  volume={16},
  number={},
  pages={3974-3987},}

@inproceedings{chen2021crossvit,
  title={Crossvit: Cross-attention multi-scale vision transformer for image classification},
  author={Chen, Chun-Fu Richard and Fan, Quanfu and Panda, Rameswar},
  booktitle={Proc. Int. Conf. Comp. Vis. (ICCV)},
  address={Montreal, BC, Canada},
  pages={7868--7877},
  month={Oct.},
  year={2021}
}

@inproceedings{rfdna,
author = {Pan, Qingrui and An, Zhenlin and Yang, Xueyuan and Zhao, Xiaopeng and Yang, Lei},
title = {{RF-DNA}: large-scale physical-layer identifications of {RFIDs} via dual natural attributes},
year = {2022},
booktitle = {Proc. Int. Conf. on Mobile Comput. Netw. (MobiCom)},
pages = {419–431},
numpages = {13},
location = {Sydney, NSW, Australia}
}

@inproceedings{parpart2024transformer,
  title={Transformer Masked Autoencoders for {RF} Device Fingerprinting},
  author={Parpart, Gavin and Tu, Jonathan H and Clymer, Bradley and Lee, Jung and Babcock, Jasen},
  booktitle={IEEE Mil. Commun. Conf. (MILCOM)},
  pages={859--862},
  month={Oct.},
  year={2024},
  address={Washington, DC, USA},
}

@ARTICLE{length2023,
  author={Shen, Guanxiong and Zhang, Junqing and Marshall, Alan and Valkama, Mikko and Cavallaro, Joseph R.},
  journal=IEEE_J_IFS,
  title={Toward Length-Versatile and Noise-Robust Radio Frequency Fingerprint Identification}, 
  year={2023},
  volume={18},
  number={},
  pages={2355-2367},}

@ARTICLE{rmse2020,
  author={Essien, Aniekan and Giannetti, Cinzia},
  journal={IEEE Trans. Ind. Informatics}, 
  title={A Deep Learning Model for Smart Manufacturing Using Convolutional {LSTM} Neural Network Autoencoders}, 
  year={2020},
  volume={16},
  number={9},
  pages={6069-6078}}

@article{zhang2020new,
  title={A new frontier for {IoT} security emerging from three decades of key generation relying on wireless channels},
  author={Zhang, Junqing and Li, Guyue and Marshall, Alan and Hu, Aiqun and Hanzo, Lajos},
  journal={IEEE Access},
  volume={8},
  pages={138406--138446},
  year={2020}
}
\begin{IEEEbiography}[{\includegraphics[width=1in,height=1.25in,clip,keepaspectratio]{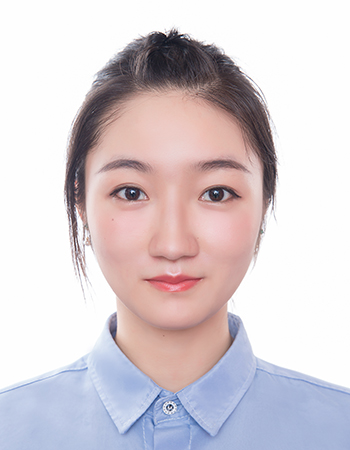}}]{Jie Ma} (Member, IEEE) received the B.Eng. degree from Harbin Engineering University in 2017, the M.Eng. degree from the Beijing University of Posts and Telecommunications in 2020, and the Ph.D. degree in Electrical and Electronic Engineering from the University of Liverpool in 2025. 
She is currently a Research Associate at Newcastle University. Her research interests include wireless communications, Internet of Things, deep learning, and health AI.
\end{IEEEbiography}

\begin{IEEEbiography}
[{\includegraphics[width=1in,height=1.25in,clip,keepaspectratio]{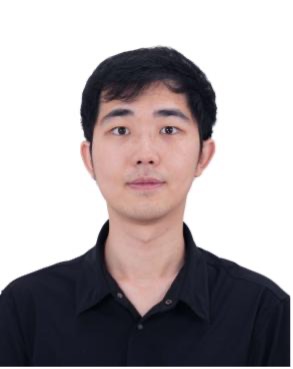}}]{Guanxiong Shen} (Member, IEEE) is an Associate Professor at Southeast University, China. He received the B.Eng. degree from Xidian University, China, in 2019, and the Ph.D. degree from the Department of Electrical Engineering and Electronics at the University of Liverpool, UK, in 2023. He also holds a bachelor’s degree from Heriot-Watt University, UK. His research interests lie at the intersection of wireless communications, artificial intelligence, and cybersecurity, in particular the application of artificial intelligence to enhance wireless networking and physical layer security.
\end{IEEEbiography}

\begin{IEEEbiography}
[{\includegraphics[width=1in,height=1.25in,clip,keepaspectratio]{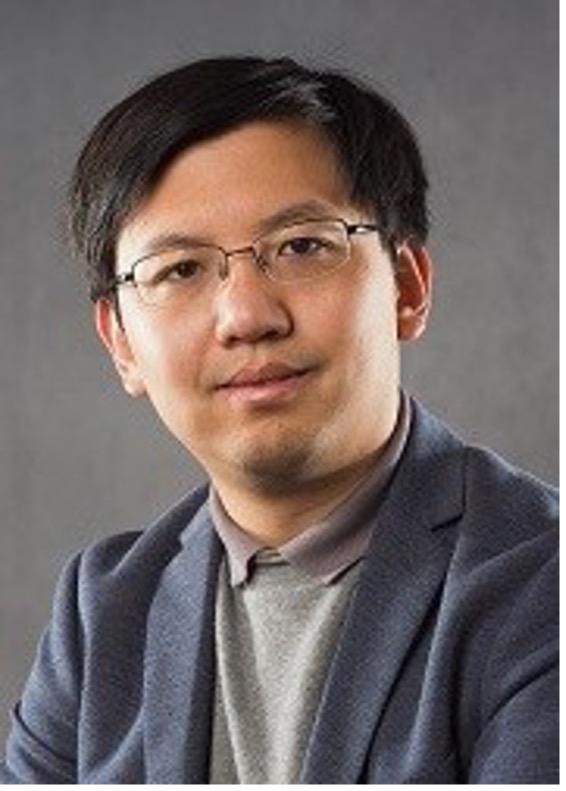}}]{Linning Peng} (Member, IEEE) received the received the B.Eng. and M.Eng. degrees from the Southeast University, Nanjing, China, in 2006 and 2010, respectively, and the Ph.D. degree from the Electronics and Telecommunications Institute of Rennes (IETR) Laboratory, National Institute of Applied Sciences (INSA), Rennes, France, in 2014. He is an Associate Professor with Southeast University, Nanjing, China, and also with the Purple Mountain Laboratories, Nanjing, China. He is ranked as the World's Top 2\% most-cited scientists 2023-2026 by Stanford University. His research interests include Internet of Things and physical layer security.
\end{IEEEbiography}

\begin{IEEEbiography}
[{\includegraphics[width=1in,height=1.25in,clip,keepaspectratio]{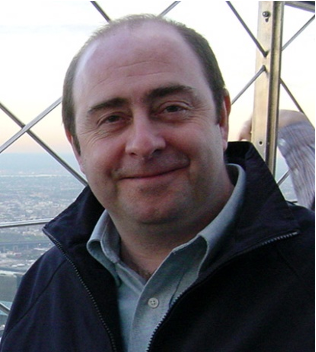}}]{Alan Marshall} (Senior Member, IEEE) holds the chair in Communications Networks at the University of Liverpool in the school where he is also director of the Advanced Networks Group.  He is a senior member of IEEE and a Fellow of the IET. He has spent over 28 years working in the Telecommunications and Defense Industries, and in Academia. He was involved in the development and standardization of 2G and 3G cellular systems throughout Europe. He has published over 300 scientific papers and holds a number of joint patents in the areas of communications and network security. He formed a successful spin-out company specializing in intrusion detection \& prevention for wireless networks.  His research interests include mobile and wireless network architectures and protocols, network security and multisensory communications including haptics and olfaction. He is the UK secretariat for the “UK-Jiangsu World class Universities” initiative. 
\end{IEEEbiography}
	
\end{document}